\documentclass[pdflatex,sn-apa]{sn-jnl}% APA Reference Style
\usepackage{graphicx}%
\usepackage{multirow}%
\usepackage{amsmath,amssymb,amsfonts}%
\usepackage{amsthm}%
\usepackage{mathrsfs}%
\usepackage[title]{appendix}%
\usepackage{xcolor}%
\usepackage{textcomp}%
\usepackage{manyfoot}%
\usepackage{booktabs}%
\usepackage{algorithm}%
\usepackage{algorithmicx}%
\usepackage{algpseudocode}%
\usepackage{listings}%
\usepackage{subcaption}
\theoremstyle{thmstyleone}%
\theoremstyle{thmstyletwo}%

\theoremstyle{thmstylethree}%
\newcommand{\expSetupImageWidth}{0.44}

\begin{document}

\title[The Interplay of Attention and Memory in Visual Enumeration]{The Interplay of Attention and Memory in \\ Visual Enumeration}
% \title[Enumeration of Objects over Large Visual Fields]{Enumeration of Objects over Large Visual Fields}


\author[1]{\fnm{B.} \sur{Sankar}}\email{sankarb@iisc.ac.in}
\equalcont{These authors contributed equally to this work.}

\author[2]{\fnm{Devottama} \sur{Sen}}\email{sendevottama@gmail.com}
\equalcont{These authors contributed equally to this work.}

\author[2]{\fnm{Dibakar} \sur{Sen}}\email{dibakar@iisc.ac.in}

\affil[1]{\orgdiv{Department of Mechanical Engineering}, \orgname{Indian Institute of Science (IISc)}, \orgaddress{\street{CV Raman Road}, \city{Bangalore}, \postcode{560012}, \state{Karnataka}, \country{India}}}

% \affil[2]{\orgdiv{Department}, \orgname{Organization}, \orgaddress{\street{Street}, \city{City}, \postcode{10587}, \state{State}, \country{Country}}}

\affil[3]{\orgdiv{Department of Design and Manufacturing (erstwhile CPDM)}, \orgname{Organization}, \orgaddress{\street{CV Raman Road}, \city{Bangalore}, \postcode{560012}, \state{Karnataka}, \country{India}}}


\abstract{Humans navigate and understand complex visual environments by subconsciously quantifying what they see, a process known as visual enumeration. However, traditional studies using flat screens fail to capture the cognitive dynamics of this process over the large visual fields of real-world scenes. To address this gap, we developed an immersive virtual reality system with integrated eye-tracking to investigate the interplay between attention and memory during complex enumeration. We conducted a two-phase experiment where participants enumerated scenes of either simple abstract shapes or complex real-world objects, systematically varying the task intent (e.g., selective vs. exhaustive counting) and the spatial layout of items. Our results reveal that task intent is the dominant factor driving performance, with selective counting imposing a significant cognitive cost that was dramatically amplified by stimulus complexity. The semantic processing required for real-world objects reduced accuracy and suppressed memory recall, while the influence of spatial layout was secondary and statistically non-significant when higher-order cognitive task intent was driving the human behaviour. We conclude that real-world enumeration is fundamentally constrained by the cognitive load of semantic processing, not just the mechanics of visual search. Our findings demonstrate that under high cognitive demand, the effort to understand what we are seeing directly limits our capacity to remember it.}




\maketitle

\section{Introduction}

\subsection{The Primacy of Vision and the Numerical World}

The human experience is fundamentally mediated through vision. Our eyes serve as the primary gateway through which we perceive, relate to, and ultimately understand the environment around us. It is through this vital sensory channel that we gather the information necessary to make countless decisions based on the counts, navigating the complexities of the world by interpreting the objects and scenes we encounter. A deeper examination of this perceptual process reveals a fascinating and often subconscious mechanism: we see the world through numbers. This numerical lens is not always applied consciously; rather, it operates as an underlying framework for comprehension.

\begin{figure}[ht!]
    \centering
    \includegraphics[width=0.85\textwidth]{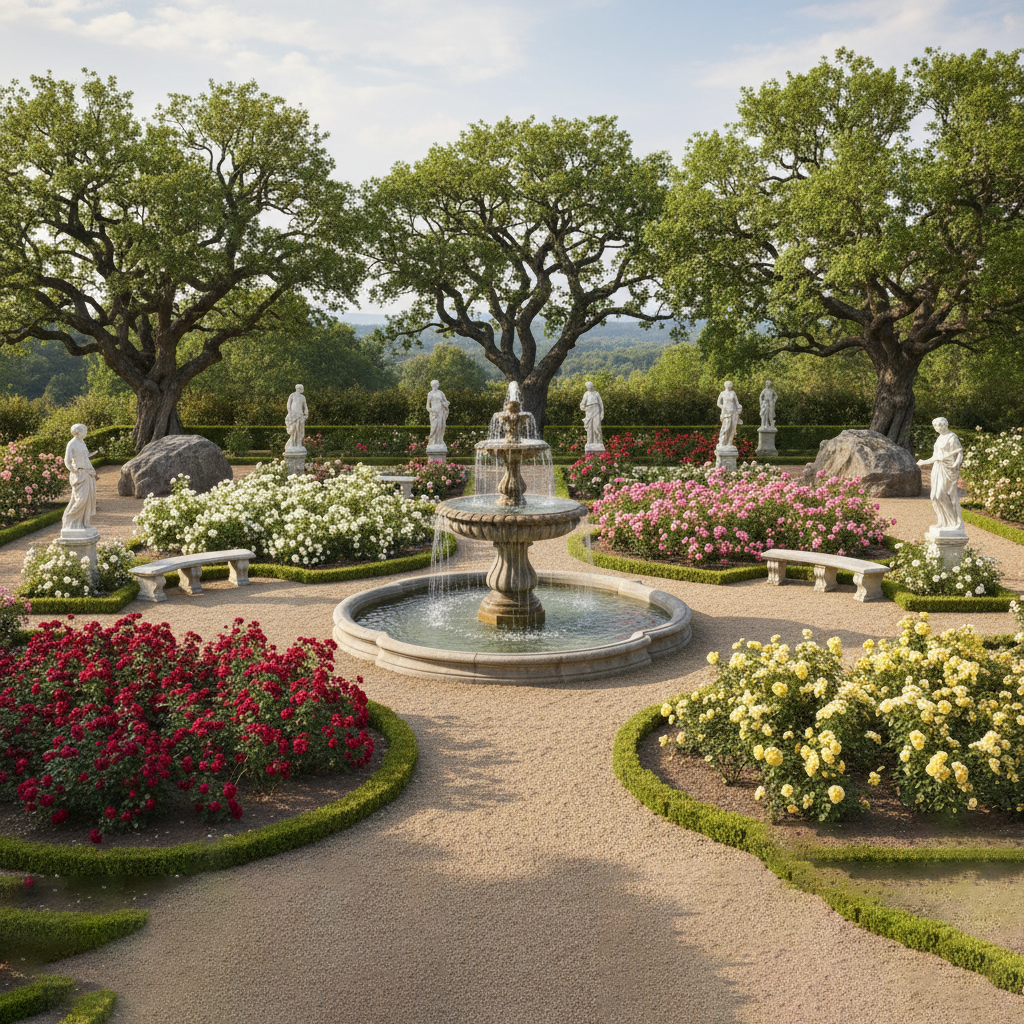}
    \caption{A typical real-world scene, a formal garden, which humans subconsciously quantify to build a mental model. Describing this scene naturally involves enumerating key features like fountains, benches, and statues.}
    \label{fig:garden}
\end{figure}

Consider, for instance, the task of a person describing a garden visited by him/her on a pleasant afternoon, as depicted in Figure~\ref{fig:garden}. An initial description might focus on qualitative aspects—its beauty, the sense of tranquillity, the fragrance of the flowers. However, a more detailed and concrete description inevitably gravitates towards quantification. One might recall that there was a \textit{single}, central fountain, surrounded by \textit{four} distinct plots of rose plantation of yellow, white, maroon, and pink colours. The description might continue by noting the presence of \textit{two} large rocks, \textit{three} stone benches, and \textit{perhaps five to seven} mature oak trees, and \textit{about seven to ten} classical statues arranged throughout the landscape. 

This seemingly simple recollection highlights a profound cognitive process. The mind, even when not engaged in a deliberate act of counting, is continuously performing a form of subconscious enumeration to structure and make sense of the visual input. The description itself reveals nuances in this process. Certain quantities, like the '1' fountain, '4' rose plots, and '3' benches, are recalled with a high degree of firmness and confidence. In contrast, the quantities of trees and statues are expressed as ranges—'5 to 7' and '8 to 10'—indicating a degree of uncertainty or approximation. This distinction between precise, instant recognition and effortful cognitive estimation points towards the different mechanisms of visual enumeration, the process by which we determine the number of items in a scene using our eyes. 

The ability to perceive and quantify the number of objects in a scene is not merely a mathematical exercise but a foundational cognitive skill that shapes our interaction with the world. It is an act that people perform almost instinctively, irrespective of their formal education, suggesting it is deeply ingrained in their memory. In everyday activities, such as counting vehicles at a junction, fruit in a basket, or screws on a workbench, these mechanisms operate seamlessly on rapid timescales yet remain sensitive to factors like clutter, object similarity, and time pressure.  While the act of counting may appear straightforward, it is psychologically supported not by a single faculty but through a complex interplay of various cognitive mechanisms (\cite{Feigenson2004-uh, Piazza2002-lu, Ansari2007-mm, Dehaene1992-xz, Trick1994}). While visual perception provides the initial sensory input, enumeration relies heavily on working memory to maintain and iteratively update a mental tally of the counted items. Furthermore, visual attention is pivotal, serving to serially guide the gaze from one object to the next, ensuring that items are not missed or double-counted. The ability to mentally index each object after inspection is important in achieving an accurate final count. This perspective underscores that enumeration is not a simple action but rather a sophisticated cognitive task that requires the seamless integration of perception, attention, and memory.

\subsection{The Two Modes of Visual Enumeration: Subitizing and Counting}

\begin{figure}[ht!]
    \centering
    % Creating subfigures using the subcaption package
    \begin{subfigure}[]{0.32\textwidth}
        \centering
        \includegraphics[width=\textwidth]{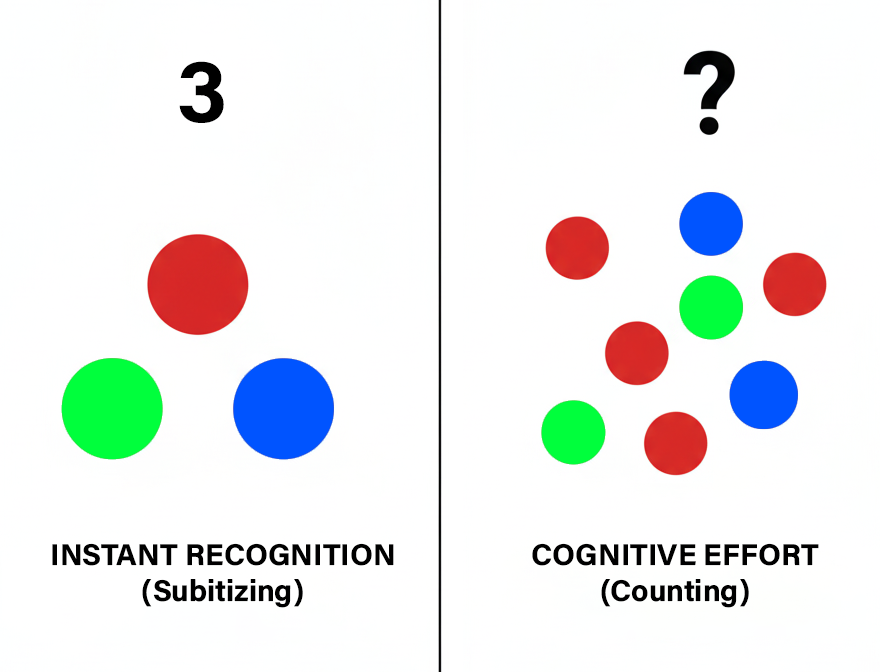}
        \caption{An illustration of subitizing versus counting using simple coloured dots. A small quantity (left) is instantly recognised as '3', while a larger, more cluttered set (right) requires deliberate, effortful counting to determine the exact number.}
        \label{fig:dots}
    \end{subfigure}
    \hfill
    \begin{subfigure}[]{0.32\textwidth}
        \centering
        \includegraphics[width=\textwidth]{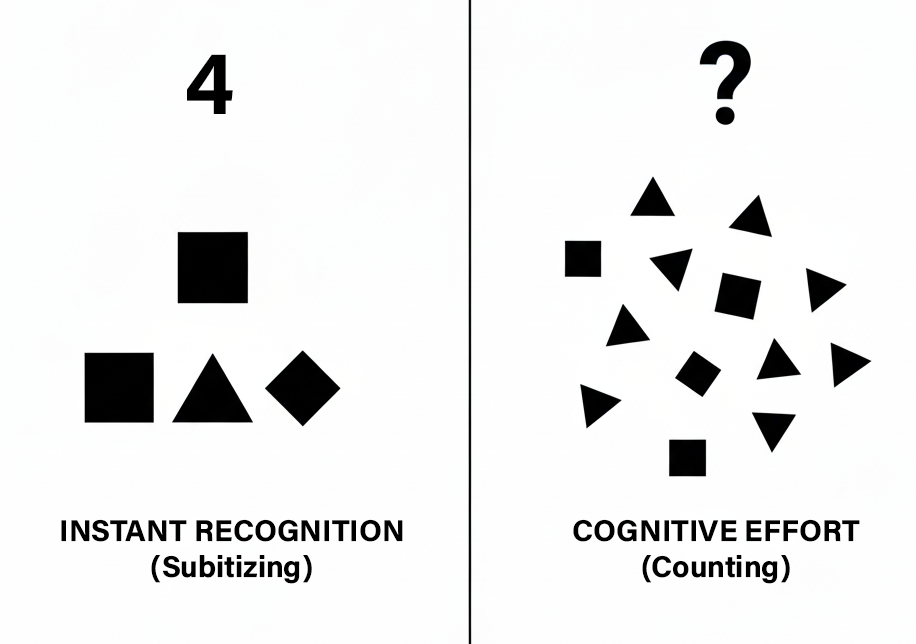}
        \caption{Enumeration complexity based on shape characteristics. The distinct shapes on the left are easily subitized as '4', whereas the jumbled collection on the right requires a more focused counting strategy to differentiate and tally each shape.}
        \label{fig:shapes}
    \end{subfigure}
    \hfill
    \begin{subfigure}[]{0.32\textwidth}
        \centering
        \includegraphics[width=\textwidth]{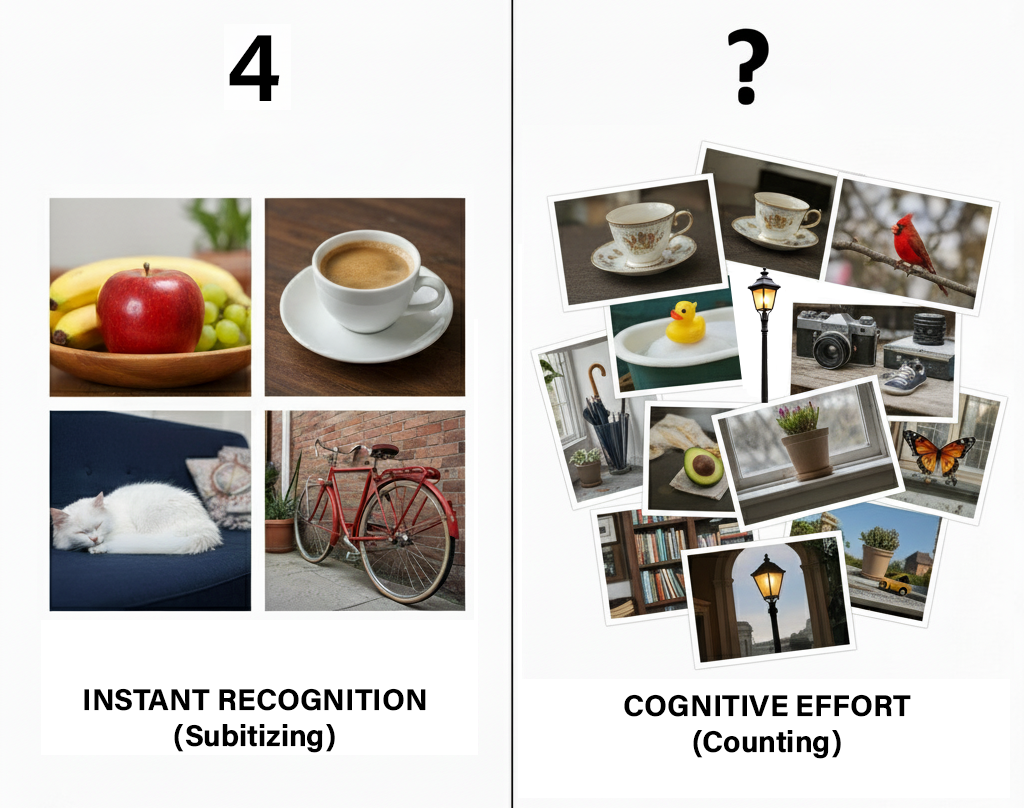}
        \caption{Enumeration of complex, real-world objects. While four distinct objects (left) can be subitized, a larger, overlapping collection (right) demands significant cognitive effort, involving not just counting but also object recognition and visual search.}
        \label{fig:objects}
    \end{subfigure}
    
    % Adding a common caption and label for the entire figure
    \caption{Comparison of subitizing and counting across different types of visual stimuli, demonstrating the transition from rapid enumeration to effortful counting as complexity increases.}
    \label{fig:subitizing_comparison}
\end{figure}

The field of cognitive science has long recognized that visual enumeration is not a monolithic process. Instead, it operates via two distinct modes, determined primarily by the number of items being observed. For small quantities, typically up to four or five items, humans can ascertain the number almost instantly and without conscious effort. This rapid, accurate, and seemingly pre-attentive process is known as \textit{subitizing}. As illustrated in the left panels of Figures~\ref{fig:dots}, \ref{fig:shapes}, and \ref{fig:objects}, whether the items are simple coloured dots, geometric shapes, or complex real-world objects, a small set allows for immediate recognition of its cardinality. During subitizing, the brain appears to process the items in parallel, apprehending the total quantity as a perceptual whole rather than as a sum of individual units based on the Gestalt's law of psychology (Law of Proximity, Similarity, Closure, Continuity, Symmetry \& Order, Common Region, Focal Point). This is why the numbers recalled with firmness in our garden example (1, 3, 4) fall squarely within this subitizing range.

% Setting up the figure environment for subfigures

% \begin{figure}[ht!]
%     \centering
%     \includegraphics[width=0.85\textwidth]{Figures/SC_Colours.png}
%     \caption{An illustration of subitizing versus counting using simple coloured dots. A small quantity (left) is instantly recognised as '3', while a larger, more cluttered set (right) requires deliberate, effortful counting to determine the exact number.}
%     \label{fig:dots}
% \end{figure}

However, as the number of items increases beyond this limited threshold, a cognitive shift occurs. The effortless process of subitizing gives way to a more deliberate and cognitively demanding procedure: \textbf{counting}. Counting is a serial process that requires the deployment of focused attention, which is sequentially shifted from one item to the next. Each item must be individually attended to, identified as a distinct unit, and added to a running tally maintained in working memory. This process is inherently slower, more effortful, and more prone to errors, such as missing an item or counting one item twice. The right panels of Figures~\ref{fig:dots}, \ref{fig:shapes}, and \ref{fig:objects} depict scenarios where the quantity of items surpasses the subitizing range, necessitating this application of cognitive effort. Literature suggests this transition point generally occurs around five items. This distinction is fundamental; while subitizing is a form of pattern recognition, counting is a methodical algorithm involving attention, memory, and sequential processing.

\subsection{Factors Influencing Enumeration Complexity}
The transition from subitizing to counting marks a significant increase in cognitive load due to the sheer complexity involved. However, the difficulty of an enumeration task is not determined solely by the number of items. A confluence of factors related to the items themselves, the observer's goal, and the structure of the scene can dramatically alter the strategy, efficiency, and accuracy of counting. We identify three primary axes of complexity: the intrinsic characteristics of the objects, the intent of the counting task, and the spatial distribution of the items.

\subsubsection{Object Characteristics}
Enumerating a set of simple, homogenous items is cognitively less demanding than counting a set of complex, heterogeneous objects. The three comparative figures illustrate this principle. In Figure~\ref{fig:dots}, the primary variation among the items is colour. In Figure~\ref{fig:shapes}, the variation is based on geometric form (squares, triangles, rhombi). Finally, in Figure~\ref{fig:objects}, the items are photographs of real-world objects, exhibiting immense variation in shape, colour, texture, and semantic meaning. As we move from simple dots to complex objects, the initial step of individuating each item, distinguishing it from its neighbours and the background, becomes more challenging. For real-world objects, this also involves an act of recognition and categorization, adding another layer of cognitive processing that must occur before the item can be added to the mental tally.

% \begin{figure}[ht!]
%     \centering
%     \includegraphics[width=0.85\textwidth]{Figures/SC_Shapes.png}
%     \caption{Enumeration complexity based on shape characteristics. The distinct shapes on the left are easily subitized as '4', whereas the jumbled collection on the right requires a more focused counting strategy to differentiate and tally each shape.}
%     \label{fig:shapes}
% \end{figure}

\subsubsection{Task Intent}
The goal or intent of the observer profoundly affects the way they enumerate a scene. The simplest intent is to count all items present, as suggested by the examples so far . However, in many real-world scenarios, the task is more nuanced. We are often required to perform selective counting based on specific criteria. This can take the form of an \textit{inclusion criterion} (e.g., "count only the apples in a fruit basket") or an \textit{exclusion criterion} (e.g., "count all the fruits except the apples"). Such tasks transform the enumeration process into a form of visual search. The observer cannot simply move their gaze from one item to the next; for each item, they must make a decision: "Does this item meet the criterion for inclusion in my count?". This decision-making process adds significant cognitive load, requiring not just attention and working memory for the tally, but also executive functions for filtering and inhibition. A familiar analogy is searching for a specific book on a cluttered shelf; our intent is to scan and reject items until we find the one that matches our target. This goal-directed filtering fundamentally alters the attentional strategies employed compared to a simple, exhaustive count.

% \begin{figure}[ht!]
%     \centering
%     \includegraphics[width=0.85\textwidth]{Figures/SC_Images.png}
%     \caption{Enumeration of complex, real-world objects. While four distinct objects (left) can be subitized, a larger, overlapping collection (right) demands significant cognitive effort, involving not just counting but also object recognition and visual search.}
%     \label{fig:objects}
% \end{figure}

\subsubsection{Spatial Distribution}
Finally, the organization or spatial distribution of items within a scene is a major determinant of the counting strategy and its resulting efficiency. Humans, by nature, tend to seek the path of least cognitive resistance, adopting strategies that allow them to count as quickly and accurately as possible. A structured arrangement, such as a grid, is one of the easiest distributions to enumerate. An observer can quickly determine the number of rows and columns and use multiplication to arrive at the total, often bypassing the need for serial, item-by-item counting altogether. However, the effectiveness of such a strategy is itself impacted by the item characteristics and the task intent. If the task is to count only the red items in a multi-coloured grid, the structural advantage is diminished. Conversely, a random, unstructured distribution offers no such shortcuts, forcing the observer to develop a more methodical scanning path to ensure all items are covered without repetition. The layout of a scene can either facilitate an efficient, systematic strategy or demand a more taxing and error-prone search.

\subsection{Research Gap: From Static Central Vision to Dynamic Large-Field Search}

\begin{figure}[ht!]
    \centering
    \includegraphics[width=0.85\textwidth]{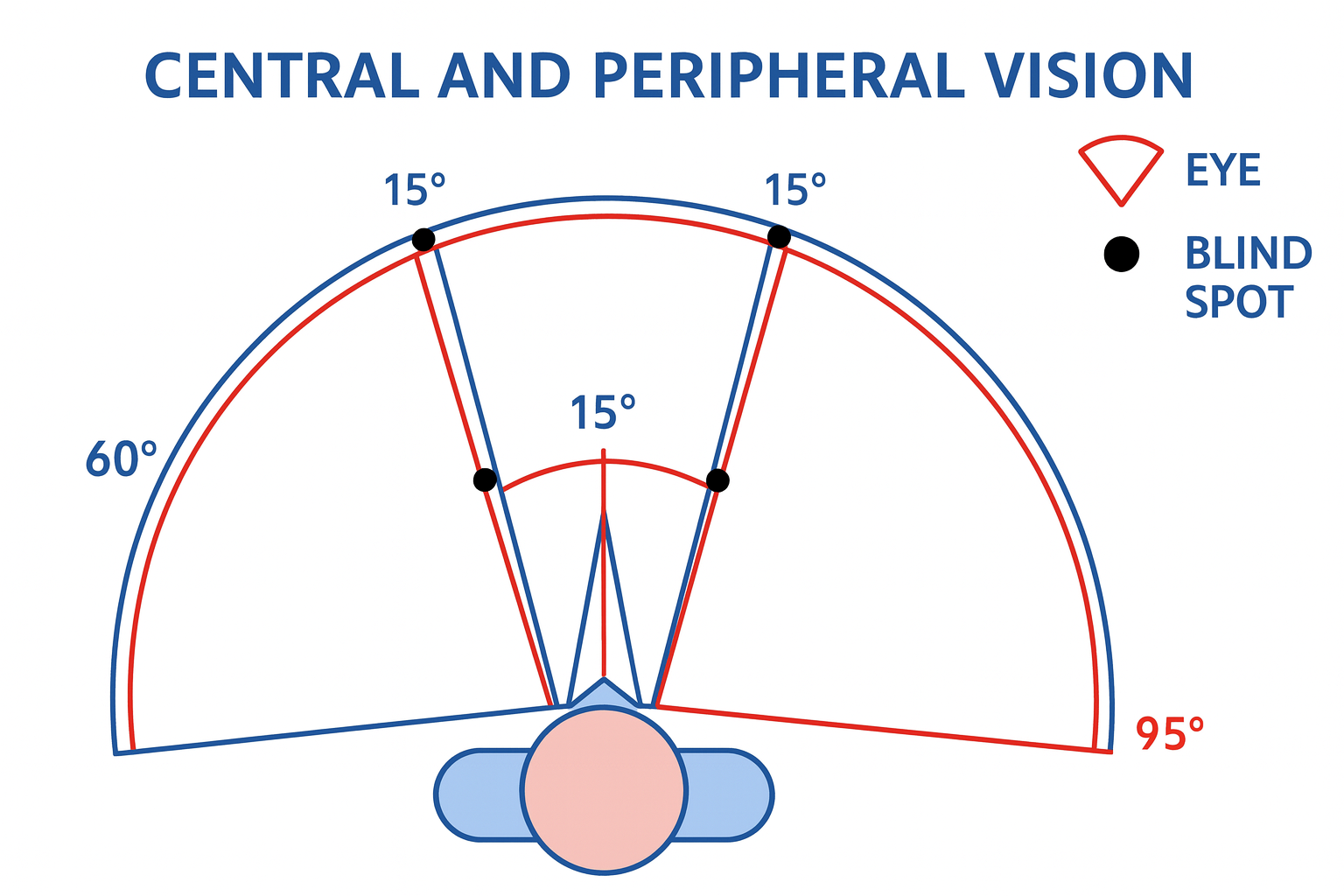}
    \caption{A diagram of the human visual field, illustrating the distinction between the narrow, high-acuity central vision (inner cones) and the broader, low-resolution peripheral vision (outer arcs). Traditional studies often confine stimuli to the central field[cite: 902].}
    \label{fig:visionfield}
\end{figure}

Despite the extensive body of research on visual enumeration, a significant portion of this work has been conducted under conditions that limit its applicability to the real world. A majority of studies have focused on enumeration within the central visual field, presenting stimuli within a small, foveal area where our visual acuity, detail, and colour perception are highest. As illustrated in Figure~\ref{fig:visionfield}, our vision is not uniform; the high-resolution central field is surrounded by a vast peripheral field, which is sensitive to motion and gross form but has much lower resolution.

When tasks are confined to this central vision, participants can often perform them with minimal eye movement and little to no head rotation. This constitutes a form of \textit{static search}. However, real-world scenes are not confined to a small area in front of us. They span a large visual field, both horizontally and vertically. To comprehend and interact with such scenes, we must engage in \textit{dynamic search}, a process that involves coordinated rotation of both the eyes and the head to actively bring different regions of the scene and objects of interest into the central field for detailed inspection. While we can detect a change or movement in our periphery, detailed analysis requires foveation. To our knowledge, no prior work has systematically investigated the enumeration of complex scenes presented over these large visual fields, where the challenges of object characteristics, task intent, and spatial distribution are compounded by the demands of dynamic search. This represents a crucial gap in our understanding of how these fundamental cognitive processes operate in ecologically valid contexts.

\subsection{The Opportunity for a New Paradigm: Immersive Virtual Reality}

The advent of immersive Virtual Reality (VR) technology presents a powerful and timely opportunity to bridge this research gap. Traditional experimental setups, with their flat-screen displays and restricted fields of view, are inherently limited in their ability to simulate the complexities of real-world perception. VR, in contrast, allows for the creation of naturalistic, three-dimensional environments where participants can freely move their head and eyes to scan a wide field of view, closely mimicking the behaviours of natural visual exploration. This capability is paramount for creating an ecologically valid experimental setting to study the cognitive and perceptual processes of visual counting in a realistic context.

This research leverages the Meta Quest Pro VR headset, with its integrated eye-tracking capabilities, to facilitate a detailed investigation into the attention mechanisms, gaze patterns, and subsequent memory recall that occur during enumeration. Unlike flat displays, VR affords rich depth cues, full peripheral stimulation, and the ability to measure natural head-eye coordination. This allows for a much richer measurement of search dynamics, saccadic planning, and fixation distributions across a large field of regard. The current study is designed to advance the understanding of the cognitive and perceptual processes involved in visual object counting within such an immersive setting, with a particular focus on enumeration beyond the subitizing range in heterogeneous scenes. We operationalize the three axes of complexity by presenting two types of stimuli—simple abstract shapes (coloured spheres) and complex real-world objects (photographs)—to vary the \textbf{characteristics}. We manipulate the \textbf{task intent} through instructions for counting all, selective inclusion, or selective exclusion. Finally, we control the \textbf{spatial distribution} by arranging stimuli in structured (grid, circular) and unstructured (random) layouts on a curved surface that reduces edge effects and enables cleaner comparisons of gaze behaviour.

\subsection{Research Questions and Contributions}

The central challenge addressed by this research is the lack of understanding of how humans enumerate complex visual scenes that extend across large visual fields, necessitating dynamic search strategies. We aim to deconstruct this process by examining how performance is shaped by the interplay of object properties, task goals, and environmental structure. This leads to our overarching research question:

\textit{How do object characteristics, task intent, and spatial distribution collectively influence human cognitive strategies—specifically visual attention and memory—during the enumeration of complex scenes over large visual fields?}

To address this broad question, we pose three specific sub-questions that guide our empirical investigation:
\begin{enumerate}
    \item \textbf{SQ1:} How does the intent of the task (i.e., counting instructions for inclusion, exclusion, or counting all) modulate the allocation of visual attention and the efficiency of the enumeration process?
    \item \textbf{SQ2:} In what ways do different spatial distributions (structured vs. unstructured) of objects across a large visual field shape the development of visual search strategies and impact counting performance?
    \item \textbf{SQ3:} What is the functional relationship between the patterns of overt visual attention (i.e., scan paths) deployed during enumeration and the subsequent fidelity of memory for both counted and uncounted objects in the scene?
\end{enumerate}

In addressing these questions, this paper makes the following contributions to the fields of cognitive science, affective computing, and human-computer interaction:
\begin{itemize}
    \item We introduce Recall and Attention during Visual EnumeratioN - Virtual Reality (RAVEN-VR), a novel VR-based experimental system designed for the detailed study of visual enumeration in an ecologically valid, large-field context.
    \item We provide a detailed empirical analysis of how specific task goals (intent) and environmental structure (distribution) jointly affect visual attention, dynamic search strategies, and overall counting performance for stimuli of varying complexity (characteristics).
    \item By integrating eye-tracking data with post-task memory assessments, this work offers direct insights into the relationship between overt visual attention and memory encoding during a complex, goal-directed cognitive task.
    \item Additionally, the framework establishes reusable protocols for stimulus construction, gaze-feature engineering, and analysis pipelines that can be extended to related domains such as crowd estimation, inventory auditing, and visual search training.
\end{itemize}

%%=========================================================

\section{Related Work} 

\subsection{Understanding Visual Enumeration: Subitizing and Counting}
The ability to quickly and accurately determine the number of items in a visual display has been a long-standing area of research in cognitive psychology and neuroscience. A central phenomenon in this domain is \textit{subitizing}, which refers to the fast and precise enumeration of small sets of objects, typically up to three or four items (\cite{Trick1994, Kaufman1949, Sophian}). This process has been recognized as distinct from counting, which involves a slower, serial enumeration of larger quantities (\cite{Trick1994, Simon1996}). Early observations by Jevons in 1871 (\cite{Jevons1871}) and later by Kaufman, Lord, Reese, and Volkmann in 1949 (\cite{Kaufman1949}) laid the foundation for the concept of subitizing, noting that small numbers of items can be apprehended instantaneously without overt counting.

A point of discussion in the literature revolves around the nature of subitizing, specifically whether it is a "pre-attentive" process (\cite{Railo2008, Pylyshyn1989}). Some researchers have claimed subitizing to be pre-attentive, implying it occurs without the need for focused attention (\cite{Railo2008}). However, existing experimental methods and results regarding this claim have shown inconsistencies, and a recent meta-analysis reports reliable attention costs in the subitizing range \cite{Chen2022}. In contrast to subitizing, counting is generally accepted to involve serial attention shifts \cite{Wilder2009, Logan2010, Simon1996}. The distinction between subitizing and counting is further supported by observations that enumeration latency, or reaction time, shows a discontinuity; it increases slowly for small numbers (the subitizing range) and then much more steeply for larger numbers (the counting range) \cite{Trick1994, Simon1996, Kaufman1949}. This discontinuity is evident in both reaction times and eye-movement behavior across various age groups \cite{Logan2010, Schleifer2011}. Studies have also demonstrated that enumeration in the subitizing range relies on a specialised mechanism that differs from counting \cite{Trick1994, Simon1996, Pylyshyn2001}. In addition, broader reviews of numerical perception place subitizing within a family of visual–numerical processes that interact with perceptual organization and decision demands \cite{Mock2016, Mazza2012}.

\subsection{The Role of Attention in Enumeration}
The debate surrounding the pre-attentive nature of subitizing has prompted extensive research into the involvement of attention. A systematic review and meta-analysis of fourteen studies, encompassing 22 experiments and 35 comparisons, suggests that \textit{manipulations to attentional demands interfere with the enumeration of small sets} \cite{Chen2022}. These manipulations lead to slower response times, lower accuracy, and poorer Weber acuity \cite{Chen2022}. This collective finding challenges the claim that subitizing is a purely pre-attentive process, indicating that attention is indeed involved \cite{Railo2008}.

However, precisely defining the role of attention in subitizing is complicated by varied definitions of attention across studies \cite{Railo2008, Mock2016}. Some researchers view attention as a general cognitive ability (global attention process), while others define it as a specific sub-process, such as alertness, orienting, executive, temporal, sustained, divided, selective, or spatial attention \cite{Railo2008}. Despite this ambiguity, a consensus perspective suggests that attentional resources are required for subitizing \cite{Chen2022, Railo2008}. For instance, studies have shown that increasing task difficulty, such as by adding distractors, can increase response time and decrease accuracy in subitizing tasks \cite{Paul2020, Wilder2009, Mock2016}.

A proposed unifying framework for visual enumeration suggests progressively greater attentional involvement from estimation to subitizing to counting \cite{Chen2022, Trick1994}. This framework posits that there is no strict attention dichotomy between subitizing and counting; instead, it describes a difference in weak versus strong attention involvement between these two processes \cite{Railo2008, Trick1993}. Findings from some studies suggest that enumerating a single item might be relatively immune to some attention manipulations \cite{Chen2022}, which presents a challenge to the idea that subitizing universally requires attention. However, the meta-analytic evidence still shows consistent effects of attention manipulations on subitizing performance, including speed, accuracy, and Weber fractions, across various experimental designs and stimuli \cite{Chen2022}. Complementary accounts emphasize that perceptual grouping and scene organization can either aid or hinder subitizing by changing attentional set and item individuation demands \cite{Mazza2012}.

\subsection{Eye Movements and Attentional Mechanisms in Enumeration}
Eye tracking is a valuable tool for understanding the underlying cognitive processes in enumeration tasks, offering insights into how individuals deploy their attention. \textit{The discontinuity between subitizing and counting is not only observed in reaction times but also in eye movement behaviors} \cite{Logan2010}. Studies indicate that while saccadic movements may occur within the subitizing range, they are not always necessary for accurate and efficient enumeration in adults \cite{Watson2007}. For larger numbers, eye movements are a characteristic feature of counting procedures, with parameters such as saccadic frequency varying across enumeration ranges \cite{Logan2010}.

Research on visual search and counting further shows that \textit{eye movements can be voluntarily controlled and modulated to align with explicit task demands} \cite{Wilder2009, Paul2020}. This suggests that spatial and temporal patterns of eye movements are influenced by higher-level task strategies, beyond merely being driven by lower-level visual processing based on stimulus image statistics \cite{Wilder2009}. The capacity for enumeration may thus be linked to numerical strategies, such as noticing combinations of sets, rather than solely quantitative differences in numerical processing, such as memory for items or counting speed \cite{Paul2020}.

In the context of visual search, studies employing eye tracking and fixation-related potentials (FRPs) have explored cognitive load and attention allocation in various environments, including mixed reality (MR) and extended-reality settings \cite{Chiossi2024}. These studies demonstrate how eye movements and FRPs can reveal varying cognitive demands and influence visual search efficiency and mental workload \cite{Chiossi2024}. For example, increased distractors and perceptual load can lead to longer search times and higher error rates, capturing attention \cite{Mock2016, Paul2020}. The allocation of attentional resources is a continuous process involving both stimulus-driven (bottom-up) and goal-driven (top-down) mechanisms \cite{Mock2016}.

\subsection{Theoretical Frameworks: FINST and Working Memory}
Several theoretical models attempt to explain the mechanisms underlying visual enumeration. One such model is the \textit{FINST (Fingers of Instantiation) model}, which proposes a pre-attentive mechanism that "picks out" or "individuates" features in a visual display before spatial patterns or relations can be discerned \cite{Pylyshyn1989, Pylyshyn2001, Trick1993}. FINSTs are conceptualized as a limited resource, allowing for parallel assignment to a small number of target items during the pre-attentive phase of object recognition \cite{Pylyshyn2001}. The model suggests that the visual system must first establish these "pointers" to objects before any spatial relationships can be processed \cite{Pylyshyn1989}.

However, the precise role of FINSTs in enumeration is a subject of ongoing discussion. Some views suggest that the pre-attentive assignment of FINSTs might not be sufficient for enumeration itself, and that some degree of attentional processing may be required even for very small numerosities \cite{Trick1993, Railo2008, Simon1996}. This implies that access to information about the number of active FINSTs might necessitate a partly attentional enumeration process applied to the associated objects \cite{Trick1993}.

\textit{Working memory} also plays an important role in enumeration. Research differentiates various types of working memory and highlights their interaction with visual–numerical processing during enumeration \cite{Mock2016, Tai2004, Hesse2016}. Studies on enumeration often reveal distinct patterns of interference depending on the type of working memory engaged. For instance, tapping tasks, which involve spatial operations, can produce more interference in spatial enumeration (e.g., in the 1–3 item range) than articulation tasks, which are verbal—a pattern opposite to what is observed in temporal enumeration \cite{Tai2004}. This suggests that spatial working memory is particularly implicated in visual enumeration tasks.

Furthermore, there is a strong connection between attention and memory. Items that receive longer or multiple fixations during counting tasks are more likely to be recalled subsequently \cite{Herten2017}. This aligns with the understanding that deeper attentional processing of an object enhances its encoding into memory \cite{Herten2017}. Conversely, objects that are irrelevant to the counting task and are thus ignored tend to be poorly remembered, reinforcing the general principle that unattended elements are less likely to be retained in complex visual scenes \cite{Herten2017}.

\subsection{Studying Visual Enumeration through Virtual Reality (VR)}
Traditional studies of visual enumeration have predominantly relied on flat displays and restricted fields of view, which do not fully mimic real-world visual experiences. The advent of \textit{virtual reality (VR) technology} offers a new platform to investigate visual counting in more naturalistic and immersive settings \cite{Chiossi2024}. VR environments, especially with integrated eye tracking (such as in current MR/VR research platforms), allow participants to freely move their heads and eyes to scan a wide field of view, closely resembling real-world visual search behaviors \cite{Chiossi2024}.

Studying visual object counting in ecologically valid VR environments provides an opportunity to understand the cognitive and perceptual processes involved, including attention allocation and working memory engagement. This approach allows for the investigation of counting larger numbers and heterogeneous sets, moving beyond the traditional focus on subitizing \cite{Mock2016, Paul2020}. VR also enables precise control of layout, viewpoint, and depth cues while preserving natural head–eye coordination, supporting richer analyses of search dynamics and fixation distributions \cite{Chiossi2024}.

Proposed research in VR aims to explore specific hypotheses regarding visual counting:
\begin{itemize}
\item \textit{Selective Counting}: Counting a subset of objects based on category is expected to take more time and be more error-prone than counting all objects, owing to the additional requirement for selective object-based attention to filter targets from distractors. Participants are predicted to show more focused gaze on task-relevant items and quicker glances over irrelevant items \cite{Paul2020, Wilder2009, Mock2016}.
\item \textit{Layout Influence}: The arrangement of objects (e.g., grid, circular, random scatter) is hypothesized to influence counting efficiency. Structured layouts are expected to facilitate systematic scanning, leading to faster counting and fewer errors, while random scatters might result in longer search paths and a greater chance of missing or double-counting items. Perceptual grouping principles also predict benefits when items can be chunked into subsets \cite{Mazza2012, Paul2020}.
\item \textit{Attention and Memory}: A strong link is expected between visual attention and memory recall. Items that are fixated longer or multiple times during counting are predicted to be better remembered, while ignored items will be poorly recalled \cite{Herten2017}.
\end{itemize}

These advancements in VR-based research provide a rich environment for exploring the complexities of human visual enumeration, particularly how attention, eye movements, and memory interact in naturalistic counting scenarios \cite{Chiossi2024, Paul2020}.

%%=========================================================

\section{Methodology}

This section provides a comprehensive description of the experimental methodology employed to investigate the cognitive behaviours of attention and recall during visual enumeration tasks. The methodology was designed to ensure a systematic and controlled examination of the factors influencing counting performance in an immersive virtual environment.

% \subsection{Hypotheses}
% This study is guided by a set of hypotheses designed to deconstruct the visual enumeration process:
% \begin{enumerate}
%     \item H1: What is the effect of different counting instructions on the attentional behaviour of an individual during a visual enumeration task?
%     \item H2: How does the spatial layout of objects influence the attentional behaviour during a visual enumeration task?
%     \item H3: What is the relationship between the visual attention allocated during an enumeration task and the subsequent memory recall of the objects and scene details?
% \end{enumerate}

\subsection{Hypothesis Formulation}
Based on the theoretical foundations of visual enumeration, attention, and memory, and in direct correspondence with our research questions, we formulated a set of hypotheses to be tested in this study. These hypotheses are intended to apply across both experimental phases (Phase 1: Abstract Shapes and Phase 2: Real-World Objects). However, it is anticipated that the magnitude of the observed effects will be more pronounced with Real-world object images (Phase 2) due to the additional cognitive overhead required for object recognition and categorization compared to recognizing Abstract geometric shapes (Phase 1).

\subsubsection{Hypotheses for SQ1: Effect of Task Intent}
This set of hypotheses addresses how different task intents (i.e., counting instructions) are expected to modulate cognitive load, task performance, and the allocation of visual attention.

\begin{itemize}
    \item[\textbf{H1:}] \textbf{Selective counting instructions will impose a greater cognitive load and result in lower performance compared to the baseline 'CountAll' instruction.} We predict that the requirement to filter items based on category will be more mentally demanding than a simple exhaustive count.
    \begin{itemize}
        \item[(a)] \textbf{Task Completion Time:} Participants will take significantly longer to complete selective counting tasks. We expect the order of completion time to be: 'CountAll' (Exhaustive Count) $<$ 'CountSome' (Selective Inclusion) $<$ 'CountNot' (Selective Exclusion)\footnote{CountAll (CA) or Exhaustive Counting (EC) - Count All the Items, CountSome (CS) or Selective Inclusion (SI) - Count only if it contains X, CountNot (CN) or Selective Exclusion (SE) - Count only if it does not contain Y.}. The 'CountNot' task is hypothesised to be the most time-consuming due to the dual cognitive processes of identifying the excluded category and then enumerating all other items.
        \item[(b)] \textbf{Accuracy:} Participants will be less accurate in selective counting tasks. We expect the order of accuracy to be: 'CountAll' $>$ 'CountSome' $>$ 'CountNot'.
        \item[(c)] \textbf{Cognitive Load:} Self-reported mental demand ratings will be lowest for the 'CountAll' instruction and highest for the 'CountNot' instruction.
    \end{itemize}
\end{itemize}

\subsubsection{Hypotheses for SQ2: Influence of Spatial Layout}
This hypothesis addresses how the spatial arrangement of objects is expected to influence the development of visual search strategies and overall task efficiency.

\begin{itemize}
    \item[\textbf{H2:}] \textbf{Structured layouts will facilitate more systematic and efficient visual search strategies compared to an unstructured random layout.} We predict that the organization of items will guide participants' scanpaths, affecting performance and search behaviour.
    \begin{itemize}
        \item[(a)] \textbf{Task Completion Time:} We expect a trend where the 'Grid' layout yields the fastest completion times, followed by the 'Circular' layout, with the 'Random' layout being the least efficient.
        \item[(b)] \textbf{Accuracy:} In line with task completion time, we expect the 'Grid' layout to produce the highest accuracy, followed by 'Circular', and then 'Random', as unstructured layouts increase the likelihood of missing or double-counting items.
        \item[(c)] \textbf{Cognitive Load:} Self-reported mental demand ratings will be lowest for the 'Grid' layout and highest for the 'Random' layout.
        \item[(d)] \textbf{Search Strategy:} The layouts will elicit distinct gaze distribution patterns. In the 'Grid' layout, we expect systematic row-by-row or column-by-column scanning. In the 'Circular' layout, we anticipate a sweeping concentric search pattern, moving from outer to inner rings or vice-versa. In the 'Random' layout, we expect a less systematic, more exploratory distribution, likely guided by principles of proximity (Gestalt clustering).
    \end{itemize}
\end{itemize}

\subsubsection{Hypotheses for SQ3: Relationship Between Attention and Memory}
This hypothesis addresses the link between the overt deployment of visual attention during the counting task and the subsequent ability to recall items from the scene.

\begin{itemize}
    \item[\textbf{H3:}] \textbf{A strong positive relationship will exist between the visual attention allocated to an item and the likelihood of it being encoded into memory.} What participants look at will predict what they remember, and overall recall will be influenced by both task intent and layout.
    \begin{itemize}
        \item[(a)] \textbf{Recall Performance:} The number of items recalled from memory will be influenced by both the task instruction and the spatial layout. We expect recall to be highest for the 'CountAll' task, followed by 'CountSome', and lowest for 'CountNot'. Similarly, we predict that the structured 'Grid' layout will lead to better memory recall than the 'Circular' layout, with the 'Random' layout resulting in the poorest recall.
    \end{itemize}
\end{itemize}

\subsection{Experimental Design}
The study was structured as a within-subjects factorial design, where every participant was exposed to all experimental conditions. This approach allows for direct comparisons of an individual's performance across different tasks and layouts, thereby controlling for inter-participant variability and enhancing the statistical power of the analysis. The design involved three primary independent variables that were systematically manipulated:

\begin{enumerate}
    \item \textit{Stimulus Type:} This variable had two levels, corresponding to the two distinct phases of the experiment. 
    \begin{itemize}
        \item \textit{Phase 1 (Abstract Geometric Shapes):} Participants were tasked with enumerating simple coloured dots, a condition designed to measure baseline visual counting performance without the cognitive load of object recognition.
        \item \textit{Phase 2 (Real-World Object Images):} Participants were tasked with enumerating a heterogeneous set of photographic images, a condition that introduced the additional cognitive steps of visual search and object categorisation.
    \end{itemize}
    \item \textit{Counting Instruction:} This variable had three levels, designed to modulate the cognitive and attentional demands of the enumeration task.
    \begin{itemize}
        \item \textit{Count All:} A baseline condition requiring participants to count the total number of items presented in the scene, without any selective filtering.
        \item \textit{Selective Inclusion (Count Only X / CountSome):} A condition requiring participants to identify and count only those items belonging to a specified category (e.g., 'count only the red dots' or 'count only the animals').
        \item \textit{Selective Exclusion (Count Not Y / CountNot):} A condition requiring participants to count all items except for those belonging to a specified category (e.g., 'count all dots that are not blue' or 'count all images that are not vehicles').
    \end{itemize}
    \item \textit{Spatial Layout:} This variable also had three levels, representing different spatial arrangements of the stimuli within the virtual environment.
    \begin{itemize}
        \item \textit{Grid Layout:} Items were arranged in a structured matrix formation on the virtual surface.
        \item \textit{Circular Layout:} Items were arranged in a series of concentric circles on the virtual surface.
        \item \textit{Random Layout:} Items were positioned in a pseudo-random, unstructured manner across the virtual surface, with a minimum separation distance to prevent overlap.
    \end{itemize}
\end{enumerate}

The dependent variables measured in this study were comprehensive, capturing multiple facets of participant behaviour and performance. These included performance metrics (counting accuracy and task completion time), detailed eye and head-tracking data (fixations, saccades, and scan paths), and qualitative responses from a post-trial questionnaire (memory recall, confidence ratings, and self-reported strategies). The assignment of counting instructions and layouts was counterbalanced across participants using a Latin Square design to mitigate any potential ordering or learning effects.

\subsection{Participants}
A total of thirty individuals (N=30) were recruited to participate in the study. The participant pool was composed of students and staff from the local university campus, with an age range of 20 to 35 years. An effort was made to maintain a balanced representation of genders within the sample. All participants were required to have normal or corrected-to-normal vision and were screened to ensure they had no known neurological, cognitive, or visual impairments that could potentially interfere with their performance on the experimental tasks. While prior experience with virtual reality was not a prerequisite for participation, individuals were screened for susceptibility to simulator sickness. Each participant provided written informed consent before the commencement of the study and was compensated for their time. The experimental protocol was approved by the institutional ethics review board.

\subsection{Apparatus: The RAVEN-VR System}
\begin{figure}[t!]
    \centering
    
    % Column Titles
    \begin{minipage}{\expSetupImageWidth \textwidth}
        \centering
        \textbf{\small{Phase 1: Abstract Shapes \\ (Dots)}}
    \end{minipage}
    \hfill
    \begin{minipage}{\expSetupImageWidth \textwidth}
        \centering
        \textbf{\small{Phase 2: Real-World Objects (Images)}}
    \end{minipage}

    \vspace{1em} % Add some vertical space after titles

    % First Row: Grid Layout
    \begin{subfigure}[b]{\expSetupImageWidth \textwidth}
        \centering
        \includegraphics[width=\textwidth]{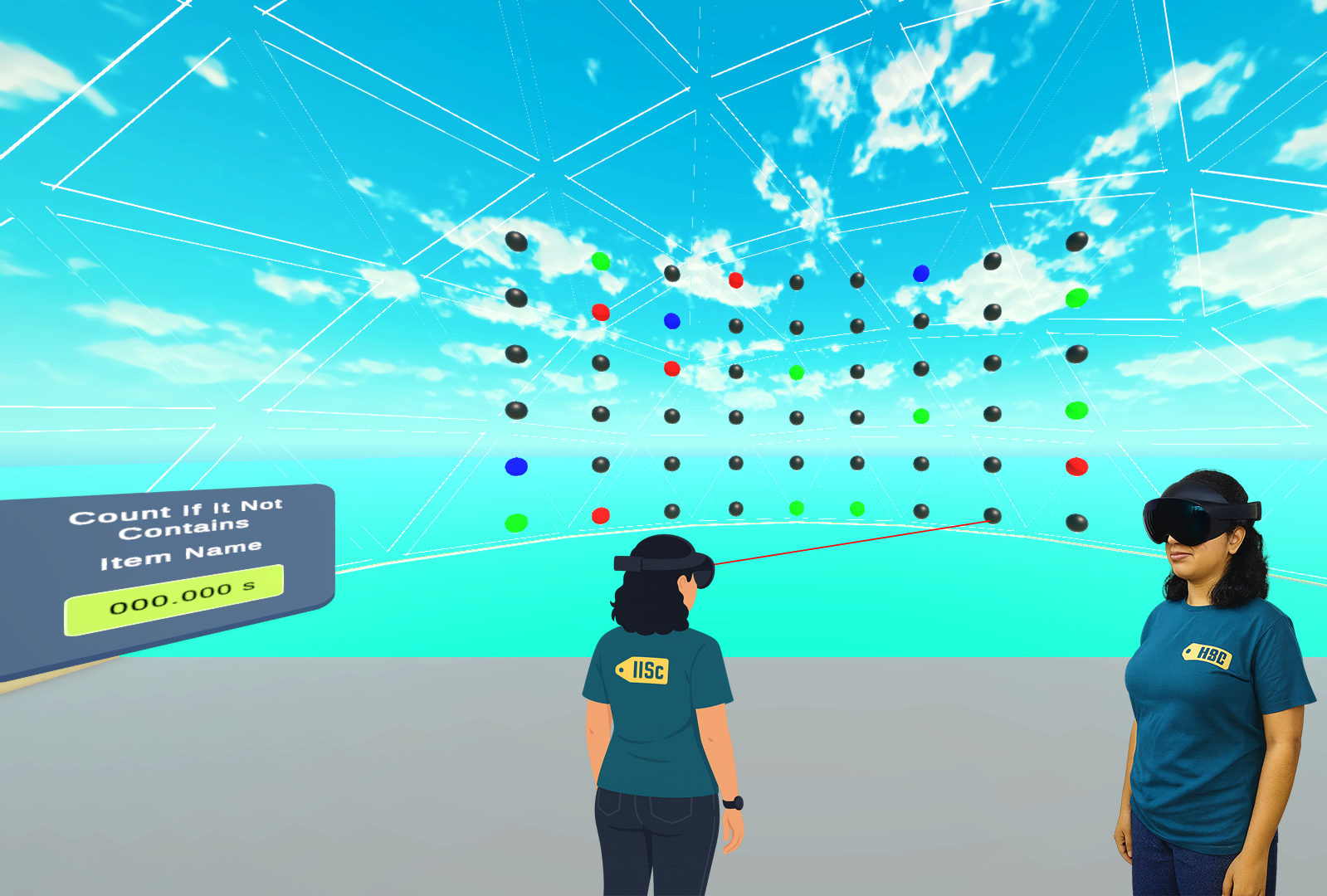}
        \caption{Grid layout with dots.}
        \label{fig:setup_dots_grid}
    \end{subfigure}
    \hfill
    \begin{subfigure}[b]{\expSetupImageWidth \textwidth}
        \centering
        \includegraphics[width=\textwidth]{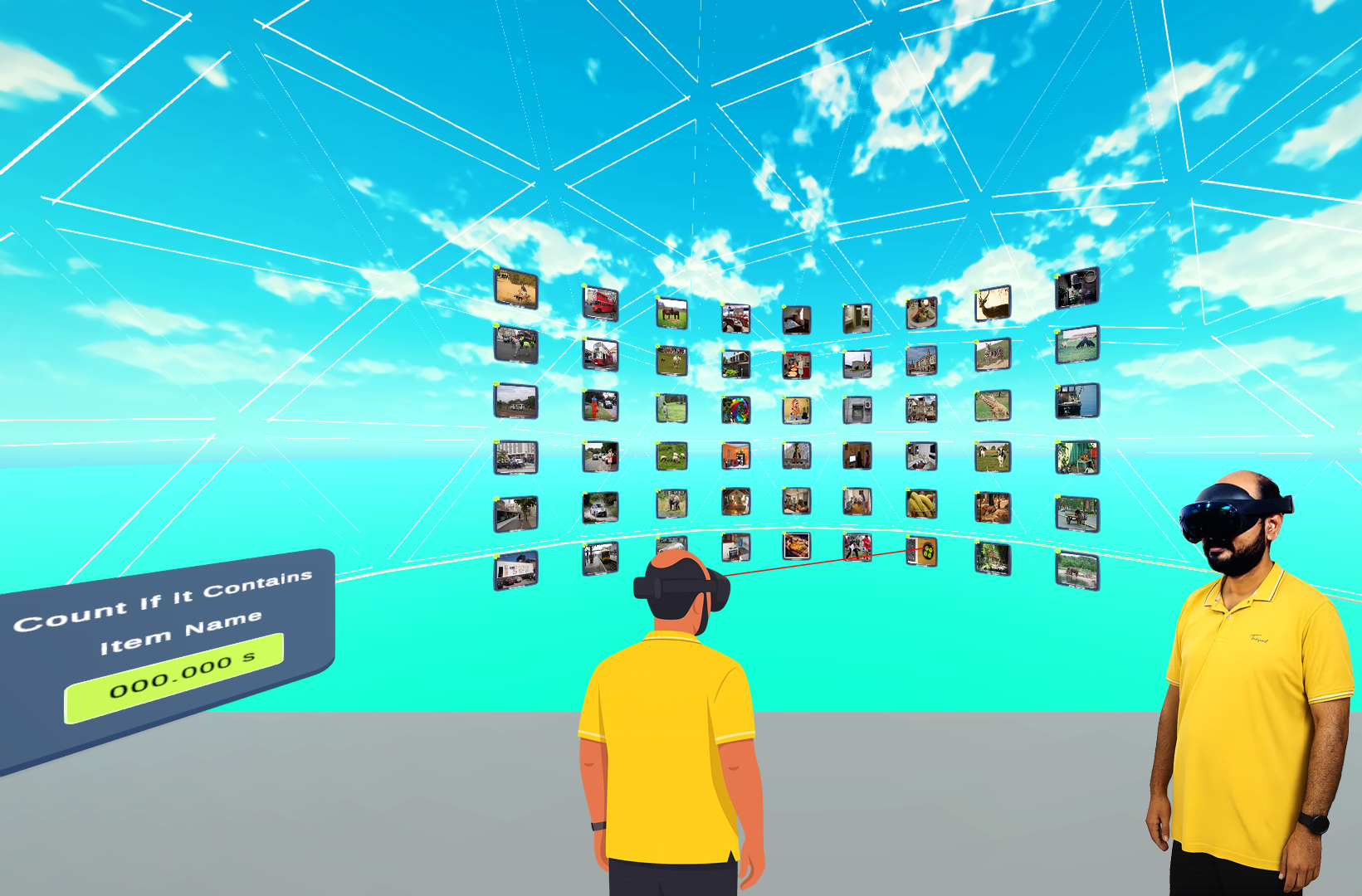}
        \caption{Grid layout with images.}
        \label{fig:setup_images_grid}
    \end{subfigure}

    \vspace{1em} % Add some vertical space between rows

    % Second Row: Circular Layout
    \begin{subfigure}[b]{\expSetupImageWidth \textwidth}
        \centering
        \includegraphics[width=\textwidth]{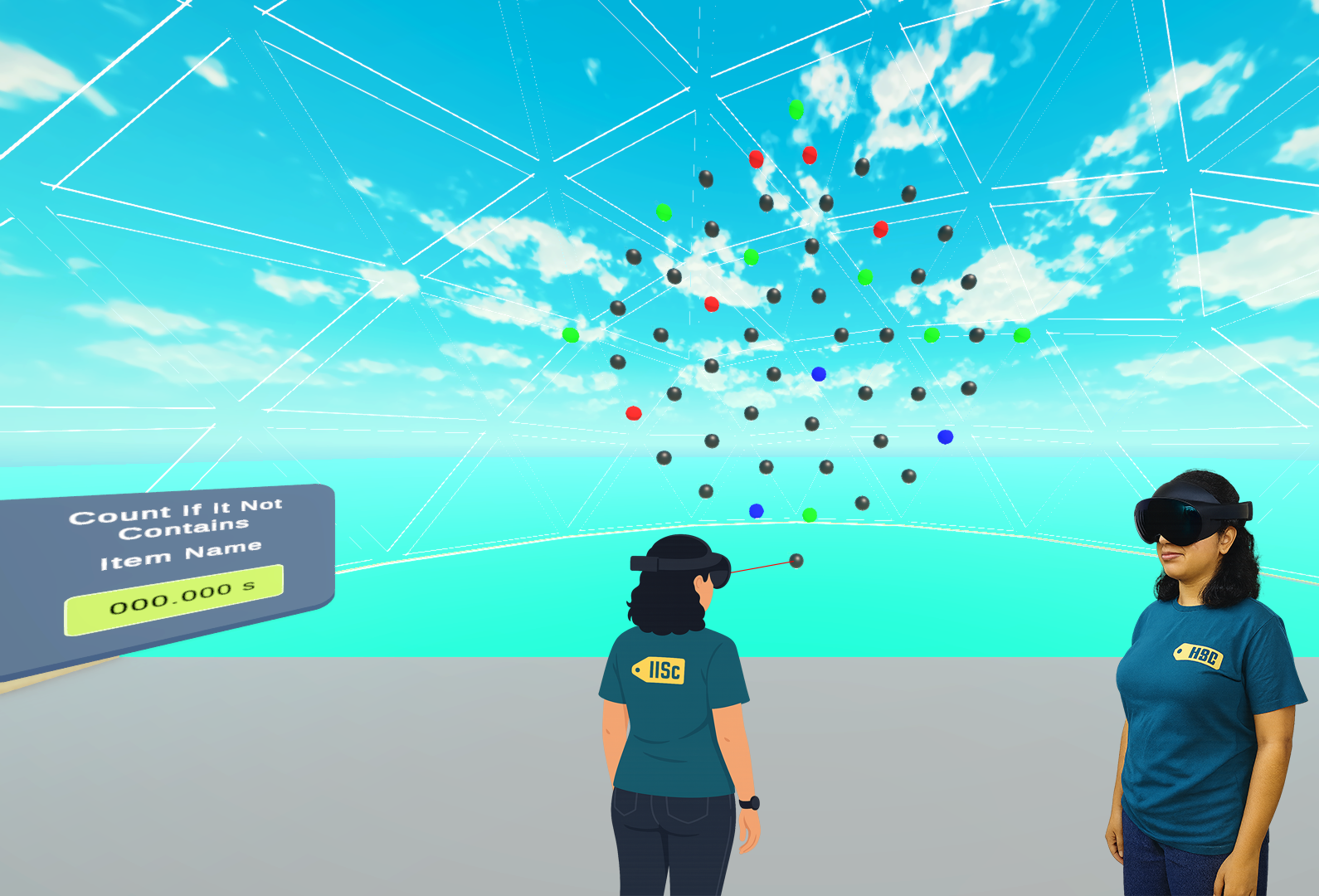}
        \caption{Circular layout with dots.}
        \label{fig:setup_dots_circular}
    \end{subfigure}
    \hfill
    \begin{subfigure}[b]{\expSetupImageWidth \textwidth}
        \centering
        \includegraphics[width=\textwidth]{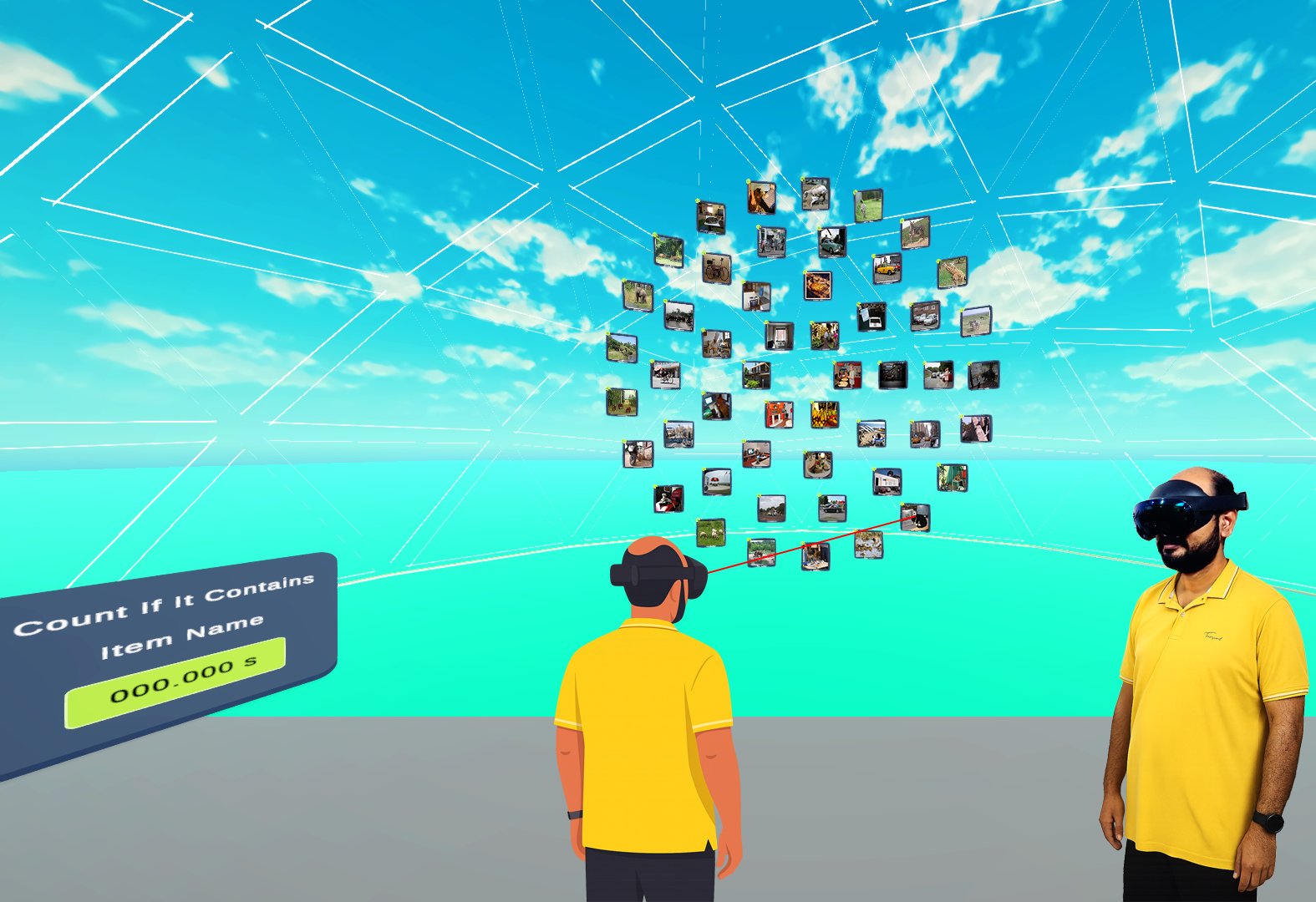}
        \caption{Circular layout with images.}
        \label{fig:setup_images_circular}
    \end{subfigure}

    \vspace{1em} % Add some vertical space between rows

    % Third Row: Random Layout
    \begin{subfigure}[b]{\expSetupImageWidth \textwidth}
        \centering
        \includegraphics[width=\textwidth]{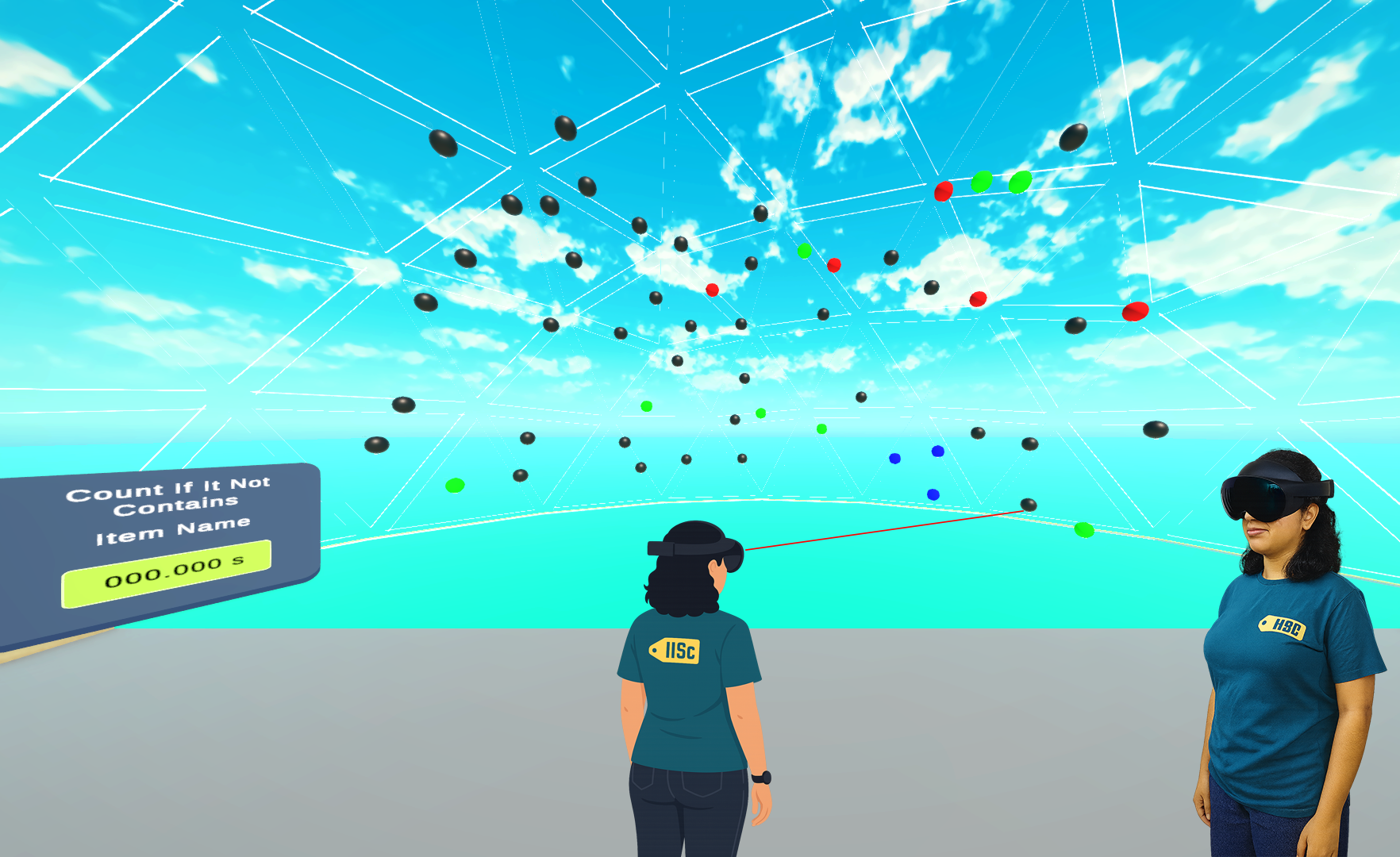}
        \caption{Random layout with dots.}
        \label{fig:setup_dots_random}
    \end{subfigure}
    \hfill
    \begin{subfigure}[b]{\expSetupImageWidth \textwidth}
        \centering
        \includegraphics[width=\textwidth]{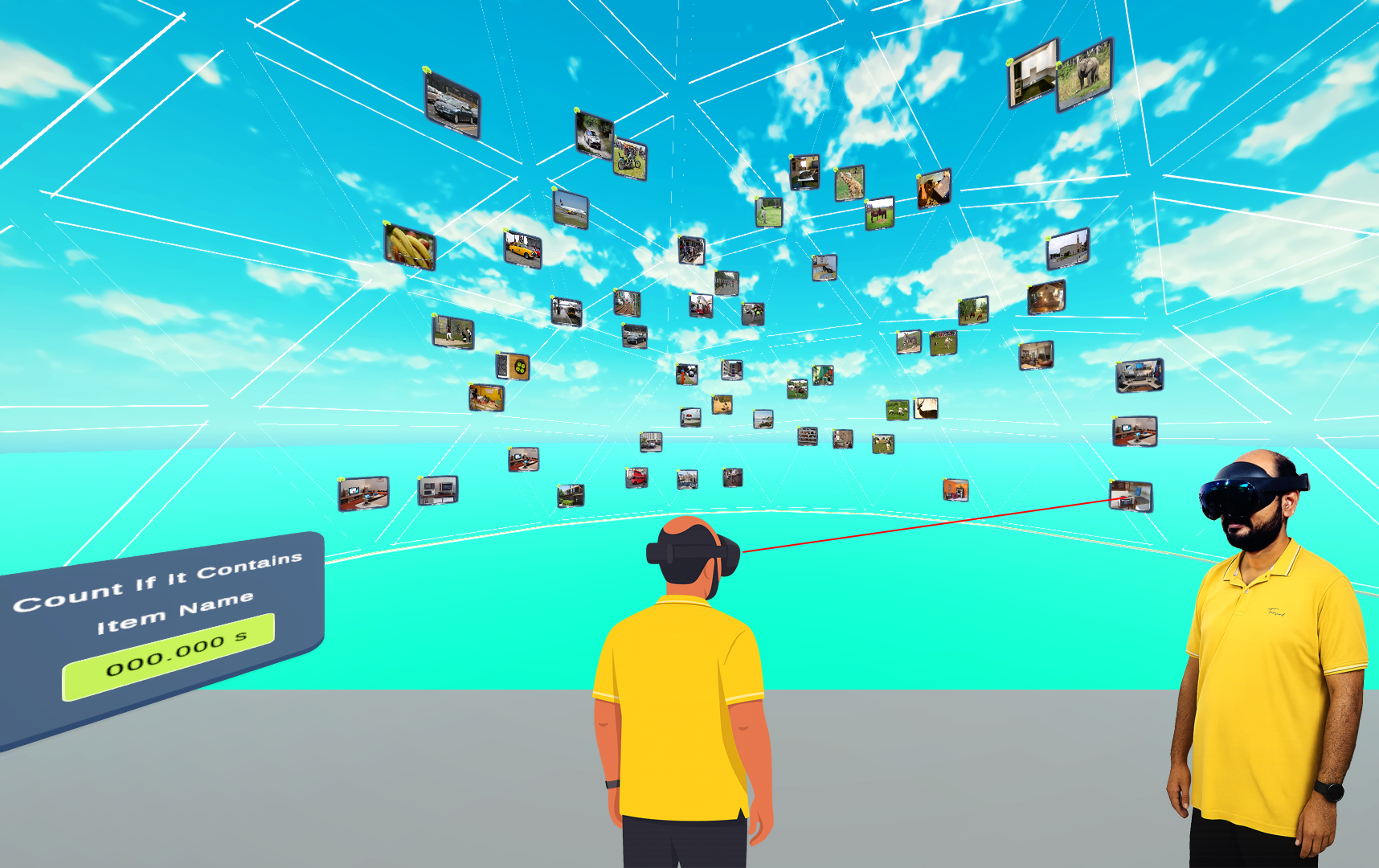}
        \caption{Random layout with images.}
        \label{fig:setup_images_random}
    \end{subfigure}

    \caption{Overview of the RAVEN-VR experimental setup. The left column (a, c, e) shows the three spatial layouts for the Phase 1 task with abstract dot stimuli. The right column (b, d, f) shows the same layouts for the Phase 2 task with complex real-world images.}
    \label{fig:experimental_setup_overview}
\end{figure}

The experiment was conducted using a custom-developed VR application named RAVEN-VR (Recall and Attention during Visual EnumeratioN - Virtual Reality). This system was specifically designed to present the stimuli, manage the experimental procedure, and log all relevant data streams. The RAVEN-VR system was designed to present stimuli across a large field of view, simulating a naturalistic viewing experience. The figures, as shown in Figure~\ref{fig:experimental_setup_overview}, provide a third-person perspective of the experimental setup, illustrating the participant's digital twin at the centre viewing the stimuli, the user interface panel displaying the task instruction on the left, and the human participant on the right. These wide-angle captures show how the different layouts and stimulus types appeared within the participant's immersive environment.

% Note: Ensure you have \usepackage{graphicx} and \usepackage{subcaption} in your document preamble.

\subsubsection{Hardware}
The primary hardware component was the Meta Quest Pro VR headset. This device was selected for its high-resolution stereoscopic displays, a wide field of view (approximately 106° horizontal and 96° vertical), and its integrated, high-fidelity eye-tracking system. The eye-tracking system captures gaze direction and point-of-regard at a high sampling rate ($\geq$90 Hz), enabling precise analysis of visual attention patterns. The headset also provides 6-degrees-of-freedom (6DoF) tracking of head position and orientation, which allowed for the accurate measurement of head movements during the task. Participants were seated on a swivel chair throughout the experiment, permitting comfortable head rotation but restricting bodily locomotion. Handheld controllers were not used during the counting tasks to minimise any potential motor interference; responses were provided verbally.

\subsubsection{Software}
The RAVEN-VR environment was developed using the Unity 3D game engine. The software was engineered to render immersive 3D scenes that displayed the visual stimuli according to the predefined layouts. A key feature of the software was its ability to log time-series data with high precision, capturing synchronised streams of gaze vectors, head orientation, and task-related event markers (e.g., scene onset, participant response). Each visual stimulus within the environment was assigned a unique identifier, which enabled the mapping of gaze data to specific objects for detailed fixation analysis. The software also included an interactive feature wherein a participant's sustained fixation on an image would cause it to responsively move closer, allowing for an enlarged and more detailed view if needed.

\subsection{Stimuli and Virtual Environment}
The visual stimuli and the environment in which they were presented were carefully designed to address the research questions of the two experimental phases.

\subsubsection{Phase 1: Abstract Geometric Shapes}
For the first phase of the experiment, the stimuli consisted of simple coloured dots. Four colours were used: Red, Green, Blue, and Black. In each scene, the black dots were predominantly featured, which served to make the other coloured dots more visually salient and provided a consistent background context. The total number of dots in each scene ranged from 48 to 54. These simple geometric shapes were chosen to establish a baseline for visual enumeration that did not require complex object recognition, thereby isolating the cognitive processes of visual scanning and counting.

\subsubsection{Phase 2: Real-World Object Images}
The second phase utilized a curated set of high-resolution, full-colour photographs of real-world objects. These images were sourced from large, publicly available annotated datasets, including MS COCO, the Open Images Dataset, and ImageNet. The selection process was governed by a strict set of criteria to ensure clarity and consistency. Each image was required to contain a single, dominant, and easily recognizable object, with minimal background clutter or ambiguity. Images containing text, symbols, or multiple countable entities were excluded. The selected images were categorised into three mutually exclusive groups:
\begin{itemize}
    \item \textit{Animals:} A diverse range of species including mammals, birds, insects, and reptiles.
    \item \textit{Vehicles:} All modes of transport, such as cars, motorcycles, aeroplanes, and boats.
    \item \textit{Other Objects:} Common, everyday items that did not fall into the previous two categories, such as furniture, tools, food items, and electronics.
\end{itemize}
All images were preprocessed to a standard aspect ratio and resolution to ensure visual consistency. The total number of images per scene in this phase also ranged from 48 to 54, with a balanced distribution across the three categories.

\subsubsection{Virtual Environment and Spatial Layouts}
All stimuli, for both phases, were presented within a consistently designed virtual environment. The environment featured a neutral grey background and uniform, bright lighting to minimise distractions and enhance the visibility of the stimuli. The items were positioned in space on the surface of a large, curved virtual cylinder that was placed in front of the participant. This surface spanned approximately 120° of the horizontal and vertical field of view, ensuring that all items were visible with minimal head rotation. The stimuli were arranged on this surface in one of the three predefined spatial layouts: Grid, Circular, or Random.

\subsection{Procedure}
The experimental session for each participant followed a standardized and structured procedure, lasting approximately 30 to 40 minutes in total.
\begin{enumerate}
    \item \textit{Onboarding and Setup:} Upon arrival, participants were welcomed, briefed on the general nature of the study, and asked to provide informed consent. They were then comfortably seated and fitted with the Meta Quest Pro headset. The experimenter ensured a proper fit for clarity and comfort. This was followed by a multi-point eye-tracking calibration procedure to ensure the accuracy of the gaze data.
    \item \textit{Practice Trials:} Before the main experiment, participants completed a series of practice trials. These trials used simplified scenes and exposed them to all three types of counting instructions and layouts. This phase served to familiarise the participants with the VR environment, the task requirements, and the process of providing verbal responses.
    \item \textit{Experimental Trial Sequence:} Each participant completed a total of six experimental trials (three for Phase 1 and three for Phase 2). The sequence for each trial was as follows:
    \begin{itemize}
        \item \textit{Instruction Display:} A screen inside the VR headset displayed the counting instruction for the upcoming trial (e.g., 'Count only the animal images'). The instruction was also read aloud by the experimenter to ensure clarity.
        \item \textit{Scene Presentation:} Following a brief fixation marker to centre the participant's gaze, the virtual scene with the stimulus layout was presented. The participant then began the counting task.
        \item \textit{Response Capture:} The participant was instructed to count silently and, upon arriving at a final count, to state the number aloud. The experimenter manually recorded the spoken count, and the VR system automatically logged the completion time from the onset of the scene to the moment of the verbal response. Once the response was given, the scene was immediately removed to prevent further inspection.
    \end{itemize}
    \item \textit{Post-Trial Questionnaire:} Immediately after each trial, to capture fresh memories, participants answered a short questionnaire. This questionnaire probed their memory of specific items, their confidence in their count (on a 1-5 scale), their perception of cognitive load, and any specific strategies they might have employed during the task.
    \item \textit{Breaks and Debriefing:} Participants were offered short breaks between trials to minimise fatigue. After the completion of all trials, they were fully debriefed about the specific research questions and hypotheses of the study. They were given an opportunity to ask questions and were then thanked and compensated for their participation.
\end{enumerate}

\subsection{Data Collection and Outcome Metrics}
A rich, multi-modal dataset was collected for each participant, synchronized via timestamps to allow for integrated analysis.
\begin{itemize}
    \item \textit{Performance Metrics:} This included the \textit{Count Accuracy}, calculated by comparing the participant's reported count to the ground truth, and the \textit{Task Completion Time}, measured in seconds from scene onset to verbal response.
    \item \textit{Statistical Tests:} To determine if the observed differences in performance were statistically meaningful, a series of non-parametric tests was conducted on the data. Given the sample size and the nature of the data, the Friedman test was selected as the primary inferential statistic for this within-subjects design. The Kruskal-Wallis test was also performed as a converging measure. For variables that showed an overall significant effect, post-hoc pairwise comparisons were conducted using the Wilcoxon signed-rank test, with p-values adjusted for multiple comparisons using the Holm-Bonferroni correction to control for the family-wise error rate.
    
    \item \textit{Eye-Tracking Metrics:} The raw gaze data were processed to extract the gaze distribution over the visual field.
    \item \textit{Head-Tracking Metrics:} Data on head orientation (yaw and pitch) was logged to analyse the extent of physical exploration of the virtual scene.
    \item \textit{Questionnaire Responses:} Qualitative and quantitative data from the post-trial questionnaires provided insights into subjective experience, including \textit{Cognitive Load} experienced, \textit{Memory Recall} (scored for accuracy), and rated \textit{Confidence}.
\end{itemize}

%==============================================================

\section{Results and Discussion}
In this section, we present and interpret the empirical findings from our study. The analysis is structured to address our research questions and evaluate the formulated hypotheses. We present the results for each phase of the experiment separately to draw a clear distinction between the cognitive behaviours associated with enumerating abstract shapes (Phase 1) and real-world objects (Phase 2).

\subsection{Phase 1: Enumeration of Abstract Geometric Shapes}
This phase established a baseline for performance and behaviour in a task that involved pure visual counting, without the additional cognitive load of complex object recognition. The stimuli consisted of simple coloured spheres.

\subsubsection{Task Completion Time}

\paragraph{Observations}
The analysis of task completion time, defined as the duration from the presentation of the scene to the participant's verbal response, revealed distinct patterns influenced by both the task intent (instruction type) and the spatial layout of the stimuli.

\begin{figure}[ht!]
    \centering
    \begin{subfigure}[b]{0.49\textwidth}
        \centering
        \includegraphics[width=\textwidth]{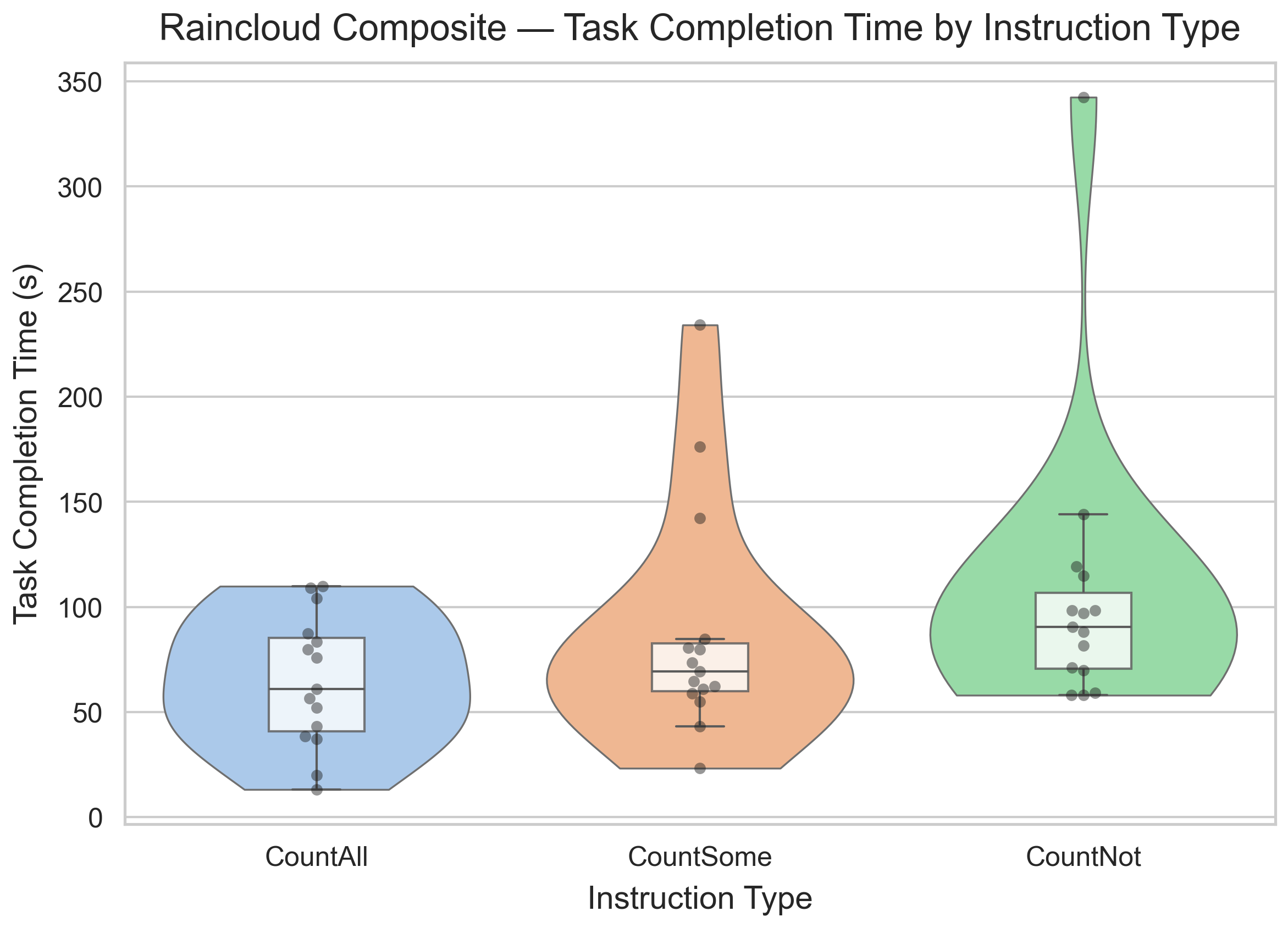}
        \caption{Raincloud composite plot illustrating the distribution of task completion times across the three instruction types for Phase 1. The plot shows the probability density (cloud), the boxplot (rain), and the individual data points (drizzle), clearly indicating an increase in completion time.}
        \label{fig:instruction_time}
    \end{subfigure}
    \hfill % Adds horizontal space between the subfigures
    \begin{subfigure}[b]{0.49\textwidth}
        \centering
        \includegraphics[width=\textwidth]{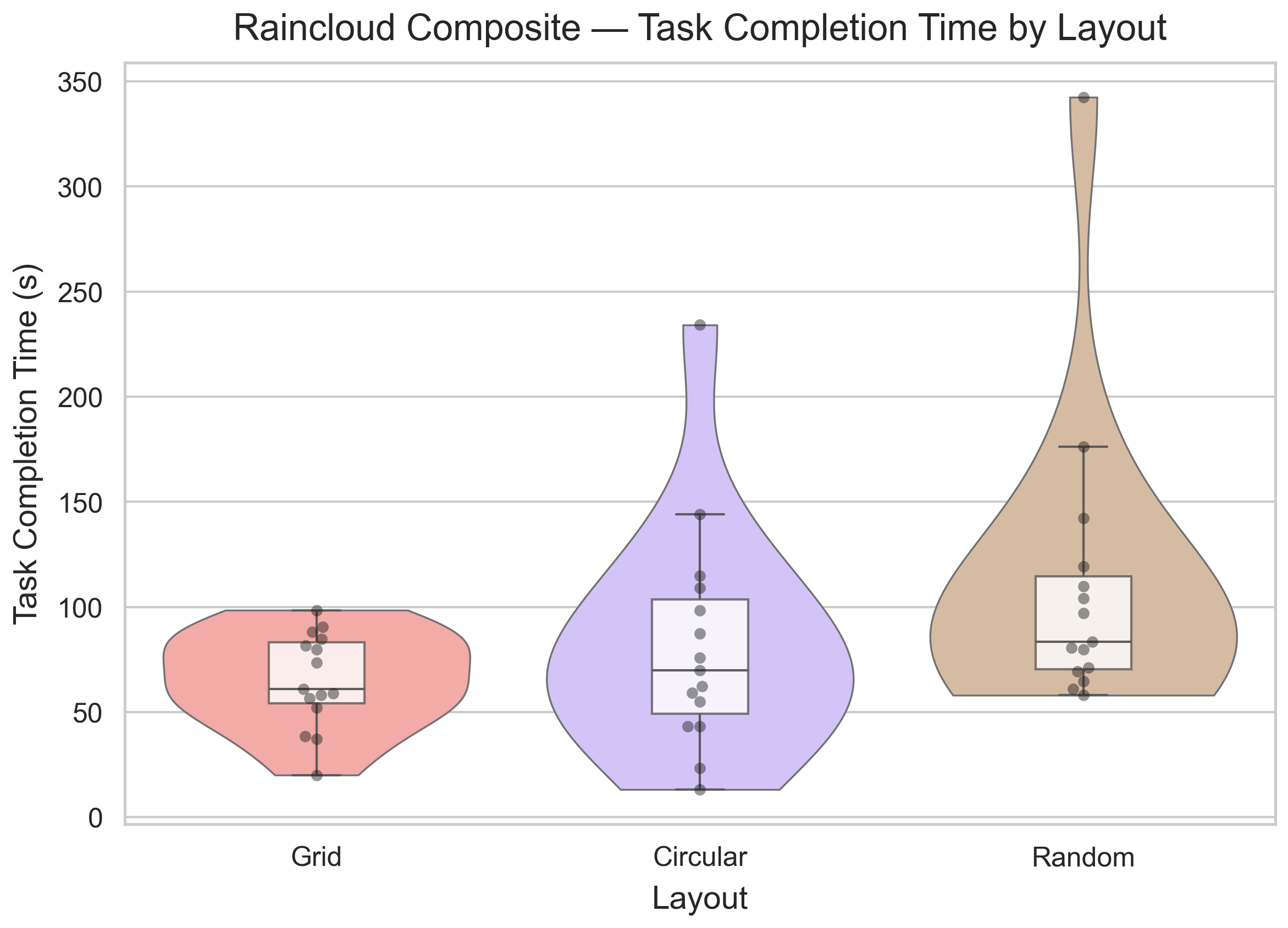}
        \caption{Raincloud composite plot showing the distribution of task completion times across the three spatial layouts for Phase 1. The 'Grid' layout was the fastest, while the 'Random' layout took the longest, highlighting the influence of environmental structure on enumeration efficiency.}
        \label{fig:layout_time}
    \end{subfigure}
    \caption{Distributions of Task Completion Time in Phase 1, analyzed by (a) Instruction Type and (b) Spatial Layout.}
    \label{fig:phase1_tct_distributions}
\end{figure}

The primary effect of the instruction type is clearly illustrated in Figure~\ref{fig:instruction_time}. The 'CountAll' instruction was, on average, the fastest to complete, followed by the 'CountSome' (selective inclusion) instruction, with the 'CountNot' (selective exclusion) instruction taking the longest. The raincloud plot shows that while the distribution for 'CountAll' is relatively compact, the distributions for 'CountSome' and particularly 'CountNot' are wider and have longer tails, indicating greater variability and difficulty among participants for these selective tasks. The within-participant spaghetti plot further reinforces this trend, showing that a majority of individuals followed this increasing time pattern as the task's cognitive demands increased.

% Note: Ensure you have \usepackage{graphicx} and \usepackage{subcaption} in your document preamble.

% \begin{figure}[ht!]
%     \centering
%     \includegraphics[width=0.85\textwidth]{Figures/Phase1/TaskCompletionTime/05_raincloud_instruction_tct_p1.png}
%     \caption{Raincloud composite plot illustrating the distribution of task completion times across the three instruction types for Phase 1. The plot shows the probability density (cloud), the boxplot (rain), and the individual data points (drizzle), clearly indicating that completion time increases from 'CountAll' to 'CountSome' to 'CountNot'.}
%     \label{fig:instruction_time}
% \end{figure}

Similarly, the spatial layout of the items had a discernible impact on task completion time, as shown in Figure~\ref{fig:layout_time}. The 'Grid' layout was consistently the fastest, followed by the 'Circular' layout. The 'Random' layout proved to be the most time-consuming. The spaghetti plot for layouts confirms that most participants were quicker in the structured layouts compared to the unstructured ones.

% \begin{figure}[ht!]
%     \centering
%     \includegraphics[width=0.85\textwidth]{Figures/Phase1/TaskCompletionTime/11_raincloud_layout_tct_p1.png}
%     \caption{Raincloud composite plot showing the distribution of task completion times across the three spatial layouts for Phase 1. The 'Grid' layout was the fastest, while the 'Random' layout took the longest, highlighting the influence of environmental structure on enumeration efficiency.}
%     \label{fig:layout_time}
% \end{figure}

\begin{figure}[ht!]
    \centering
    \begin{subfigure}[b]{0.49\textwidth}
        \centering
        \includegraphics[width=\textwidth]{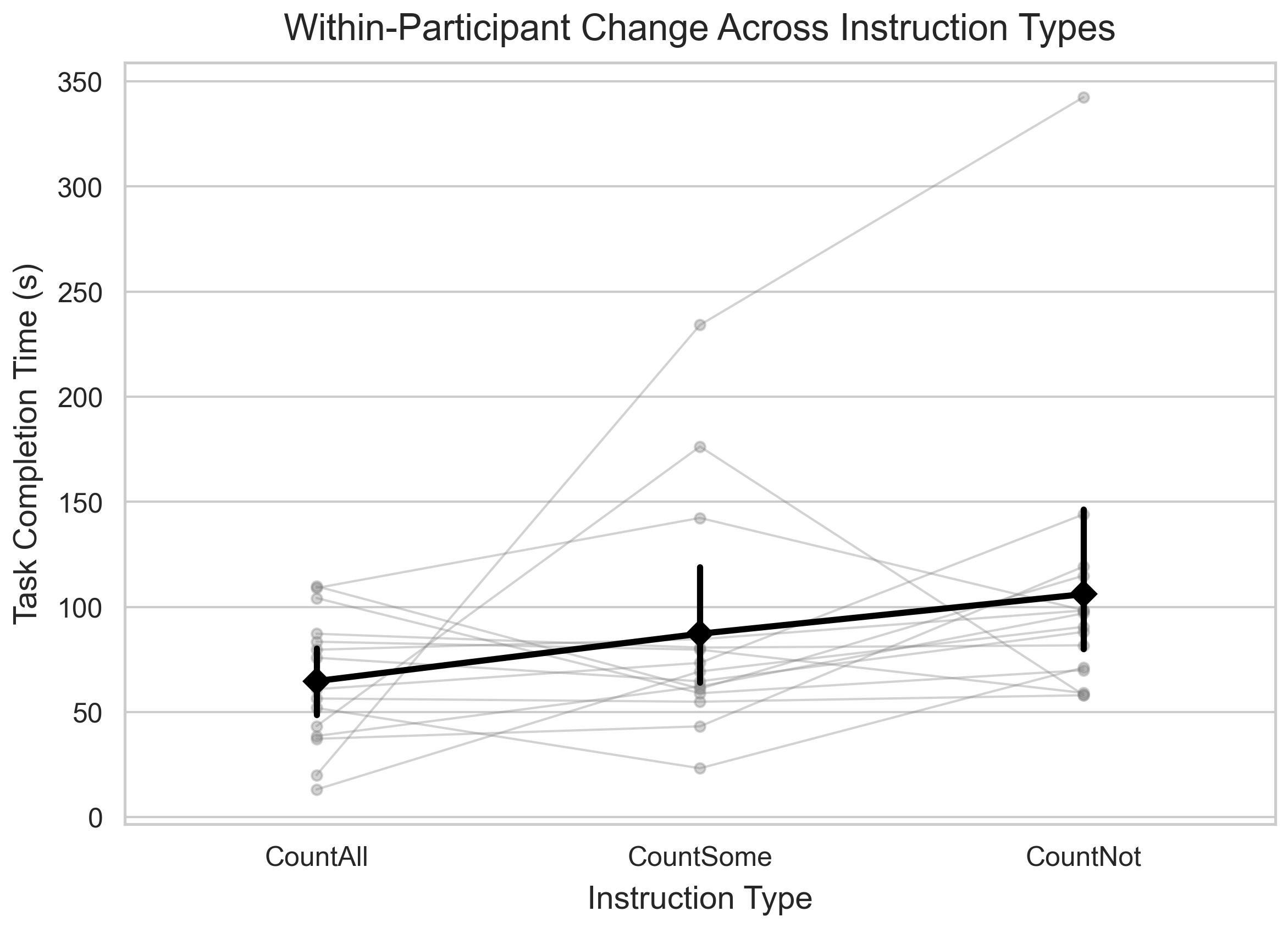}
        \caption{Change across Instruction Types. Most participants took longer as the task moved from `CountAll` to `CountNot`.}
        \label{fig:spaghetti_instruction}
    \end{subfigure}
    \hfill % Adds horizontal space between the subfigures
    \begin{subfigure}[b]{0.49\textwidth}
        \centering
        \includegraphics[width=\textwidth]{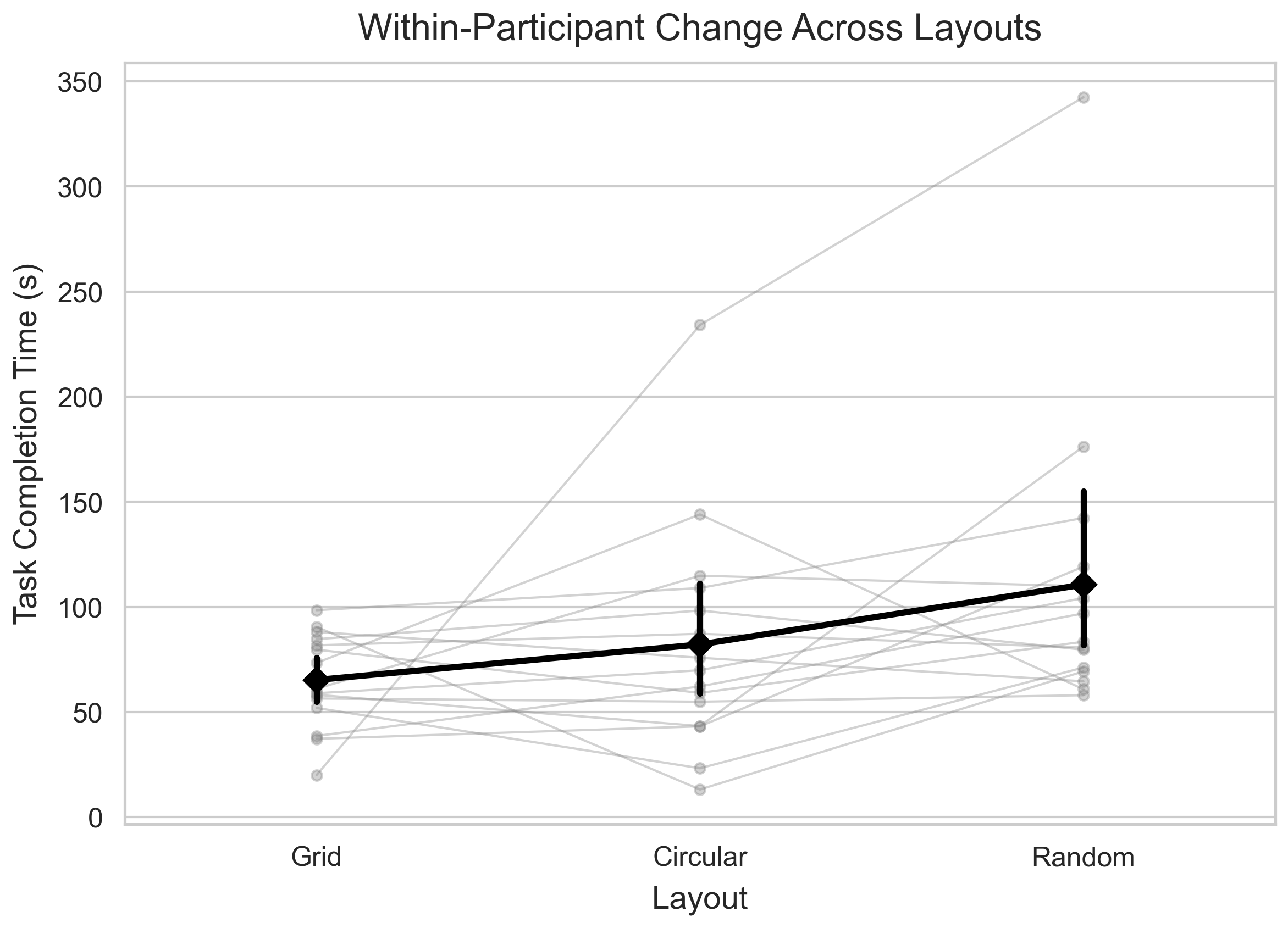}
        \caption{Change across Layouts. The mean trend shows an increase in time from `Grid` to `Random`.}
        \label{fig:spaghetti_layout}
    \end{subfigure}
    \caption{Within-participant spaghetti plots for Task Completion Time in Phase 1, illustrating individual and mean trends across (a) instruction types and (b) spatial layouts.}
    \label{fig:phase1_spaghetti_plots}
\end{figure}

The within-participant trends, visualized in the spaghetti plots in Figure~\ref{fig:phase1_spaghetti_plots}, further reinforce these main effects. A clear majority of individual participants showed a consistent increase in completion time as the task's cognitive demands grew from CountAll to CountNot (Figure~\ref{fig:spaghetti_instruction}). Similarly, most participants completed the task faster in the structured Grid layout and took progressively longer in the Circular and Random layouts (Figure~\ref{fig:spaghetti_layout}), confirming the consistency of these effects at an individual level.

% Note: Ensure you have \usepackage{graphicx} and \usepackage{subcaption} in your document preamble.

\begin{figure}[ht!]
    \centering
    \begin{subfigure}[b]{0.49\textwidth}
        \centering
        \includegraphics[width=\textwidth]{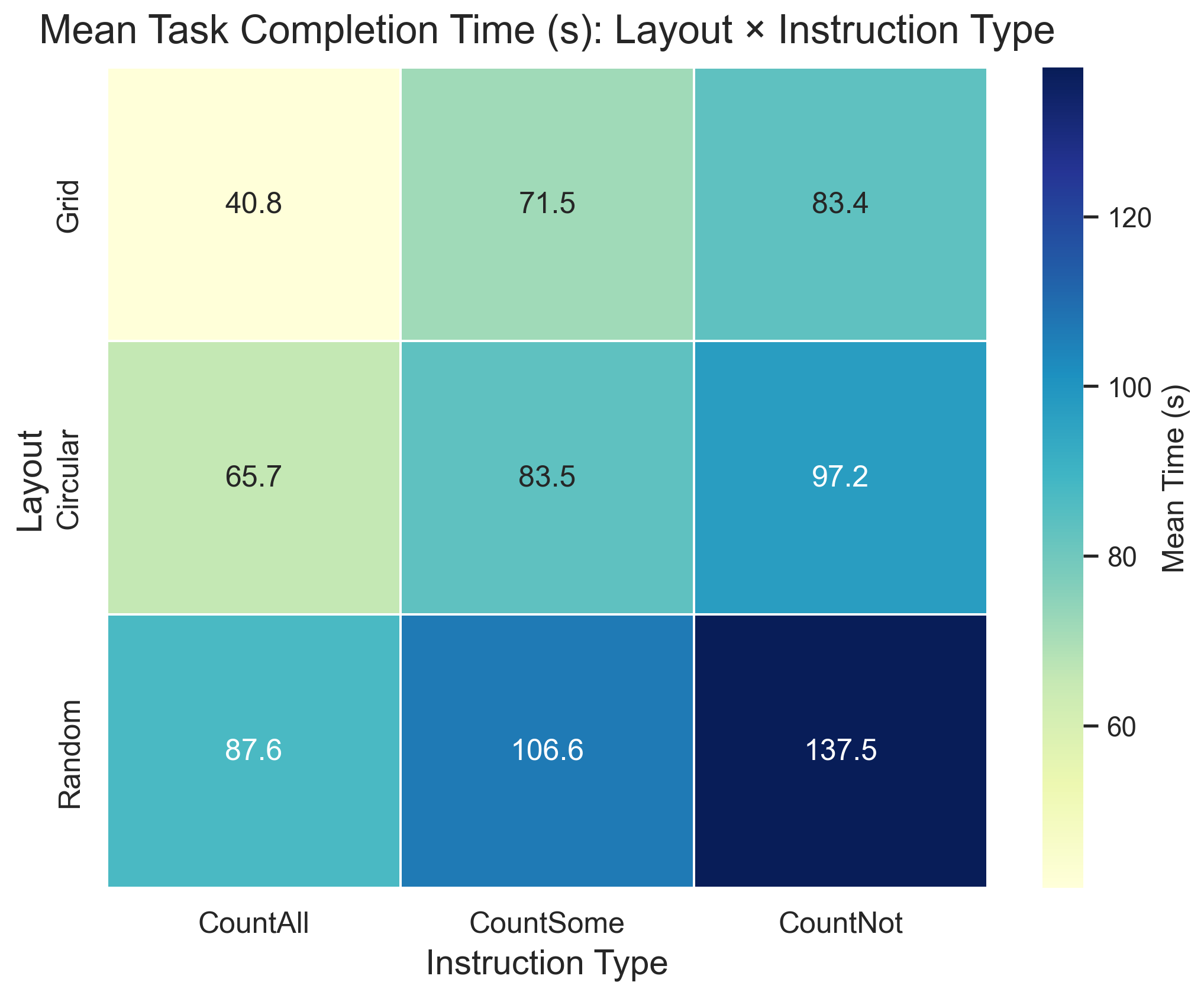}
        \caption{Heatmap of mean task completion times (in seconds) showing the interaction between spatial layout and instruction type. The colour gradient from light yellow (fastest) to dark blue (slowest) illustrates the combined effect of the two variables.}
        \label{fig:interaction_heatmap}
    \end{subfigure}
    \hfill % Adds horizontal space between the subfigures
    \begin{subfigure}[b]{0.49\textwidth}
        \centering
        \includegraphics[width=\textwidth]{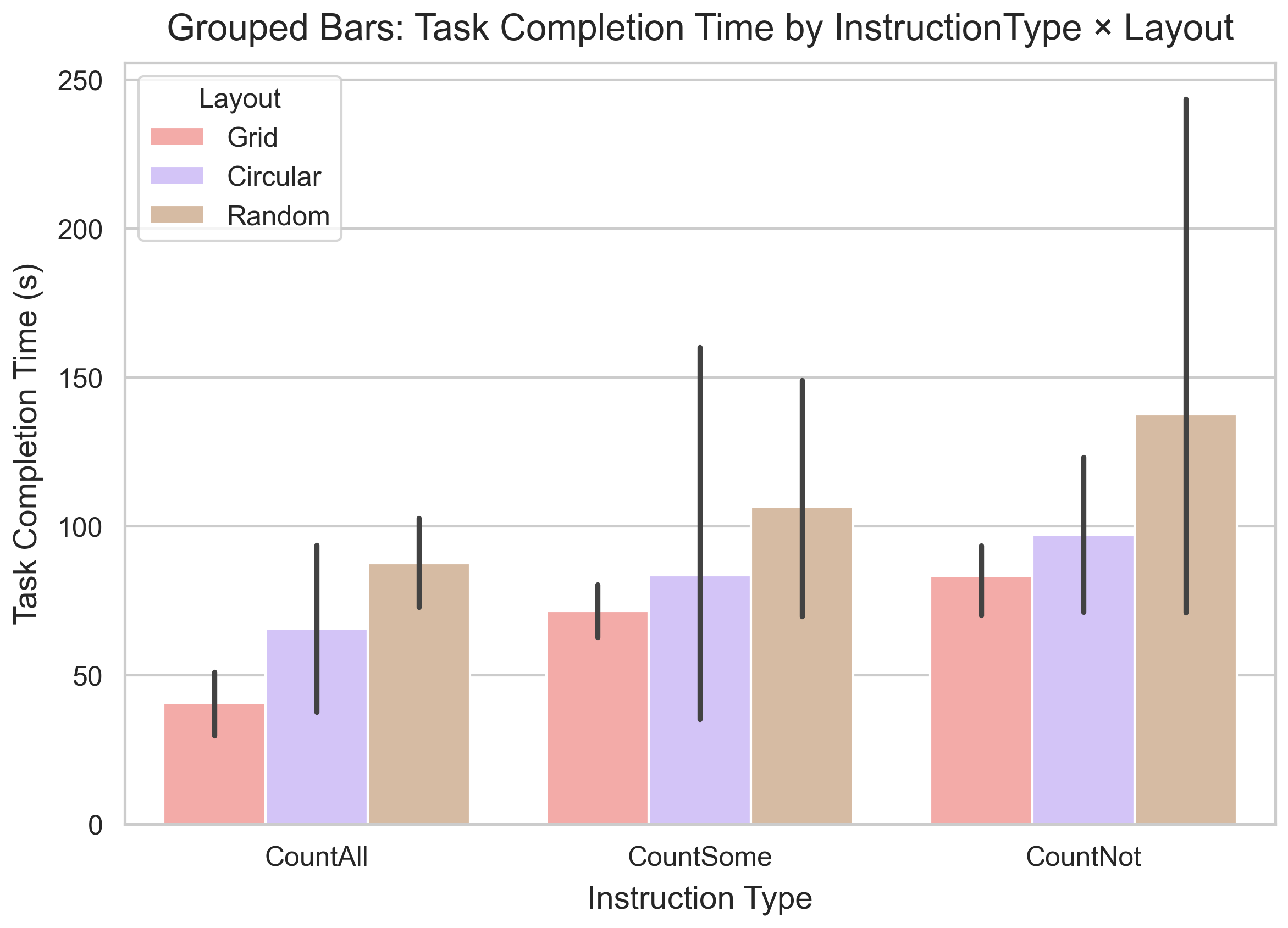}
        \caption{Grouped bar chart showing the mean task completion times with error bars for each combination of layout and instruction type. This visualization highlights the consistent trends and the increase in variance for more complex conditions.}
        \label{fig:interaction_bars}
    \end{subfigure}
    \caption{Interaction effects on Task Completion Time in Phase 1, visualized as (a) a heatmap of mean values and (b) a grouped bar chart with error bars.}
    \label{fig:phase1_tct_interactions}
\end{figure}

The interaction between these two factors provides a more granular view of performance. As depicted in the heatmap in Figure~\ref{fig:interaction_heatmap} and the bar chart in Figure~\ref{fig:interaction_bars}, the effects were compounded. The quickest condition was 'Grid' layout with the 'CountAll' instruction (mean = 40.8 s), representing the simplest task in the most structured environment. Conversely, the most time-intensive condition was the 'Random' layout with the 'CountNot' instruction (mean = 137.5 s), which combined the most complex task intent with the least structured environment. The data consistently shows that for any given layout, the completion time increases from 'CountAll' to 'CountNot', and for any given instruction, the time increases from 'Grid' to 'Random'.

% Note: Ensure you have \usepackage{graphicx} and \usepackage{subcaption} in your document preamble.

% \begin{figure}[ht!]
%     \centering
%     \includegraphics[width=0.85\textwidth]{Figures/Phase1/TaskCompletionTime/13_heatmap_means_layout_by_instruction_tct_p1.png}
%     \caption{Heatmap of mean task completion times (in seconds) showing the interaction between spatial layout and instruction type. The colour gradient from light yellow (fastest) to dark blue (slowest) illustrates the combined effect of the two variables.}
%     \label{fig:interaction_heatmap}
% \end{figure}

% \begin{figure}[ht!]
%     \centering
%     \includegraphics[width=0.85\textwidth]{Figures/Phase1/TaskCompletionTime/14_grouped_bars_interaction_tct_p1.png}
%     \caption{Grouped bar chart showing the mean task completion times with error bars for each combination of layout and instruction type. This visualization highlights the consistent trends and the increase in variance for more complex conditions.}
%     \label{fig:interaction_bars}
% \end{figure}

\paragraph{Discussion}
The observed results for task completion time in Phase 1 strongly support our initial hypotheses. The clear ordering of completion times based on instruction type ('CountAll' $<$ 'CountSome' $<$ 'CountNot') validates \textbf{H1(a)}. The 'CountAll' task served as a simple baseline of serial enumeration. The introduction of a selective inclusion criterion in the 'CountSome' task added a layer of cognitive processing; for each item, a decision had to be made ("is this the target colour?"), which naturally consumed additional time. The 'CountNot' task was the most demanding as it required participants to hold the excluded category in working memory while simultaneously performing an exhaustive search and tally of all other items. This dual-process nature of filtering one category while counting multiple others imposes the highest cognitive load and thus results in the longest completion times.

The findings related to spatial layout provide support for the trend predicted in \textbf{H2(a)}. The efficiency of the 'Grid' layout can be attributed to the low cognitive effort required to devise a search strategy. Its regular structure allows for a simple, systematic scanpath (e.g., left-to-right, top-to-bottom) that minimises the working memory load associated with keeping track of which items have already been counted. The 'Circular' layout also offers a degree of structure that can be systematically followed (e.g., a concentric, spiralling path). The 'Random' layout, lacking any inherent structure, forces participants to generate and maintain their own, more complex search path. This increases the cognitive overhead required to ensure complete coverage without repetition, leading to significantly longer task times.

The interaction effects demonstrate that top-down cognitive factors (task intent) and bottom-up environmental factors (spatial layout) are not independent but rather compound to determine overall task difficulty. The combination of a complex cognitive filter ('CountNot') with an unstructured visual field ('Random') creates a worst-case scenario for enumeration efficiency, as observed in the results.

\subsubsection{Accuracy}

\paragraph{Observations}
The accuracy of enumeration, calculated as the percentage agreement between the participant's reported count and the ground truth, was significantly influenced by the task intent, with a more subtle trend observed for the spatial layout.

\begin{figure}[ht!]
    \centering
    \begin{subfigure}[b]{0.49\textwidth}
        \centering
        \includegraphics[width=\textwidth]{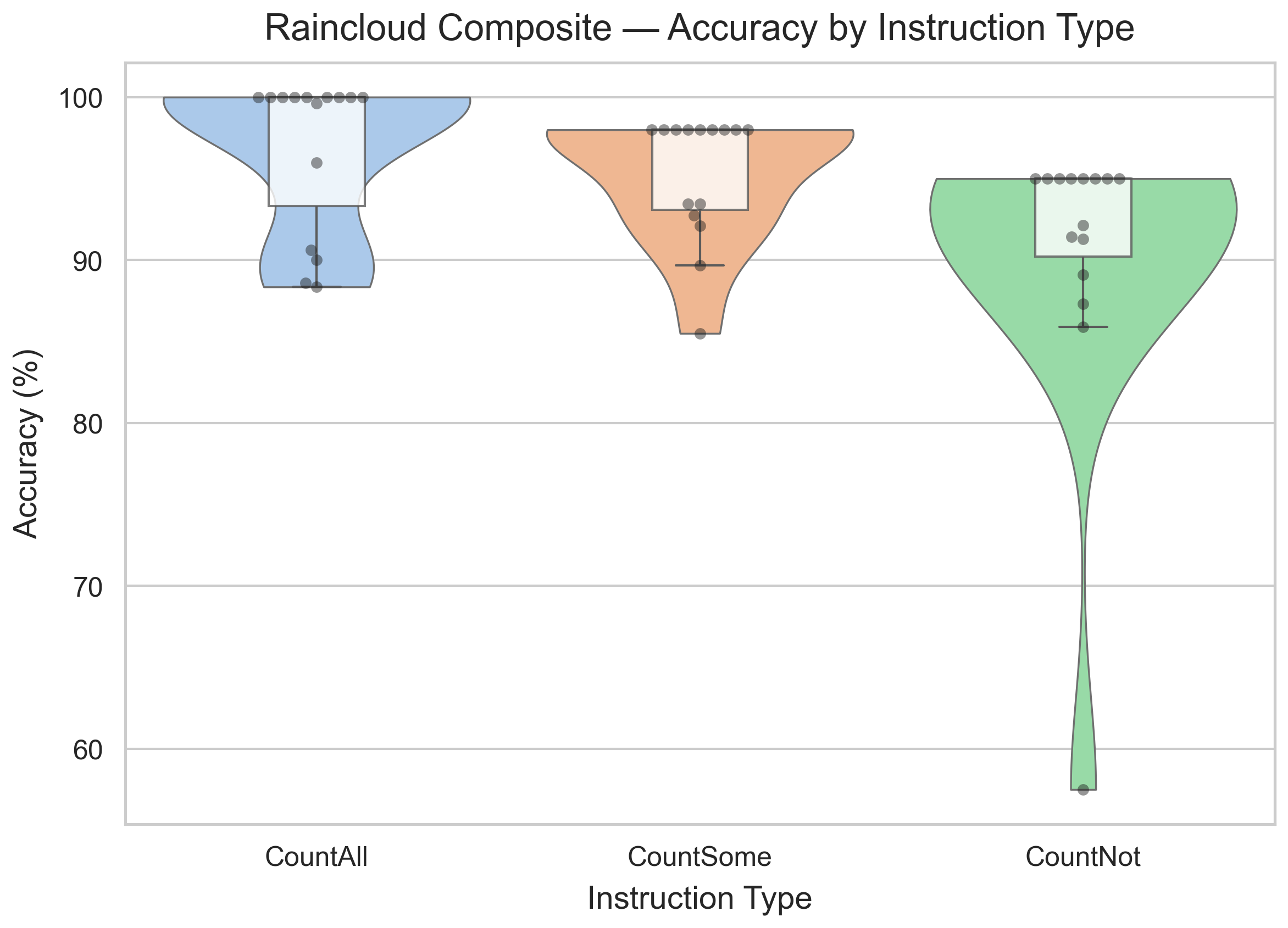}
        \caption{Raincloud composite plot of counting accuracy across the three instruction types. Accuracy was highest and most consistent for 'CountAll', decreasing for 'CountSome', and showing the lowest mean and widest variance for 'CountNot'.}
        \label{fig:instruction_acc}
    \end{subfigure}
    \hfill % Adds horizontal space between the subfigures
    \begin{subfigure}[b]{0.49\textwidth}
        \centering
        \includegraphics[width=\textwidth]{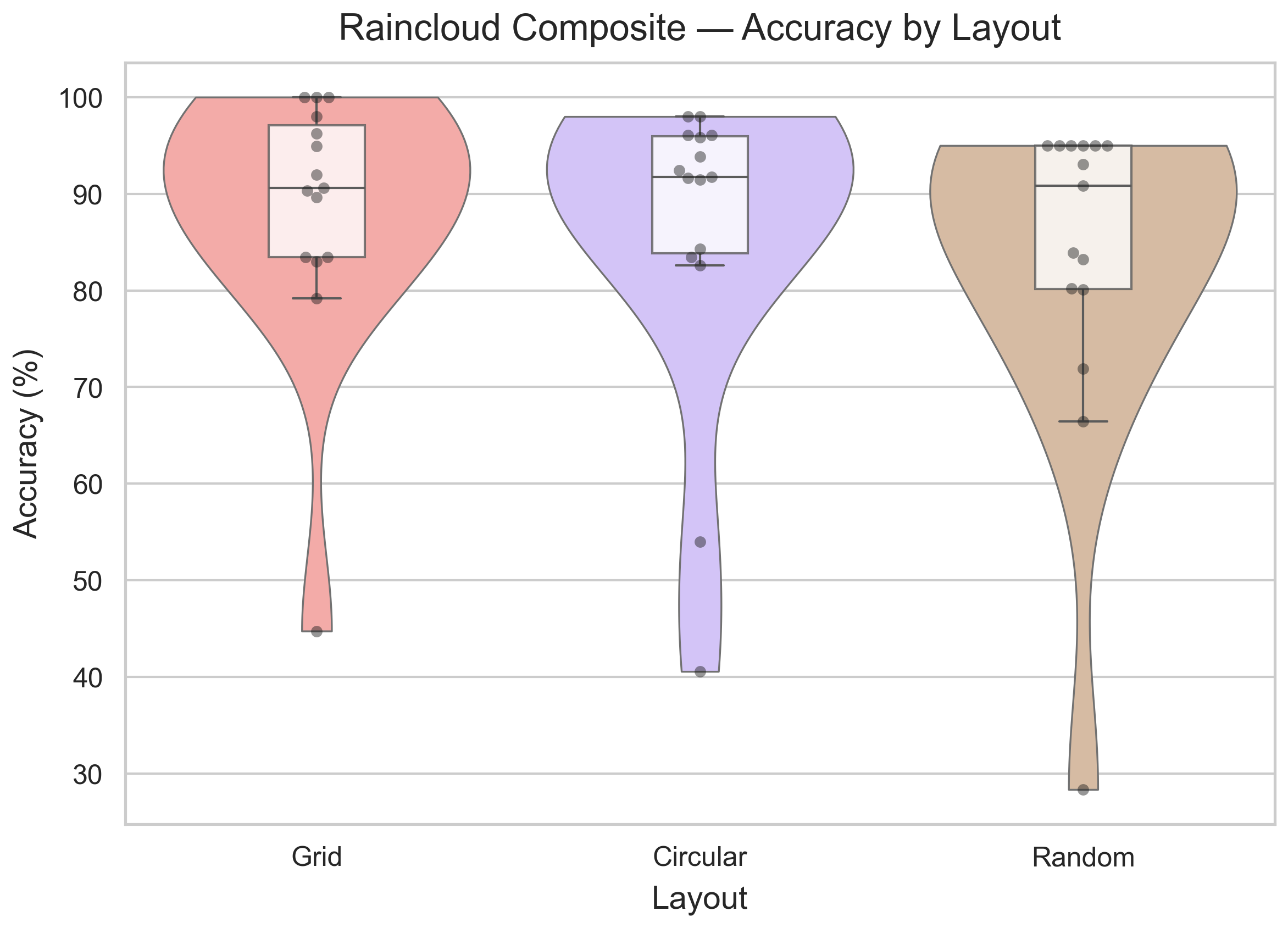}
        \caption{Raincloud composite plot of counting accuracy across the three spatial layouts. A slight downward trend in mean accuracy is visible from 'Grid' to 'Random', but the distributions show significant overlap and variance.}
        \label{fig:layout_acc}
    \end{subfigure}
    \caption{Distributions of Counting Accuracy in Phase 1, analyzed by (a) Instruction Type and (b) Spatial Layout.}
    \label{fig:phase1_accuracy_distributions}
\end{figure}

The most prominent effect was that of the instruction type, as shown in Figure~\ref{fig:instruction_acc}. Participants achieved the highest accuracy in the 'CountAll' condition, with a slight decrease in the 'CountSome' condition. The 'CountNot' condition yielded the lowest mean accuracy and, notably, the highest variability. The raincloud plot for 'CountNot' features a long lower tail, indicating that while many participants remained accurate, a subset found this task exceptionally challenging, leading to substantial undercounts. The within-participant plot confirms that the vast majority of individuals were most accurate in the 'CountAll' task and least accurate in the 'CountNot' task.

% Note: Ensure you have \usepackage{graphicx} and \usepackage{subcaption} in your document preamble.

% \begin{figure}[ht!]
%     \centering
%     \includegraphics[width=0.85\textwidth]{Figures/Phase1/Accuracy/05_raincloud_instruction_acc_p1.png}
%     \caption{Raincloud composite plot of counting accuracy across the three instruction types. Accuracy was highest and most consistent for 'CountAll', decreasing for 'CountSome', and showing the lowest mean and widest variance for 'CountNot'.}
%     \label{fig:instruction_acc}
% \end{figure}

The influence of spatial layout on accuracy was less pronounced, as depicted in Figure~\ref{fig:layout_acc}. On average, there was a slight downward trend in accuracy from the 'Grid' layout, to the 'Circular', and finally to the 'Random' layout. However, the distributions, particularly as seen in the raincloud plot, show considerable overlap. The within-participant spaghetti plot for layout reveals high individual variance; while the mean trend shows a slight decline, individual participant performance did not consistently decrease with less structure, unlike the clear pattern seen for instruction type.

% \begin{figure}[ht!]
%     \centering
%     \includegraphics[width=0.85\textwidth]{Figures/Phase1/Accuracy/11_raincloud_layout_acc_p1.png}
%     \caption{Raincloud composite plot of counting accuracy across the three spatial layouts. A slight downward trend in mean accuracy is visible from 'Grid' to 'Random', but the distributions show significant overlap and variance.}
%     \label{fig:layout_acc}
% \end{figure}

\begin{figure}[ht!]
    \centering
    \begin{subfigure}[b]{0.49\textwidth}
        \centering
        \includegraphics[width=\textwidth]{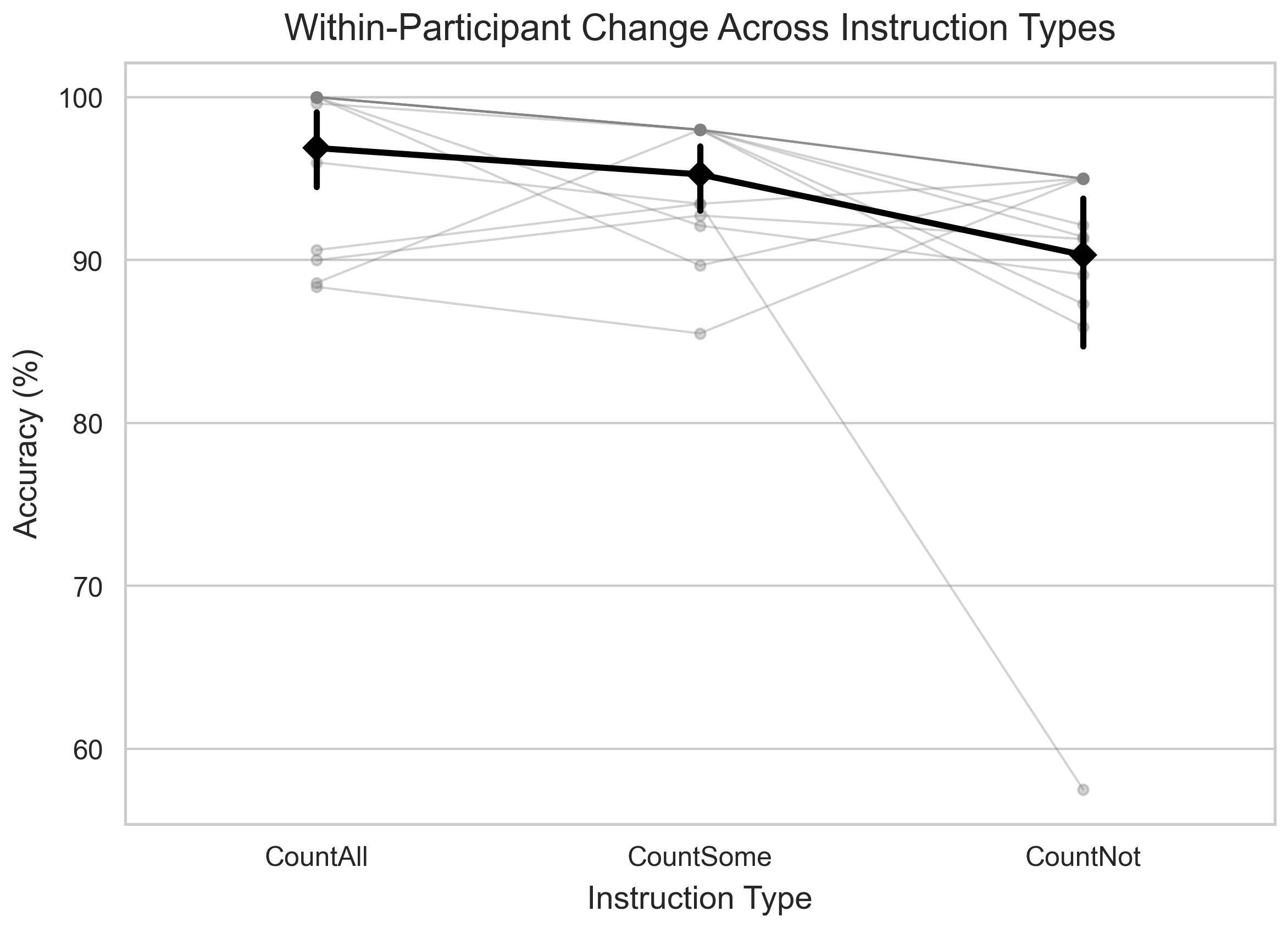}
        \caption{Change across Instruction Types. Most participants' accuracy decreased as task complexity increased.}
        \label{fig:spaghetti_instruction_acc}
    \end{subfigure}
    \hfill % Adds horizontal space between the subfigures
    \begin{subfigure}[b]{0.49\textwidth}
        \centering
        \includegraphics[width=\textwidth]{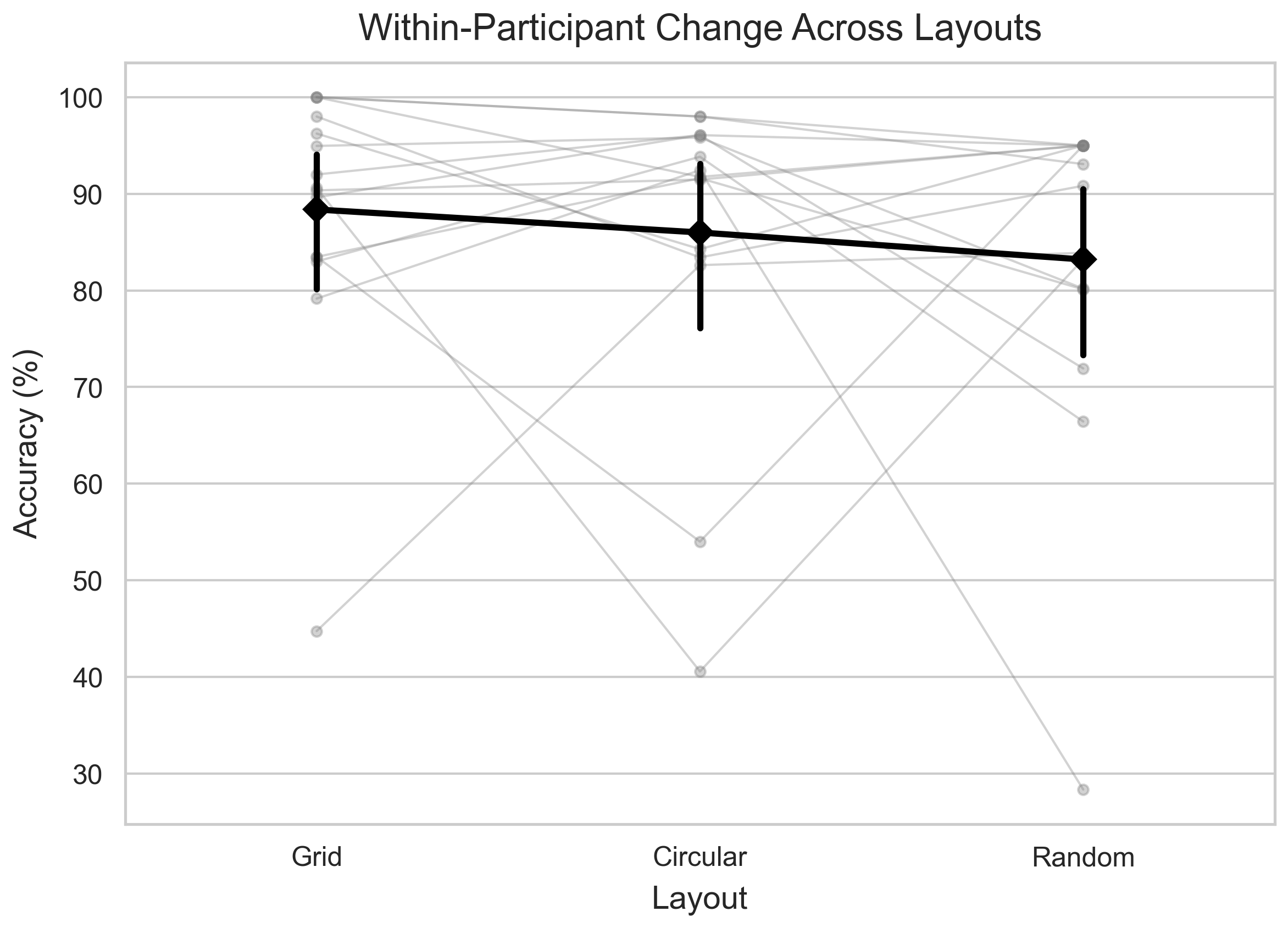}
        \caption{Change across Layouts. Individual performance was highly variable, with a nearly flat mean trend.}
        \label{fig:spaghetti_layout_acc}
    \end{subfigure}
    \caption{Within-participant spaghetti plots for Accuracy in Phase 1, illustrating individual and mean trends across (a) instruction types and (b) spatial layouts.}
    \label{fig:phase1_spaghetti_plots_acc}
\end{figure}

The within-participant trends, shown in Figure~\ref{fig:phase1_spaghetti_plots_acc}, highlight the differing impact of the two variables. The spaghetti plot for instruction type (Figure~\ref{fig:spaghetti_instruction_acc}) shows a clear and consistent pattern, with most individuals exhibiting a decline in accuracy as the task complexity increased from CountAll to CountNot. In contrast, the plot for layout (Figure~\ref{fig:spaghetti_layout_acc}) reveals a highly inconsistent and variable individual performance, with a nearly flat mean trend, underscoring the weak influence of layout on counting accuracy in this phase.

% Note: Ensure you have \usepackage{graphicx} and \usepackage{subcaption} in your document preamble.

\begin{figure}[ht!]
    \centering
    \begin{subfigure}[b]{0.49\textwidth}
        \centering
        \includegraphics[width=\textwidth]{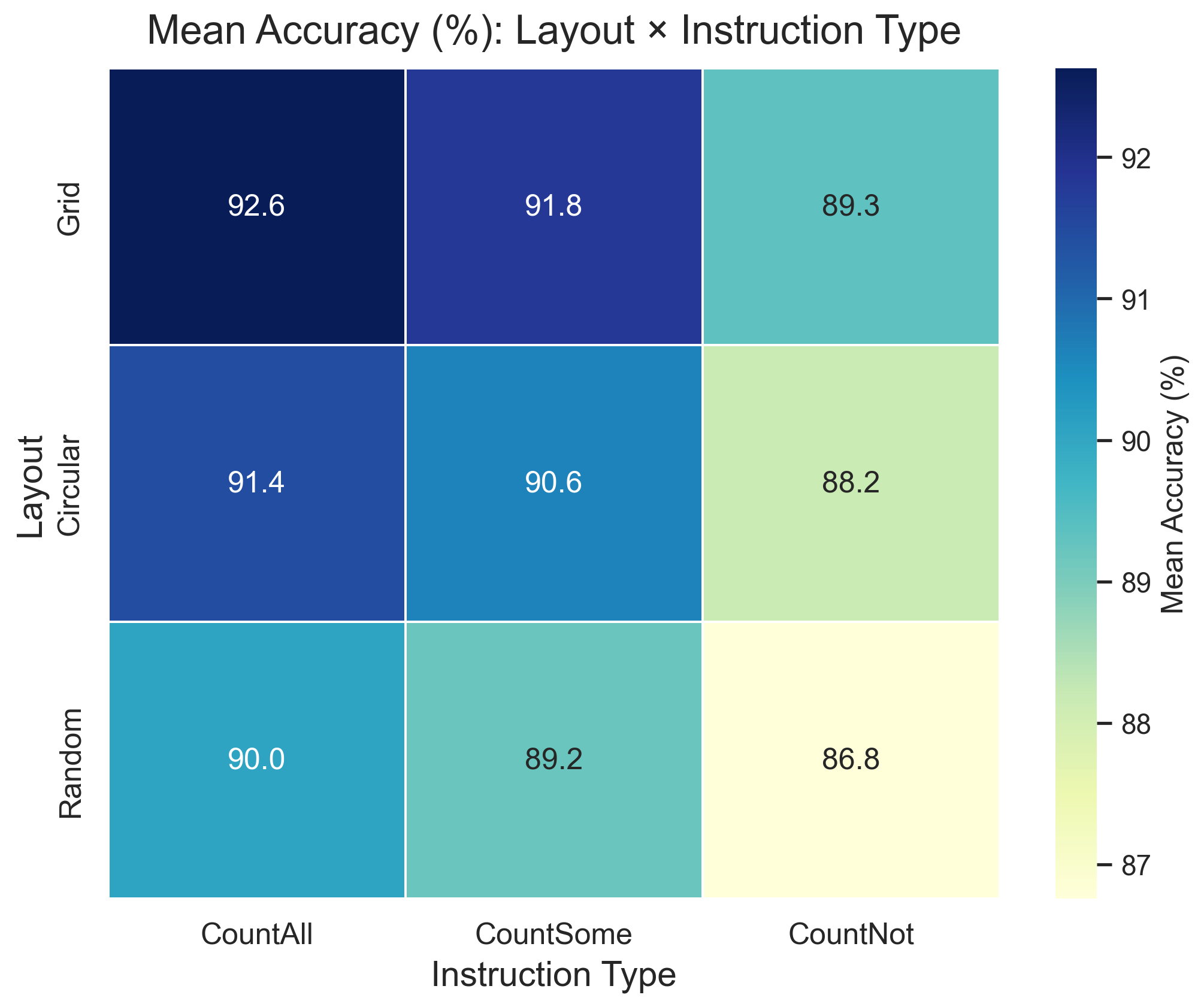}
        \caption{Heatmap of mean accuracy (\%) showing the interaction between spatial layout and instruction type. The highest accuracy is in the top-left ('Grid-CountAll'), and it generally decreases with more complex instructions and less structured layouts.}
        \label{fig:interaction_acc_heatmap}
    \end{subfigure}
    \hfill % Adds horizontal space between the subfigures
    \begin{subfigure}[b]{0.49\textwidth}
        \centering
        \includegraphics[width=\textwidth]{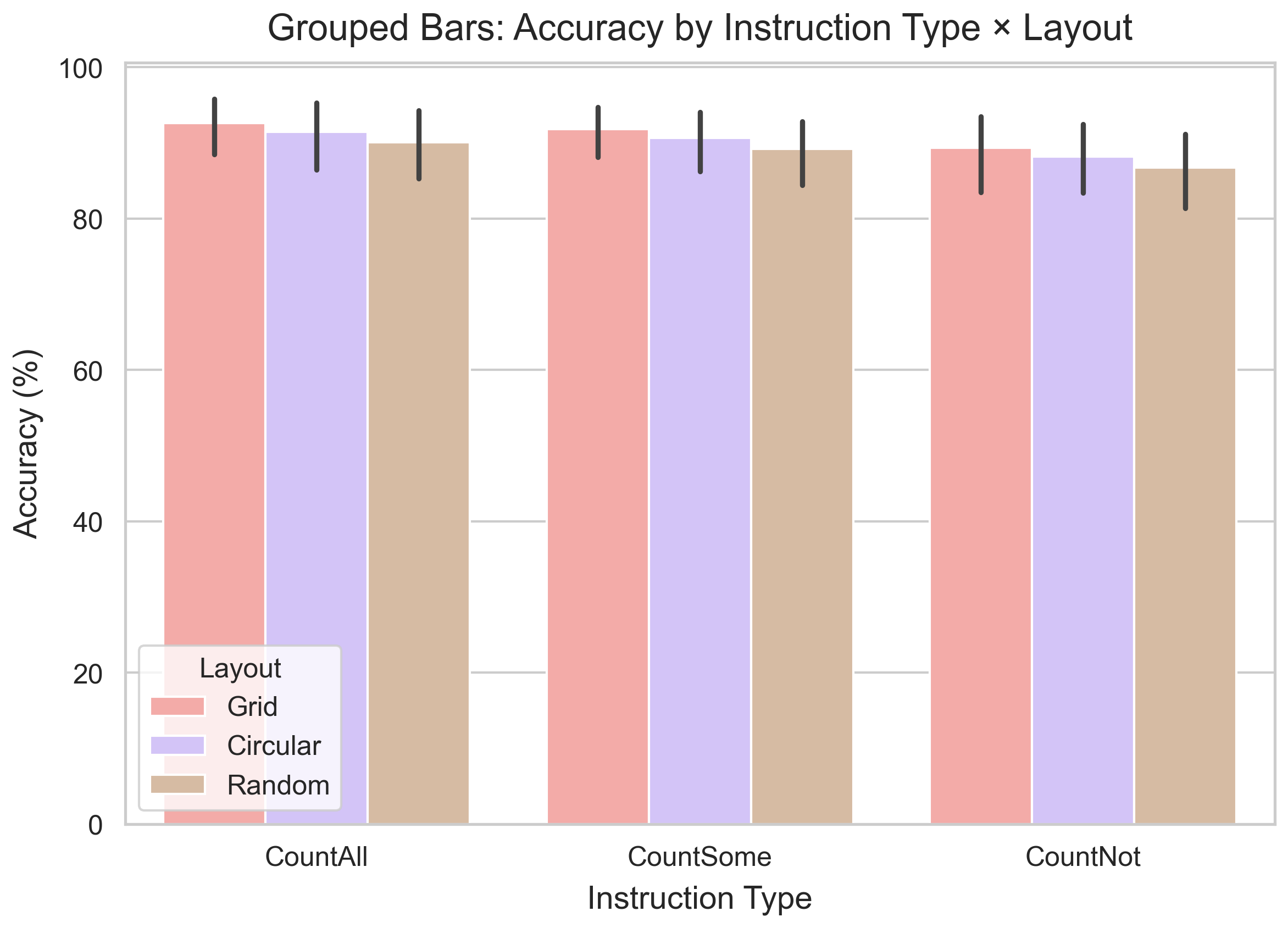}
        \caption{Grouped bar chart showing mean accuracy with error bars for each condition. The chart visually confirms that the effect of instruction type on accuracy is more pronounced than the effect of layout.}
        \label{fig:interaction_acc_bars}
    \end{subfigure}
    \caption{Interaction effects on Counting Accuracy in Phase 1, visualized as (a) a heatmap of mean values and (b) a grouped bar chart with error bars.}
    \label{fig:phase1_accuracy_interactions}
\end{figure}

The interaction between the two factors, detailed in Figure~\ref{fig:interaction_acc_heatmap} and Figure~\ref{fig:interaction_acc_bars}, confirms these primary observations. The highest average accuracy was achieved in the 'Grid-CountAll' condition (92.6\%), while the lowest was in the 'Random-CountNot' condition (86.8\%). The bar chart visualization underscores that within any given instruction type, the differences in accuracy between the three layouts are small and the error bars show substantial overlap. In contrast, the drop in accuracy from 'CountAll' to 'CountNot' is a consistent trend observed across all three layouts.

% Note: Ensure you have \usepackage{graphicx} and \usepackage{subcaption} in your document preamble.

% \begin{figure}[ht!]
%     \centering
%     \includegraphics[width=0.85\textwidth]{Figures/Phase1/Accuracy/13_heatmap_means_layout_by_instruction_acc_p1.png}
%     \caption{Heatmap of mean accuracy (\%) showing the interaction between spatial layout and instruction type. The highest accuracy is in the top-left ('Grid-CountAll'), and it generally decreases with more complex instructions and less structured layouts.}
%     \label{fig:interaction_acc_heatmap}
% \end{figure}

% \begin{figure}[ht!]
%     \centering
%     \includegraphics[width=0.85\textwidth]{Figures/Phase1/Accuracy/14_grouped_bars_interaction_acc_p1.png}
%     \caption{Grouped bar chart showing mean accuracy with error bars for each condition. The chart visually confirms that the effect of instruction type on accuracy is more pronounced than the effect of layout.}
%     \label{fig:interaction_acc_bars}
% \end{figure}

\paragraph{Discussion}
The accuracy results provide strong validation for \textbf{H1(b)}, which predicted that selective counting instructions would lead to lower performance. The decline in accuracy from 'CountAll' to 'CountSome', and further to 'CountNot', is a direct reflection of increasing cognitive load. The 'CountAll' task, being a straightforward enumeration, is less prone to error. The 'CountSome' task introduces decision-making at each item, creating opportunities for two types of errors: failing to identify a target (omission) or incorrectly identifying a non-target (commission). The 'CountNot' task exacerbates this challenge. It requires a continuous and effortful inhibition of the excluded category while maintaining a running tally of all other items. This complex demand on executive function and working memory makes it the most error-prone condition, as reflected by the data.

The trend observed for spatial layouts, while more subtle, aligns with the prediction in \textbf{H2(b)}. The 'Grid' layout's structure provides a strong scaffold for the visual search process. This external organisation reduces the load on internal working memory for tracking visited locations, thereby minimising the chances of skipping an item or counting it twice. Conversely, the 'Random' layout offers no such external support, placing the full burden of path planning and progress monitoring on the participant. This increased demand on spatial working memory can lead to a slightly higher rate of errors. The high individual variability suggests that while structure is generally helpful, some participants are adept at developing effective personal strategies for navigating even unstructured environments, mitigating the potential drop in accuracy. The results clearly indicate that while both task intent and spatial layout influence counting accuracy, the cognitive demands imposed by the task's filtering requirements ('CountNot' vs. 'CountAll') are a more powerful determinant of performance than the structure of the environment itself.

\subsubsection{Statistical Significance}

\paragraph{Effect of Task Intent (Instruction Type)}
The statistical analysis revealed that the task intent had a significant impact on participant performance. 
\begin{itemize}
    \item For \textbf{Task Completion Time}, both the Friedman test ($\chi^2(2)=6.933, p=0.031$) and the Kruskal-Wallis test ($H=8.618, p=0.013$) showed a statistically significant effect of the instruction type. This confirms that the differences in time taken to complete the 'CountAll', 'CountSome', and 'CountNot' tasks were not due to random chance. 
    \item For \textbf{Accuracy}, the effect of the instruction type was even more pronounced. The Friedman test ($\chi^2(2)=10.800, p=0.005$) and the Kruskal-Wallis test ($H=11.106, p=0.004$) were both highly significant. The post-hoc analysis further clarified this result; the Wilcoxon signed-rank test revealed a statistically significant difference in accuracy between the 'CountAll' and 'CountNot' conditions, even after applying the Holm correction ($p_{holm}=0.030$). This provides strong evidence that participants were significantly less accurate when performing the most complex selective exclusion task compared to the baseline counting task.
\end{itemize}

\paragraph{Effect of Spatial Layout}
In contrast to the strong effect of the instruction type, the spatial layout of the items did not have a statistically significant influence on performance in this phase.
\begin{itemize}
    \item For \textbf{Task Completion Time}, neither the Friedman test ($\chi^2(2)=1.733, p=0.420$) nor the Kruskal-Wallis test ($H=3.419, p=0.181$) found a significant difference among the 'Grid', 'Circular', and 'Random' layouts. 
    \item Similarly, for \textbf{Accuracy}, the Friedman test ($\chi^2(2)=1.733, p=0.420$) and the Kruskal-Wallis test ($H=1.201, p=0.549$) were also not significant. The post-hoc tests confirmed that none of the pairwise comparisons between layouts reached statistical significance for either time or accuracy.
\end{itemize}

\paragraph{Discussion}
These statistical results provide clear and robust support for \textbf{Hypothesis H1}, which posited that selective counting instructions would impose a greater cognitive load and result in lower performance. The significant p-values confirm that the changes in task completion time and accuracy based on the instruction type are reliable effects. However, the results do not provide statistical support for \textbf{Hypothesis H2}, which predicted that structured layouts would lead to more efficient performance. While the descriptive data showed trends in the expected direction (i.e., 'Grid' being faster and more accurate than 'Random'), the variability in performance was too high relative to the mean differences for the effect to be statistically significant. This suggests that for the task of enumerating simple abstract shapes, the top-down cognitive demands of the task's \textit{intent} are a much stronger determinant of performance than the bottom-up influence of the \textit{spatial layout}.

\subsubsection{Gaze Distribution}

\begin{figure}[ht!]
    \centering
    \includegraphics[width=0.85\textwidth]{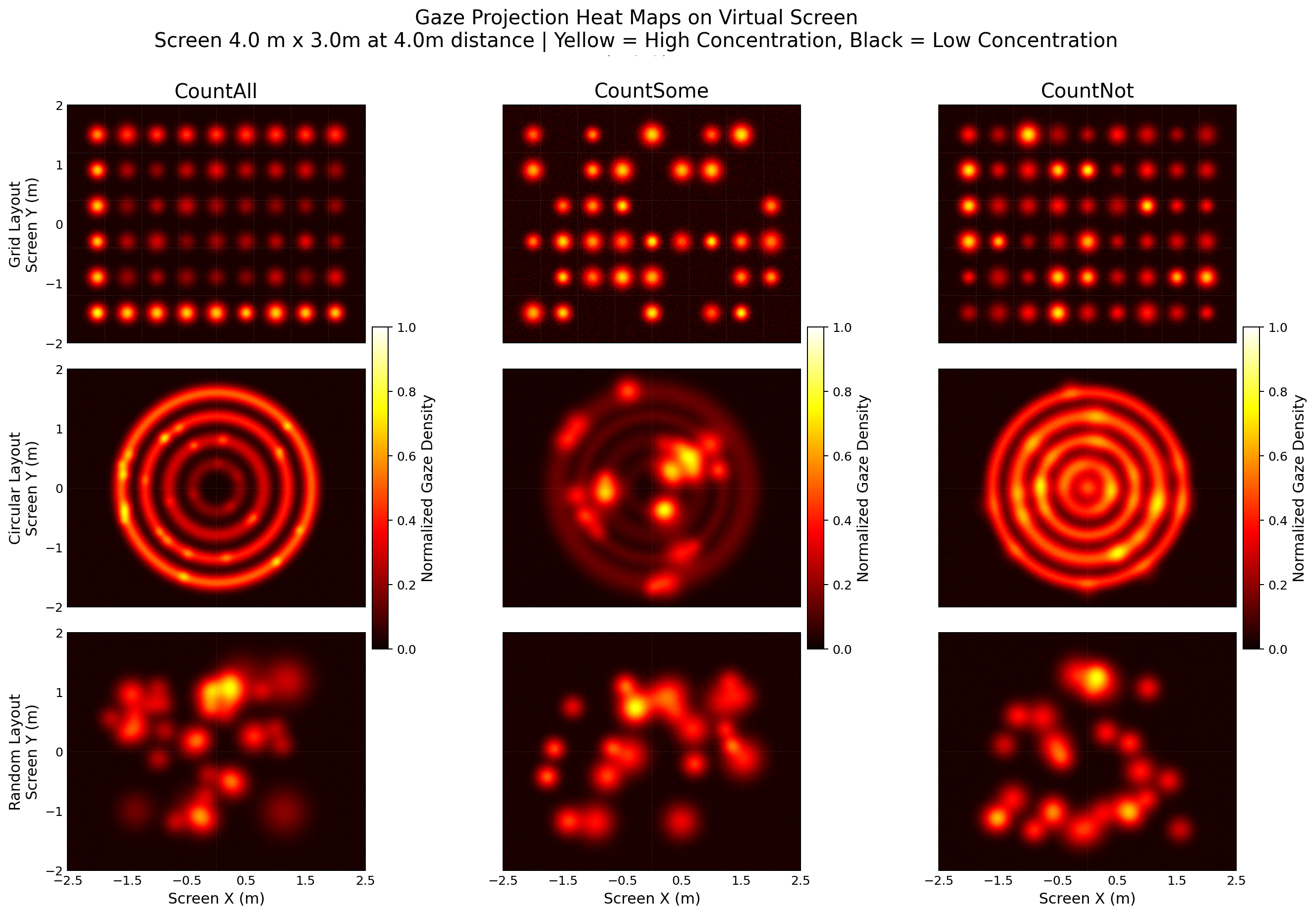}
    \caption{Aggregated gaze density heatmaps for Phase 1, showing the interaction between layout (rows) and instruction type (columns). Yellow indicates high gaze concentration. The 'CountSome' instruction consistently produced sparse, focal gaze patterns, while 'CountAll' and 'CountNot' produced exhaustive gaze coverage, with the underlying pattern being shaped by the spatial layout.}
    \label{fig:gaze_heatmaps}
\end{figure}

\paragraph{Observations}
To understand the visual search strategies employed by participants, we aggregated and visualised the gaze data from all trials into gaze density heatmaps. These heatmaps, presented in Figure~\ref{fig:gaze_heatmaps}, illustrate the collective allocation of visual attention across the different combinations of spatial layout and task instruction. Yellow and bright red areas indicate high gaze concentration, while dark red and black areas indicate locations that received little to no foveal attention.

A striking and consistent pattern emerged based on the task intent. Across all three layouts ('Grid', 'Circular', and 'Random'), the gaze distribution for the 'CountSome' (selective inclusion) instruction was markedly different from the other two. In the 'CountSome' condition, the heatmaps show a sparse pattern of distinct, high-intensity hotspots, with large intervening areas remaining dark. This indicates that participants focused their visual attention only on a subset of the available item locations. In stark contrast, the heatmaps for both the 'CountAll' and 'CountNot' conditions show a much more comprehensive and uniform pattern of gaze coverage. In these conditions, nearly every potential item location is illuminated, signifying that participants visually inspected almost all the items present in the scene.

The underlying spatial layout clearly structured these gaze patterns. 
\begin{itemize}
    \item In the \textbf{Grid layout}, the 'CountAll' and 'CountNot' conditions resulted in a complete, illuminated grid, whereas the 'CountSome' condition produced a pattern resembling a sparse checkerboard, with only the target item positions being strongly activated.
    \item In the \textbf{Circular layout}, the 'CountAll' and 'CountNot' conditions show that participants' gaze thoroughly traced the paths of the concentric rings. The 'CountSome' condition, however, shows these rings as broken or dotted, with gaze concentrated only on specific points along the circular paths.
    \item In the \textbf{Random layout}, a similar logic applies. While the 'CountAll' and 'CountNot' conditions show broad coverage of the scattered item locations, the 'CountSome' condition reveals a sparser constellation of attended locations.
\end{itemize}

\paragraph{Discussion}
The analysis of gaze distribution provides powerful visual evidence that validates \textbf{Hypothesis H2}. The data clearly demonstrate that participants' attentional strategies were systematically and profoundly modulated by both the task intent and the spatial layout.

The most important insight is the difference in visual strategy driven by the instruction type. The focal gaze patterns in the 'CountSome' condition are a direct manifestation of top-down attentional filtering. Participants successfully used the task goal to guide their search, allocating their limited foveal processing resources primarily to target items while efficiently ignoring distractors. This is a classic demonstration of selective attention in a visual search task.

Conversely, the remarkable similarity between the 'CountAll' and 'CountNot' heatmaps reveals that, despite having different objectives, they demand an almost identical underlying visual search strategy: exhaustive inspection. To be certain of what \textit{not} to count in the 'CountNot' task, participants had to fixate on nearly every item to check its identity against the exclusion criterion, just as they had to fixate on every item in the 'CountAll' task to include it in their tally. This explains why both tasks resulted in similarly long completion times and shows that the cognitive load in the 'CountNot' task comes not from a different search pattern, but from the more complex decision-making required at each fixation.

Finally, the heatmaps also support \textbf{H2(c)}, which predicted that the layouts would elicit distinct scanpath patterns. The structure of the gaze patterns—be it a grid, concentric circles, or a random scatter—directly mirrored the structure of the stimulus layout. This indicates that participants' search behaviour was strongly scaffolded by the bottom-up features of the environment. In summary, the gaze data reveals a clear division of labour: the task intent dictated \textit{what} to look at (selectivity), while the spatial layout dictated \textit{how} to look for it (strategy).

\subsubsection{Cognitive Load}

\paragraph{Observations}
To assess the subjective difficulty of the tasks, participants were asked to rate the perceived mental demand of each trial on a scale of 1 (very low) to 5 (very high). The analysis of these ratings reveals that participants' subjective experience of cognitive load was systematically influenced by both the task intent and the spatial layout, mirroring the objective performance metrics of time and accuracy.

\begin{figure}[ht!]
    \centering
    \begin{subfigure}[b]{0.49\textwidth}
        \centering
        \includegraphics[width=\textwidth]{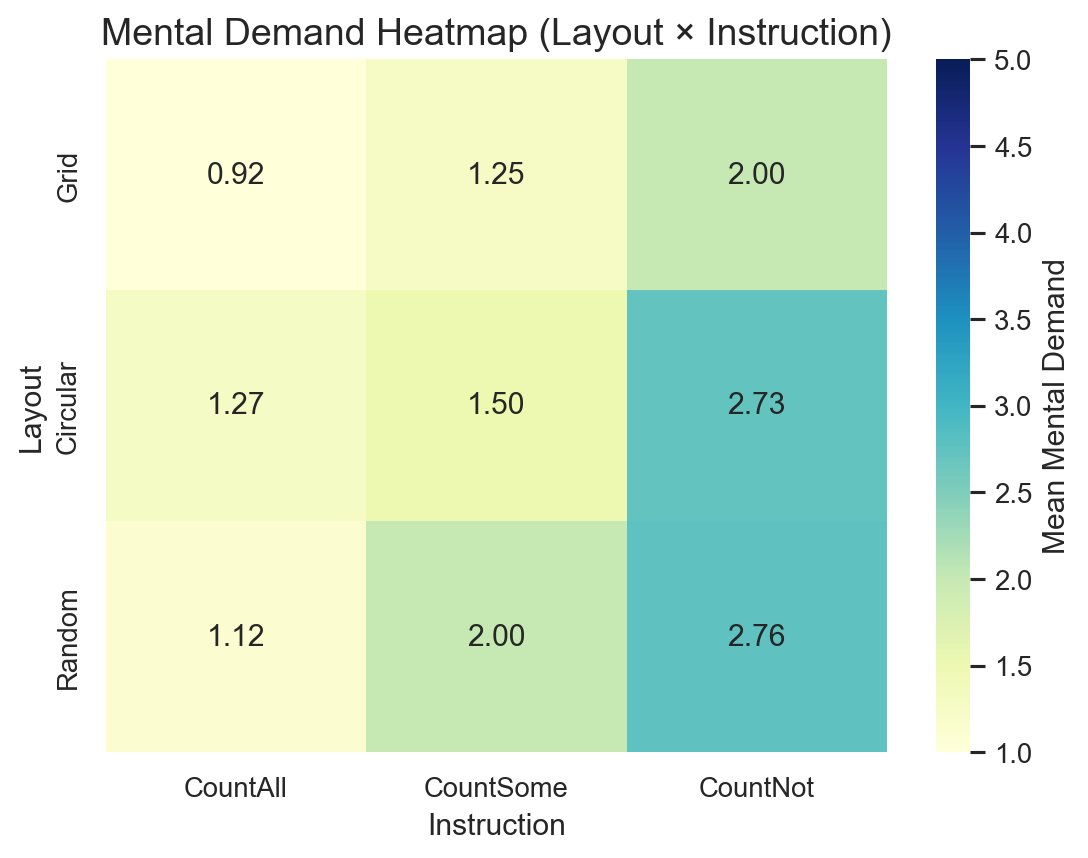}
        \caption{Heatmap of mean self-reported mental demand ratings (1-5 scale). The data shows a clear increase in cognitive load with more complex instructions (left to right) and, to a lesser extent, with less structured layouts (top to bottom).}
        \label{fig:mental_demand_heatmap}
    \end{subfigure}
    \hfill % Adds horizontal space between the subfigures
    \begin{subfigure}[b]{0.49\textwidth}
        \centering
        \includegraphics[width=\textwidth]{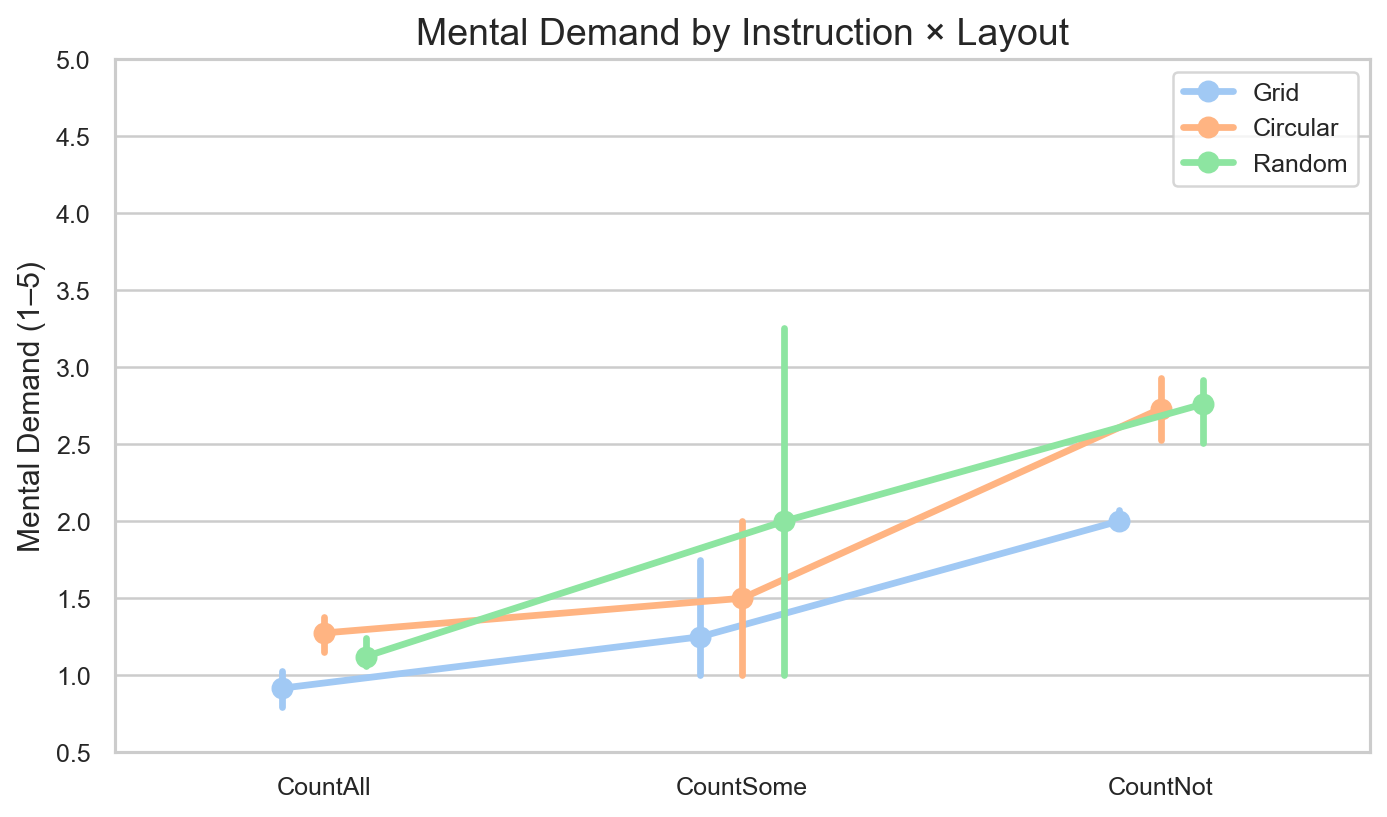}
        \caption{Interaction line plot showing mean mental demand by instruction type for each layout. All layouts show a consistent increase in perceived difficulty from 'CountAll' to 'CountNot', with the 'Grid' layout consistently rated as the least demanding.}
        \label{fig:mental_demand_lines}
    \end{subfigure}
    \caption{Visualizations of self-reported cognitive load in Phase 1, showing the interaction between instruction type and spatial layout as (a) a heatmap and (b) a line plot.}
    \label{fig:phase1_cognitiveload_visuals}
\end{figure}

The most substantial effect was driven by the instruction type. As clearly visualized in the interaction plot in Figure~\ref{fig:mental_demand_lines}, there was a strong, monotonic increase in reported mental demand as the task's filtering requirements became more complex. For all three layouts, the 'CountAll' instruction was rated as the least demanding, followed by 'CountSome', with 'CountNot' being rated as the most mentally demanding. The heatmap in Figure~\ref{fig:mental_demand_heatmap} quantifies this trend, showing that the lowest-rated condition was 'Grid-CountAll' (mean = 0.92), while the highest-rated was 'Random-CountNot' (mean = 2.76).

The spatial layout also had a consistent effect. The 'Grid' layout was perceived as the least mentally demanding across all instruction types. The 'Circular' and 'Random' layouts were rated as more demanding, with the 'Random' layout generally perceived as requiring the most mental effort, particularly in the more complex 'CountSome' and 'CountNot' conditions. Furthermore, the error bars in the line plot (Figure~\ref{fig:mental_demand_lines}) are visibly larger for the 'CountSome' and 'CountNot' conditions, especially in the 'Random' layout, indicating greater inter-participant variability in the perceived difficulty of these more challenging tasks.

% Note: Ensure you have \usepackage{graphicx} and \usepackage{subcaption} in your document preamble.

% \begin{figure}[ht!]
%     \centering
%     \includegraphics[width=0.85\textwidth]{Figures/Phase1/CognitiveLoad/02_heatmap_mental_p1.png}
%     \caption{Heatmap of mean self-reported mental demand ratings (1-5 scale). The data shows a clear increase in cognitive load with more complex instructions (left to right) and, to a lesser extent, with less structured layouts (top to bottom).}
%     \label{fig:mental_demand_heatmap}
% \end{figure}

% \begin{figure}[ht!]
%     \centering
%     \includegraphics[width=0.85\textwidth]{Figures/Phase1/CognitiveLoad/01_interaction_lines_mental_p1.png}
%     \caption{Interaction line plot showing mean mental demand by instruction type for each layout. All layouts show a consistent increase in perceived difficulty from 'CountAll' to 'CountNot', with the 'Grid' layout consistently rated as the least demanding.}
%     \label{fig:mental_demand_lines}
% \end{figure}

\paragraph{Discussion}
The subjective ratings of cognitive load provide strong convergent evidence that our experimental manipulations successfully modulated task difficulty as intended, and they lend robust support to our hypotheses. The findings directly validate \textbf{H1(c)}, which predicted that self-reported mental demand would increase with the complexity of the filtering instruction. The lowest ratings for 'CountAll' reflect a baseline enumeration effort. The intermediate ratings for 'CountSome' correspond to the added cognitive step of making a categorical decision for each item. The highest ratings for 'CountNot' align with the dual-task nature of this condition, which requires both the inhibition of a specific category and the exhaustive enumeration of all remaining items, placing the greatest demand on working memory and executive control.

Furthermore, the results align with the trends predicted for layout efficiency in \textbf{Hypothesis H2}. Participants' perception that the 'Grid' layout was the least mentally demanding is consistent with the idea that environmental structure can offload cognitive work. A highly organized layout provides an external scaffold for the search process, reducing the internal cognitive burden of planning a scanpath and maintaining a memory of visited locations. Conversely, the 'Random' layout provides no such external support, forcing the participant to rely entirely on their own cognitive resources to manage the search, an effort that is correctly perceived as being more mentally demanding. In essence, the cognitive load data acts as a bridge between the objective performance metrics (time, accuracy) and the underlying mental experience, confirming that the tasks that took longer and resulted in more errors were indeed the ones that participants consciously experienced as being the most difficult.

\subsubsection{Recall Performance}

\paragraph{Observations}
The final measure of performance we assessed was memory recall, defined as the number of items participants could correctly remember from the scene in the post-trial questionnaire. The results indicate that memory performance was strongly influenced by both the task intent and, to a lesser degree, the spatial layout.

\begin{figure}[ht!]
    \centering
    \begin{subfigure}[b]{0.49\textwidth}
        \centering
        \includegraphics[width=\textwidth]{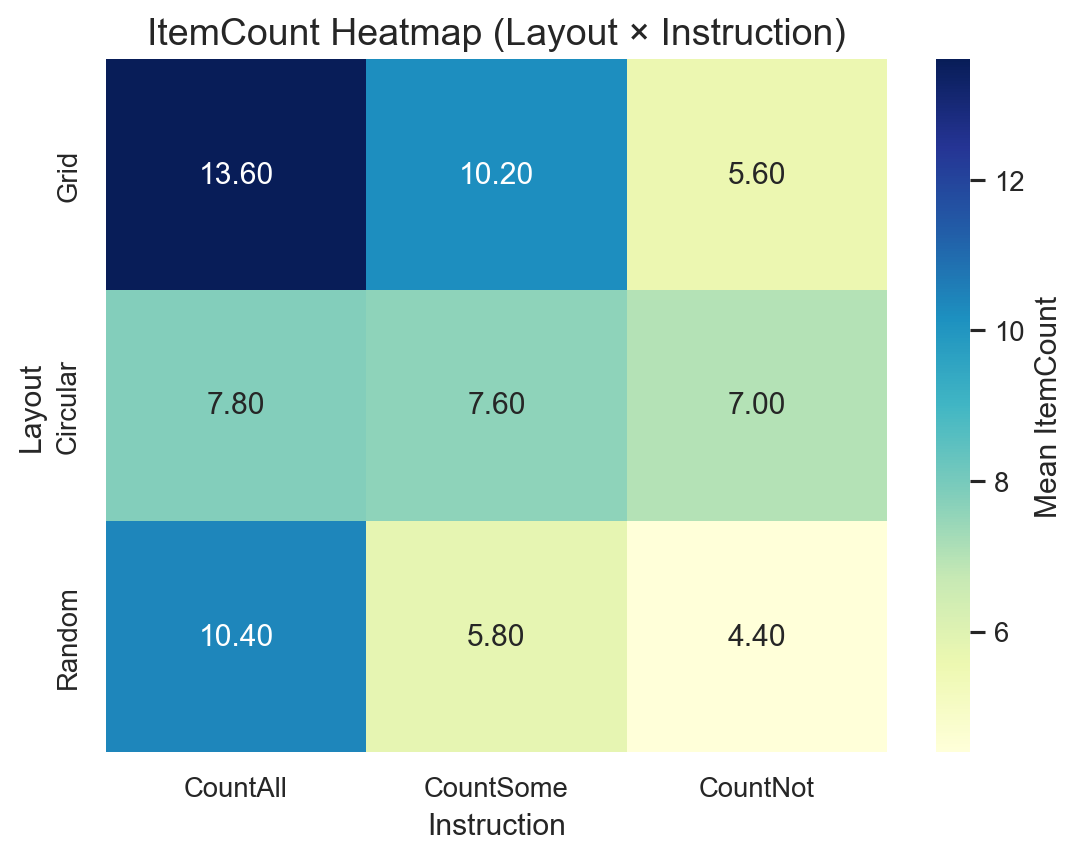}
        \caption{Heatmap of mean item recall, showing a strong decrease in the number of remembered items as the instruction type becomes more selective. The `Grid` layout generally led to the highest recall.}
        \label{fig:recall_heatmap}
    \end{subfigure}
    \hfill % Adds horizontal space between the subfigures
    \begin{subfigure}[b]{0.49\textwidth}
        \centering
        \includegraphics[width=\textwidth]{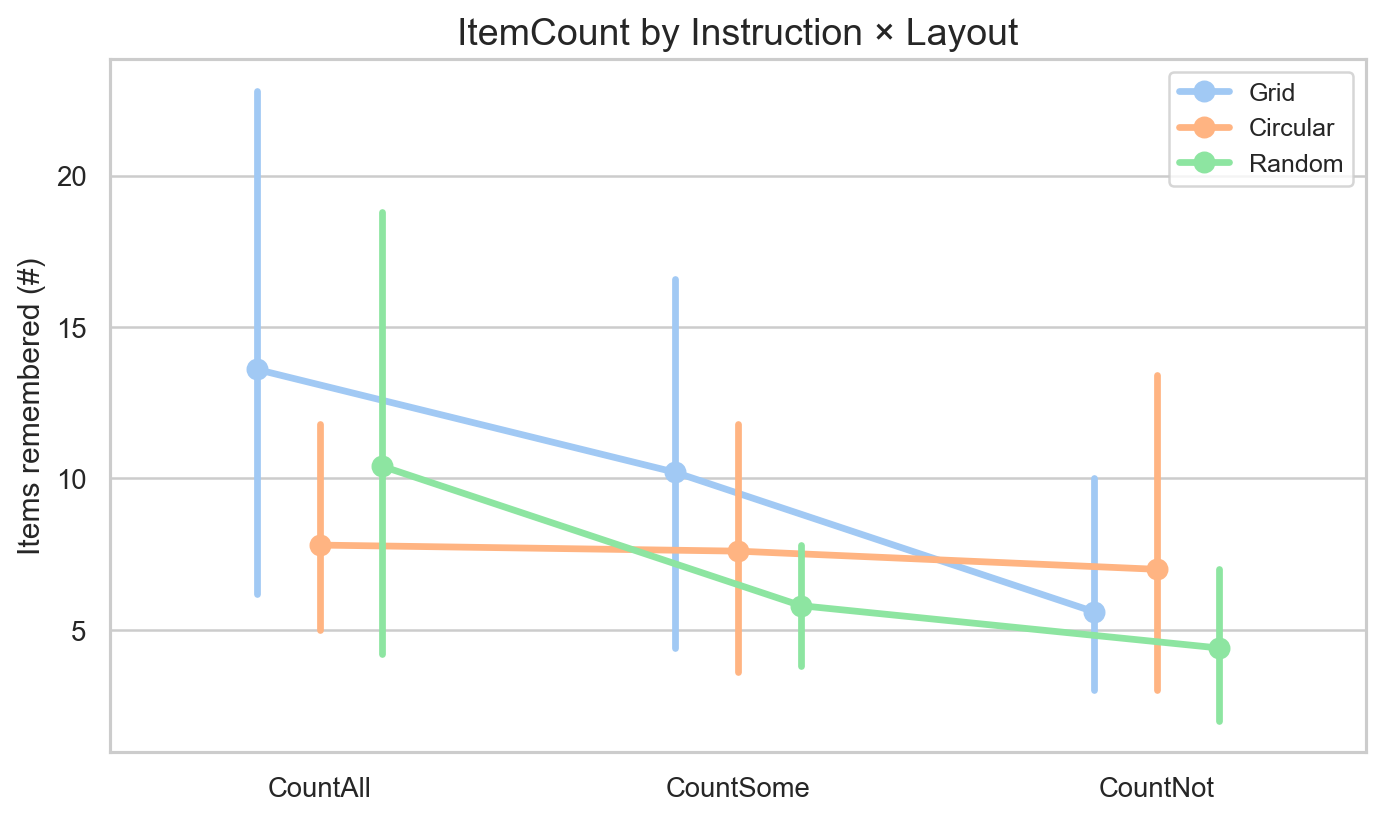}
        \caption{Interaction line plot showing mean recall performance. The downward slope of all lines confirms that memory recall diminishes with increasing task complexity, with the `Grid` layout consistently yielding the best performance.}
        \label{fig:recall_lines}
    \end{subfigure}
    \caption{Visualizations of memory recall performance in Phase 1, showing the interaction between instruction type and spatial layout as (a) a heatmap and (b) a line plot.}
    \label{fig:phase1_recall_visuals}
\end{figure}

The most significant factor affecting recall was the instruction type. As shown in the interaction plot in Figure~\ref{fig:recall_lines}, there is a clear and steep decline in the number of recalled items as the task moves from 'CountAll' to 'CountSome', and further to 'CountNot'. This trend holds true for all three spatial layouts. The heatmap in Figure~\ref{fig:recall_heatmap} quantifies this effect, showing that the highest recall was achieved in the 'Grid-CountAll' condition, where participants remembered an average of 13.60 items. In contrast, the lowest recall was in the 'Random-CountNot' condition, with an average of only 4.40 items remembered.

The spatial layout also had a discernible effect on recall. The 'Grid' layout consistently produced the highest memory recall scores across all three instruction types. The 'Random' layout was generally associated with the poorest recall performance, particularly in the selective counting conditions. The interaction plot (Figure~\ref{fig:recall_lines}) also highlights the high degree of inter-participant variability in memory performance, as indicated by the large error bars, especially in the 'CountAll' condition.

% Note: Ensure you have \usepackage{graphicx} and \usepackage{subcaption} in your document preamble.

% \begin{figure}[ht!]
%     \centering
%     \includegraphics[width=0.85\textwidth]{Figures/Phase1/MemoryRecall/01_heatmap_itemcount_p1.png}
%     \caption{Heatmap of mean item recall, showing a strong decrease in the number of remembered items as the instruction type becomes more selective. The 'Grid' layout generally led to the highest recall.}
%     \label{fig:recall_heatmap}
% \end{figure}

% \begin{figure}[ht!]
%     \centering
%     \includegraphics[width=0.85\textwidth]{Figures/Phase1/MemoryRecall/02_interaction_itemcount_p1.png}
%     \caption{Interaction line plot showing mean recall performance. The downward slope of all lines confirms that memory recall diminishes with increasing task complexity, with the 'Grid' layout consistently yielding the best performance.}
%     \label{fig:recall_lines}
% \end{figure}

\paragraph{Discussion}
The results for recall performance provide compelling support for \textbf{Hypothesis H3}, which posited a strong relationship between visual attention and memory encoding. The observed patterns directly validate both sub-points of the hypothesis.

The dramatic decline in recall from 'CountAll' to the selective tasks, as predicted in \textbf{H3(a)}, serves as a powerful illustration of the principle that attention is a gateway to memory. In the 'CountAll' condition, the exhaustive gaze strategy (as seen in the heatmaps) ensured that every item was fixated upon, giving each one an opportunity to be encoded into memory. In the 'CountSome' condition, attention was narrowed, focused primarily on target items. Consequently, the non-target items received little to no deep processing and were poorly remembered, leading to a lower overall recall count. The 'CountNot' condition resulted in the worst recall because the high cognitive load associated with the filtering task likely interfered with the memory consolidation process, leading to shallower encoding of all items, even those that were attended to.

The influence of the spatial layout on memory, also predicted in \textbf{H3(a)}, can be understood through the lens of cognitive resource allocation. The highly structured 'Grid' layout simplified the search process, reducing the cognitive load required for navigation and spatial working memory. This likely freed up more cognitive resources to be dedicated to encoding the identity of the items themselves, resulting in better memory performance. Conversely, the demanding and unstructured nature of the 'Random' layout required a greater allocation of resources to the search process itself, leaving fewer resources available for memory encoding. In conclusion, the memory data strongly suggests that what we remember from a scene is not simply what our eyes pass over, but what our mind is able to deeply process. This deep processing is facilitated when attention is broad and when the cognitive load of the primary task is low.

%%%===============================================
\subsection{Phase 2: Enumeration of Real-World Object Images}
In the second phase of the experiment, participants were tasked with enumerating real-world object images. This phase was designed to build upon the baseline established in Phase 1 by introducing the additional cognitive demands of object recognition and semantic categorisation, thereby simulating a more complex and realistic visual enumeration scenario.

\subsubsection{Task Completion Time}

\paragraph{Observations}
The introduction of real-world object images as stimuli resulted in a substantial increase in task completion times across all conditions compared to Phase 1. However, the fundamental patterns of influence from both task intent and spatial layout remained consistent, albeit with a greatly amplified magnitude.

\begin{figure}[ht!]
    \centering
    \begin{subfigure}[b]{0.49\textwidth}
        \centering
        \includegraphics[width=\textwidth]{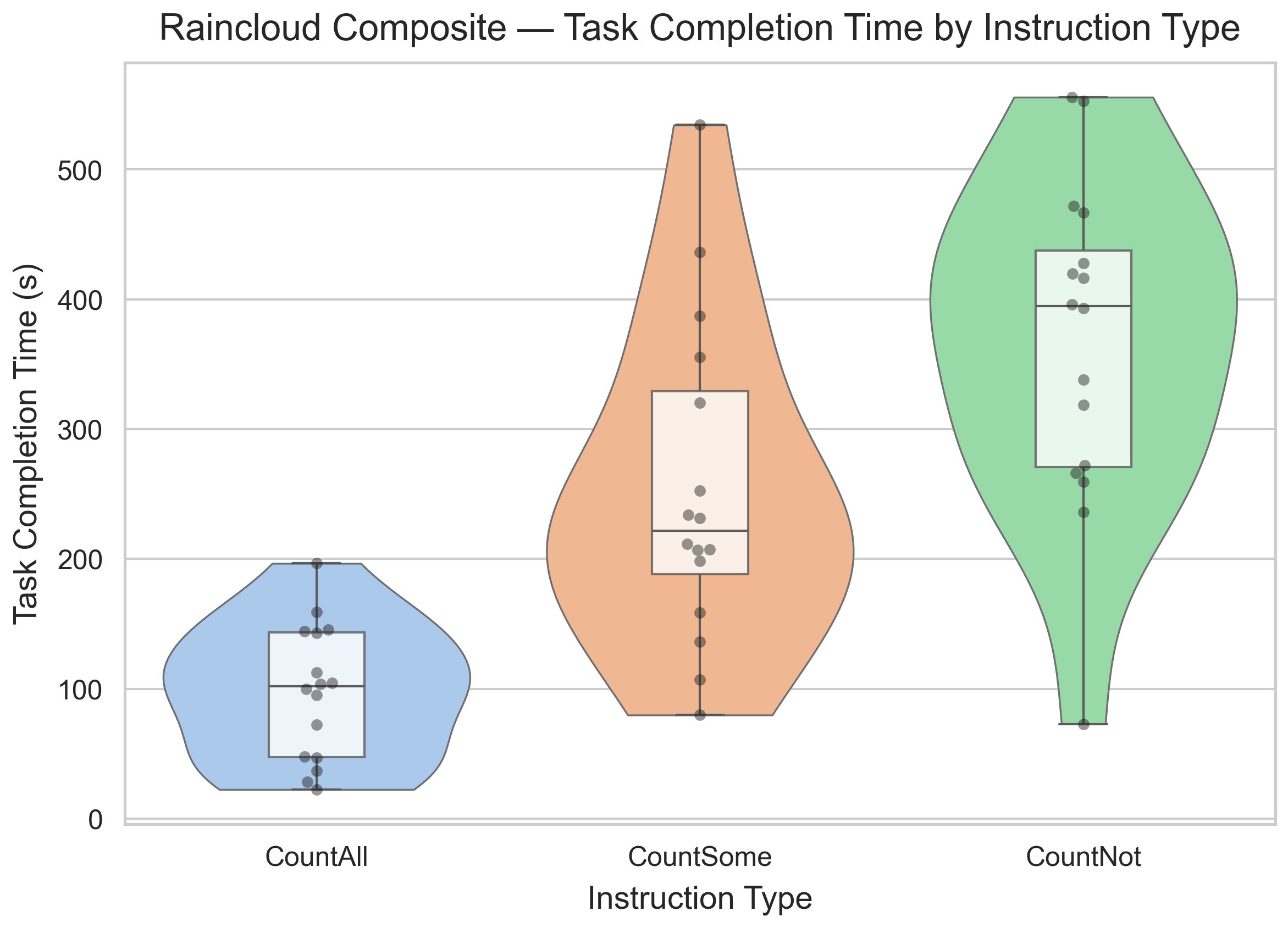}
        \caption{Raincloud composite plot of task completion times for Phase 2. The introduction of real-world objects led to substantially longer completion times overall, while maintaining the clear trend of increasing time with more complex instructions.}
        \label{fig:instruction_time_p2}
    \end{subfigure}
    \hfill % Adds horizontal space between the subfigures
    \begin{subfigure}[b]{0.49\textwidth}
        \centering
        \includegraphics[width=\textwidth]{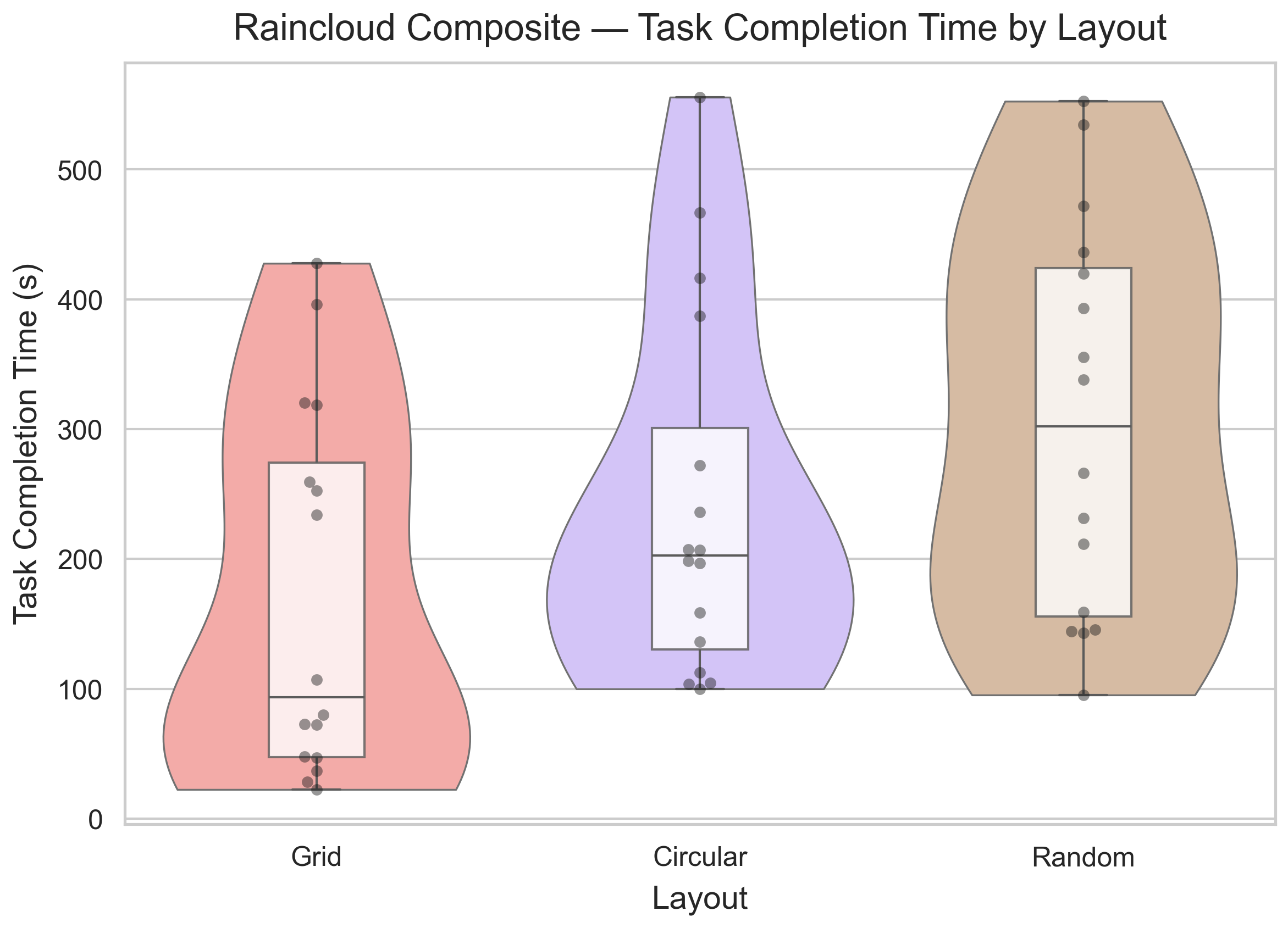}
        \caption{Raincloud composite plot of task completion times by layout for Phase 2. The 'Grid' layout remained the most efficient, but the time penalty for the 'Circular' and 'Random' layouts increased considerably compared to Phase 1.}
        \label{fig:layout_time_p2}
    \end{subfigure}
    \caption{Distributions of Task Completion Time in Phase 2, analyzed by (a) Instruction Type and (b) Spatial Layout.}
    \label{fig:phase2_tct_distributions}
\end{figure}

The effect of the instruction type was once again a dominant factor. As shown in the raincloud and spaghetti plots in Figure~\ref{fig:instruction_time_p2}, the trend of 'CountAll' $<$ 'CountSome' $<$ 'CountNot' was not only replicated but was also far more pronounced. The distributions are visibly wider, and the mean completion times are significantly longer than in the abstract shapes phase. For instance, the mean time for the 'CountAll' task was over 100 seconds, a duration that was closer to the most difficult conditions in Phase 1.

% Note: Ensure you have \usepackage{graphicx} and \usepackage{subcaption} in your document preamble.

% \begin{figure}[ht!]
%     \centering
%     \includegraphics[width=0.85\textwidth]{Figures/Phase2/TaskCompletionTime/05_raincloud_instruction_tct_p2.png}
%     \caption{Raincloud composite plot of task completion times for Phase 2. The introduction of real-world objects led to substantially longer completion times overall, while maintaining the clear trend of increasing time with more complex instructions.}
%     \label{fig:instruction_time_p2}
% \end{figure}

The effect of spatial layout also followed the same pattern observed previously ('Grid' $<$ 'Circular' $<$ 'Random'), but with larger absolute differences between the conditions, as seen in Figure~\ref{fig:layout_time_p2}. The benefit of the 'Grid' layout and the penalty of the 'Random' layout were both magnified when the items being counted were complex images.

% \begin{figure}[ht!]
%     \centering
%     \includegraphics[width=0.85\textwidth]{Figures/Phase2/TaskCompletionTime/11_raincloud_layout_tct_p2.png}
%     \caption{Raincloud composite plot of task completion times by layout for Phase 2. The 'Grid' layout remained the most efficient, but the time penalty for the 'Circular' and 'Random' layouts increased considerably compared to Phase 1.}
%     \label{fig:layout_time_p2}
% \end{figure}

\begin{figure}[ht!]
    \centering
    \begin{subfigure}[b]{0.49\textwidth}
        \centering
        \includegraphics[width=\textwidth]{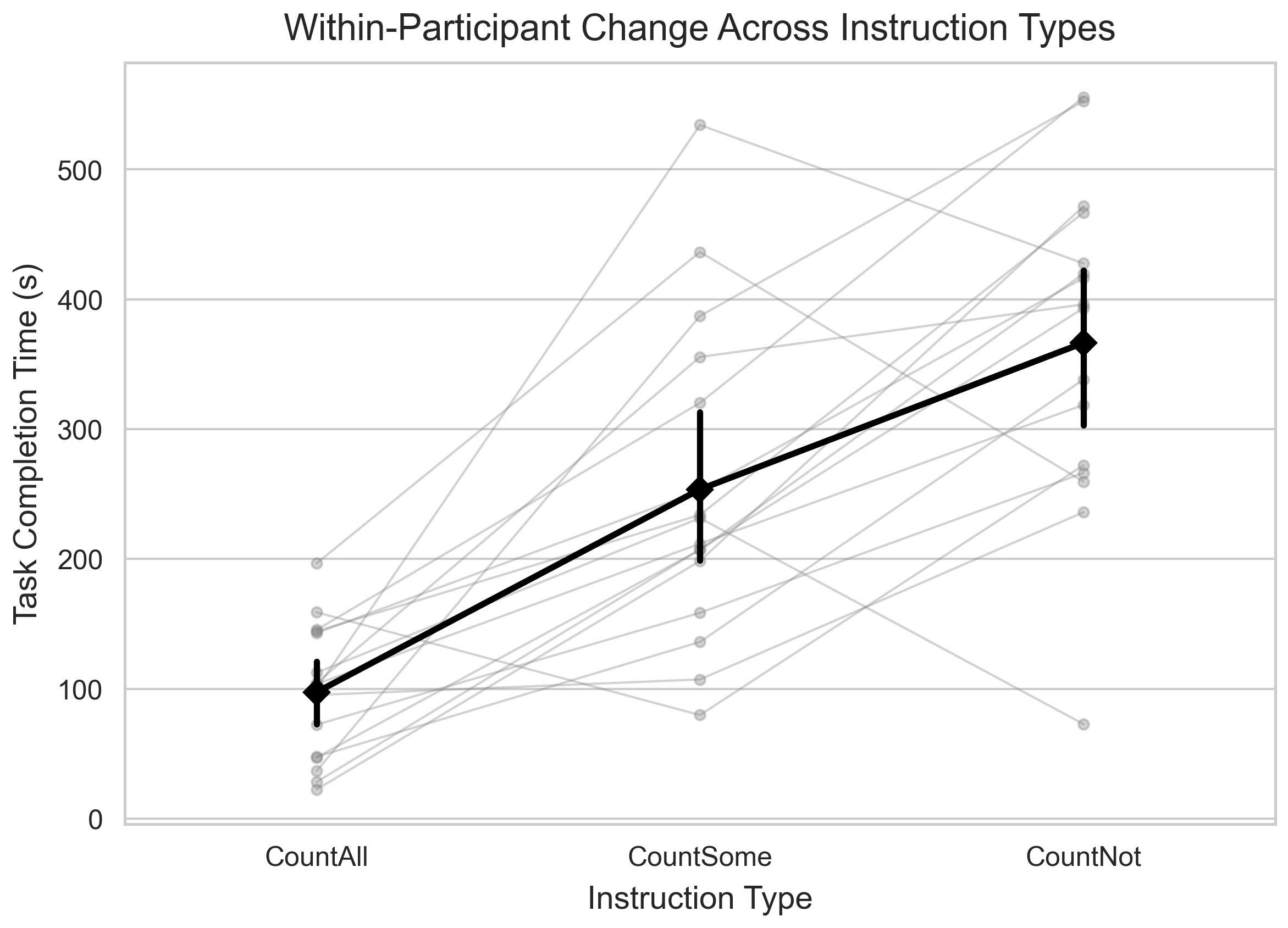}
        \caption{Change across Instruction Types. The steep upward slope shows a strong and consistent increase in time with task complexity.}
        \label{fig:spaghetti_instruction_p2}
    \end{subfigure}
    \hfill % Adds horizontal space between the subfigures
    \begin{subfigure}[b]{0.49\textwidth}
        \centering
        \includegraphics[width=\textwidth]{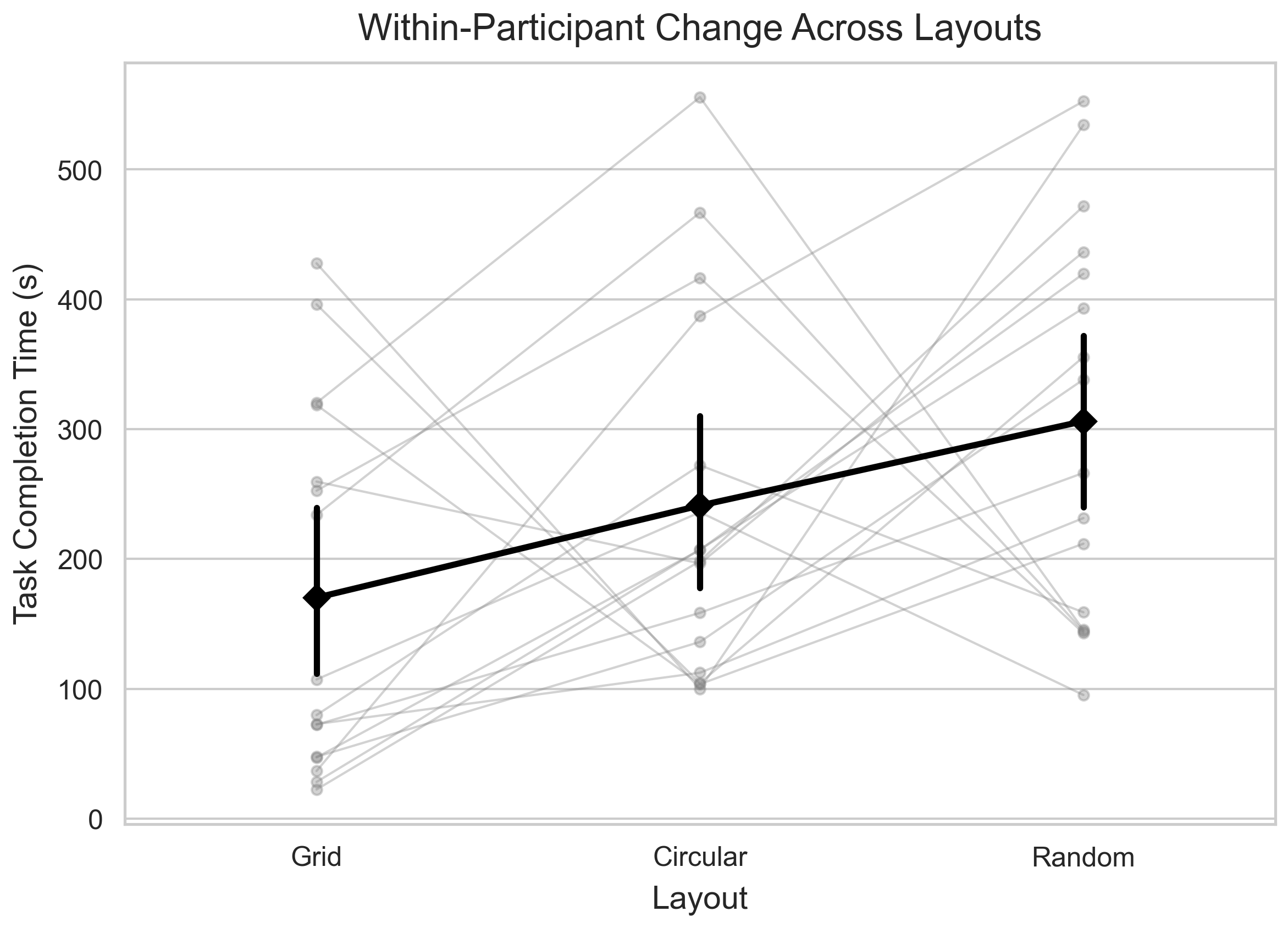}
        \caption{Change across Layouts. The mean trend shows a clear increase in time from the `Grid` to the `Random` layout.}
        \label{fig:spaghetti_layout_p2}
    \end{subfigure}
    \caption{Within-participant spaghetti plots for Task Completion Time in Phase 2, illustrating the amplified individual and mean trends across (a) instruction types and (b) spatial layouts.}
    \label{fig:phase2_spaghetti_plots}
\end{figure}

The within-participant spaghetti plots for Phase 2, shown in Figure~\ref{fig:phase2_spaghetti_plots}, illustrate the amplified effects of the experimental manipulations. The plot for instruction type (Figure~\ref{fig:spaghetti_instruction_p2}) shows a steep and consistent increase in completion time for almost every participant as the task's semantic filtering demands grew. Similarly, the plot for layout (Figure~\ref{fig:spaghetti_layout_p2}) shows a clear mean trend of increasing task time with less structured environments, confirming that the performance costs for both complex instructions and layouts were magnified when dealing with real-world objects.

% Note: Ensure you have \usepackage{graphicx} and \usepackage{subcaption} in your document preamble.

\begin{figure}[ht!]
    \centering
    \begin{subfigure}[b]{0.49\textwidth}
        \centering
        \includegraphics[width=\textwidth]{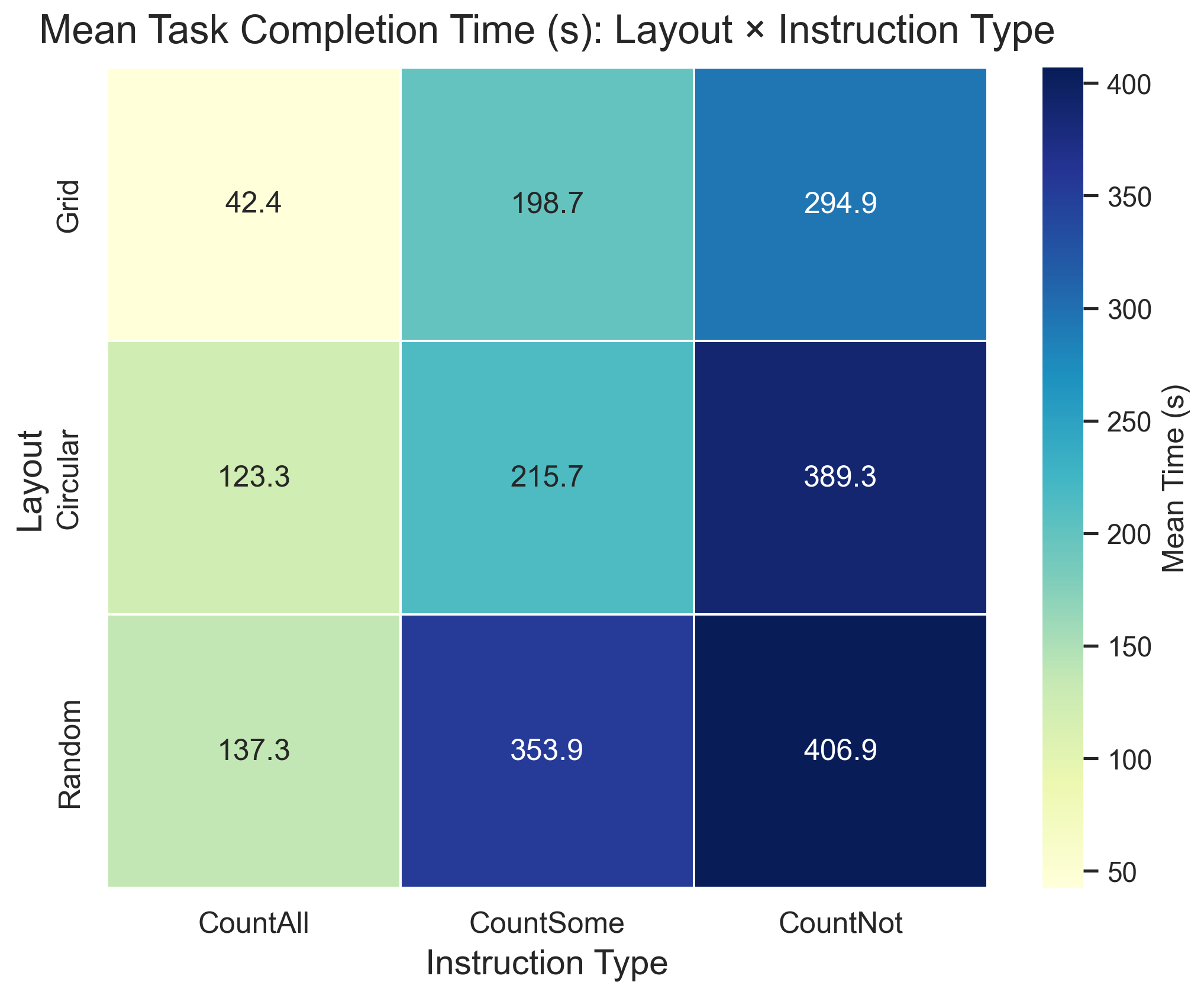}
        \caption{Heatmap of mean task completion times (s) for Phase 2. The range of times is much larger than in Phase 1, with the 'Random-CountNot' condition taking over 400 seconds on average.}
        \label{fig:interaction_heatmap_p2}
    \end{subfigure}
    \hfill % Adds horizontal space between the subfigures
    \begin{subfigure}[b]{0.49\textwidth}
        \centering
        \includegraphics[width=\textwidth]{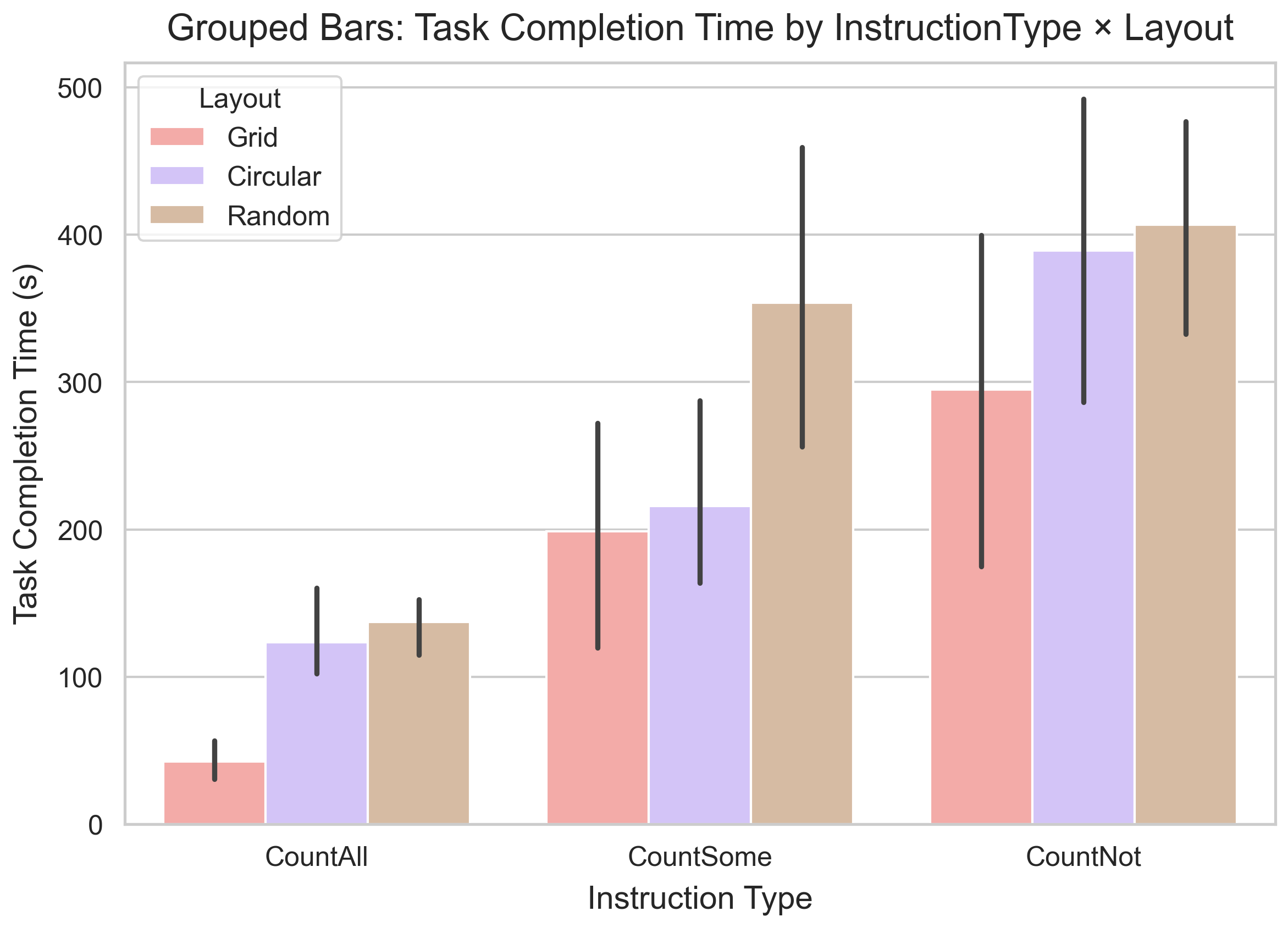}
        \caption{Grouped bar chart for Phase 2, visually representing the amplified effect of both instruction and layout on task completion time. The error bars also indicate a substantial increase in inter-participant variance.}
        \label{fig:interaction_bars_p2}
    \end{subfigure}
    \caption{Interaction effects on Task Completion Time in Phase 2, visualized as (a) a heatmap of mean values and (b) a grouped bar chart with error bars.}
    \label{fig:phase2_tct_interactions}
\end{figure}

The interaction between the two factors, presented in the heatmap (Figure~\ref{fig:interaction_heatmap_p2}) and the bar chart (Figure~\ref{fig:interaction_bars_p2}), vividly illustrates this amplification. The time taken for the easiest condition, 'Grid-CountAll', was 42.4 seconds, which is comparable to the same condition in Phase 1. However, the time for the most difficult condition, 'Random-CountNot', surged to an average of 406.9 seconds—nearly five times longer than the most difficult selective task in Phase 1. This demonstrates a dramatic compounding effect of task and environmental complexity.

% Note: Ensure you have \usepackage{graphicx} and \usepackage{subcaption} in your document preamble.

% \begin{figure}[ht!]
%     \centering
%     \includegraphics[width=0.85\textwidth]{Figures/Phase2/TaskCompletionTime/13_heatmap_means_layout_by_instruction_tct_p2.png}
%     \caption{Heatmap of mean task completion times (s) for Phase 2. The range of times is much larger than in Phase 1, with the 'Random-CountNot' condition taking over 400 seconds on average.}
%     \label{fig:interaction_heatmap_p2}
% \end{figure}

% \begin{figure}[ht!]
%     \centering
%     \includegraphics[width=0.85\textwidth]{Figures/Phase2/TaskCompletionTime/14_grouped_bars_interaction_tct_p2.png}
%     \caption{Grouped bar chart for Phase 2, visually representing the amplified effect of both instruction and layout on task completion time. The error bars also indicate a substantial increase in inter-participant variance.}
%     \label{fig:interaction_bars_p2}
% \end{figure}

\paragraph{Discussion}
The task completion time results from Phase 2 offer further, stronger validation for our hypotheses. The consistent replication of the performance patterns confirms the robustness of the effects, while the significant increase in overall times highlights the profound impact of stimulus complexity. The findings again provide unequivocal support for \textbf{H1(a)} and the trend predicted in \textbf{H2(a)}.

The core reason for the dramatically longer completion times is the introduction of a mandatory, time-consuming cognitive step that precedes the actual act of counting: object recognition and semantic categorisation. In Phase 1, identifying a dot's colour was a rapid, low-level feature recognition task. In Phase 2, determining if an image contained an "animal" or was "not a vehicle" required a much deeper level of semantic processing. This act of recognition and classification had to be performed on each item inspected, adding a substantial time cost to every fixation, particularly in the selective conditions.

This additional cognitive load served to amplify the effects observed in Phase 1. The time difference between 'CountAll' and 'CountNot' became much larger because the filtering decision itself was more complex and time-consuming. Similarly, the time penalty for an inefficient search path in the 'Random' layout was magnified. When each fixation and decision takes longer, the time lost to re-fixating on an already-counted item or searching for the next target in a cluttered scene becomes much more significant. The benefit of the 'Grid' layout in reducing this search overhead is therefore even more valuable in a complex task. In essence, the Phase 2 results demonstrate that in realistic scenarios, visual enumeration is not just a process of counting, but a two-stage process of "search-then-count." The time cost of the initial "search and identify" stage is a critical and substantial component of the total effort, and it is this stage that is most heavily impacted by the complexity of the objects in our environment.

\subsubsection{Accuracy}

\paragraph{Observations}
The introduction of complex, real-world object images in Phase 2 had a profound and detrimental effect on counting accuracy, particularly in the selective counting conditions. While the general trends mirrored those of Phase 1, the magnitude of the performance decline and the inter-participant variability were substantially greater.

\begin{figure}[ht!]
    \centering
    \begin{subfigure}[b]{0.49\textwidth}
        \centering
        \includegraphics[width=\textwidth]{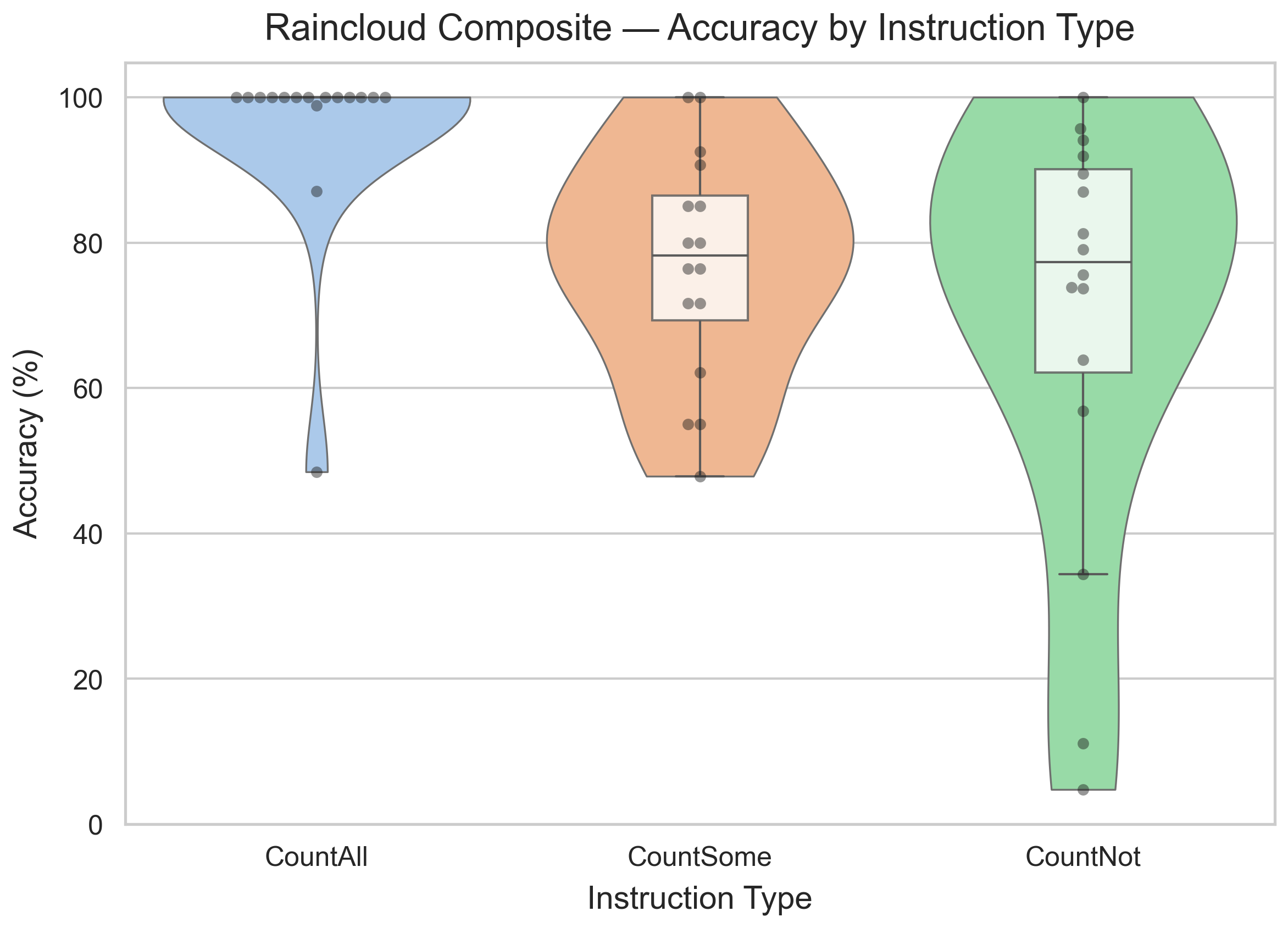}
        \caption{Raincloud composite plot of accuracy for Phase 2 by instruction type. The plot highlights a severe drop in accuracy and a massive increase in variance for the 'CountNot' instruction.}
        \label{fig:instruction_acc_p2}
    \end{subfigure}
    \hfill % Adds horizontal space between the subfigures
    \begin{subfigure}[b]{0.49\textwidth}
        \centering
        \includegraphics[width=\textwidth]{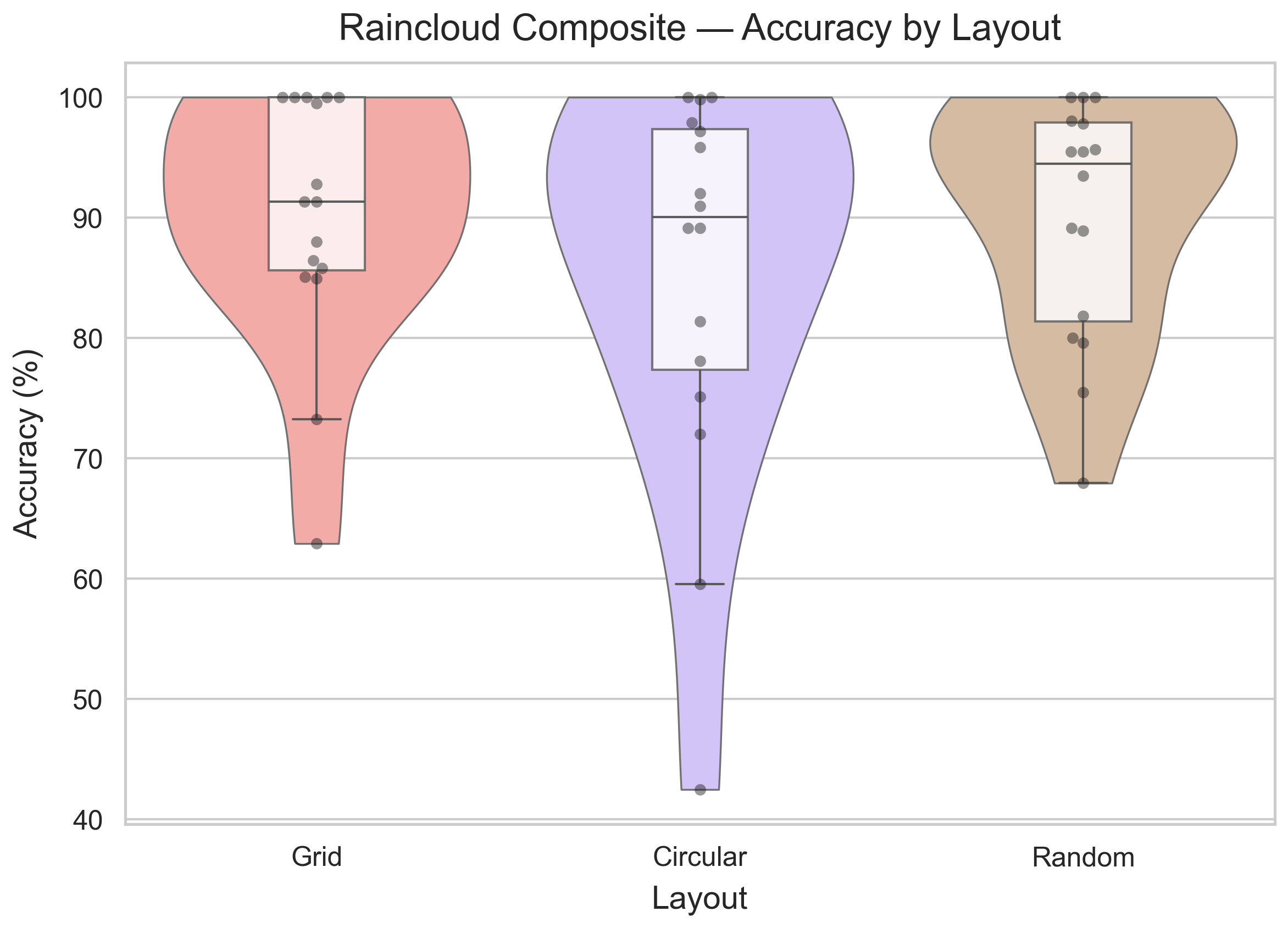}
        \caption{Raincloud composite plot of accuracy across layouts for Phase 2. The highly similar and overlapping distributions indicate that spatial layout had no consistent effect on participant accuracy.}
        \label{fig:layout_acc_p2}
    \end{subfigure}
    \caption{Distributions of Counting Accuracy in Phase 2, analyzed by (a) Instruction Type and (b) Spatial Layout.}
    \label{fig:phase2_accuracy_distributions}
\end{figure}

The analysis of the raincloud plots reveals that the instruction type was the primary determinant of accuracy. As shown in Figure~\ref{fig:instruction_acc_p2}, participants achieved very high and consistent accuracy in the 'CountAll' condition. However, performance dropped in the 'CountSome' condition and then fell dramatically in the 'CountNot' condition. The distribution for 'CountNot' is particularly notable for its immense variance, featuring an extremely long lower tail with outliers approaching zero accuracy. This indicates that while some participants managed the task, a significant subset experienced a near-complete failure to perform it correctly. In stark contrast, the raincloud plot for spatial layouts (Figure~\ref{fig:layout_acc_p2}) shows heavily overlapping distributions with very similar medians, suggesting that the layout had a negligible and inconsistent effect on accuracy.

% Note: Ensure you have \usepackage{graphicx} and \usepackage{subcaption} in your document preamble.

% \begin{figure}[ht!]
%     \centering
%     \includegraphics[width=0.8\textwidth]{Figures/Phase2/Accuracy/05_raincloud_instruction_acc_p2.png}
%     \caption{Raincloud composite plot of accuracy for Phase 2 by instruction type. The plot highlights a severe drop in accuracy and a massive increase in variance for the 'CountNot' instruction.}
%     \label{fig:instruction_acc_p2}
% \end{figure}

% \begin{figure}[ht!]
%     \centering
%     \includegraphics[width=0.8\textwidth]{Figures/Phase2/Accuracy/11_raincloud_layout_acc_p2.png}
%     \caption{Raincloud composite plot of accuracy across layouts for Phase 2. The highly similar and overlapping distributions indicate that spatial layout had no consistent effect on participant accuracy.}
%     \label{fig:layout_acc_p2}
% \end{figure}

\begin{figure}[ht!]
    \centering
    \begin{subfigure}[b]{0.49\textwidth}
        \centering
        \includegraphics[width=\textwidth]{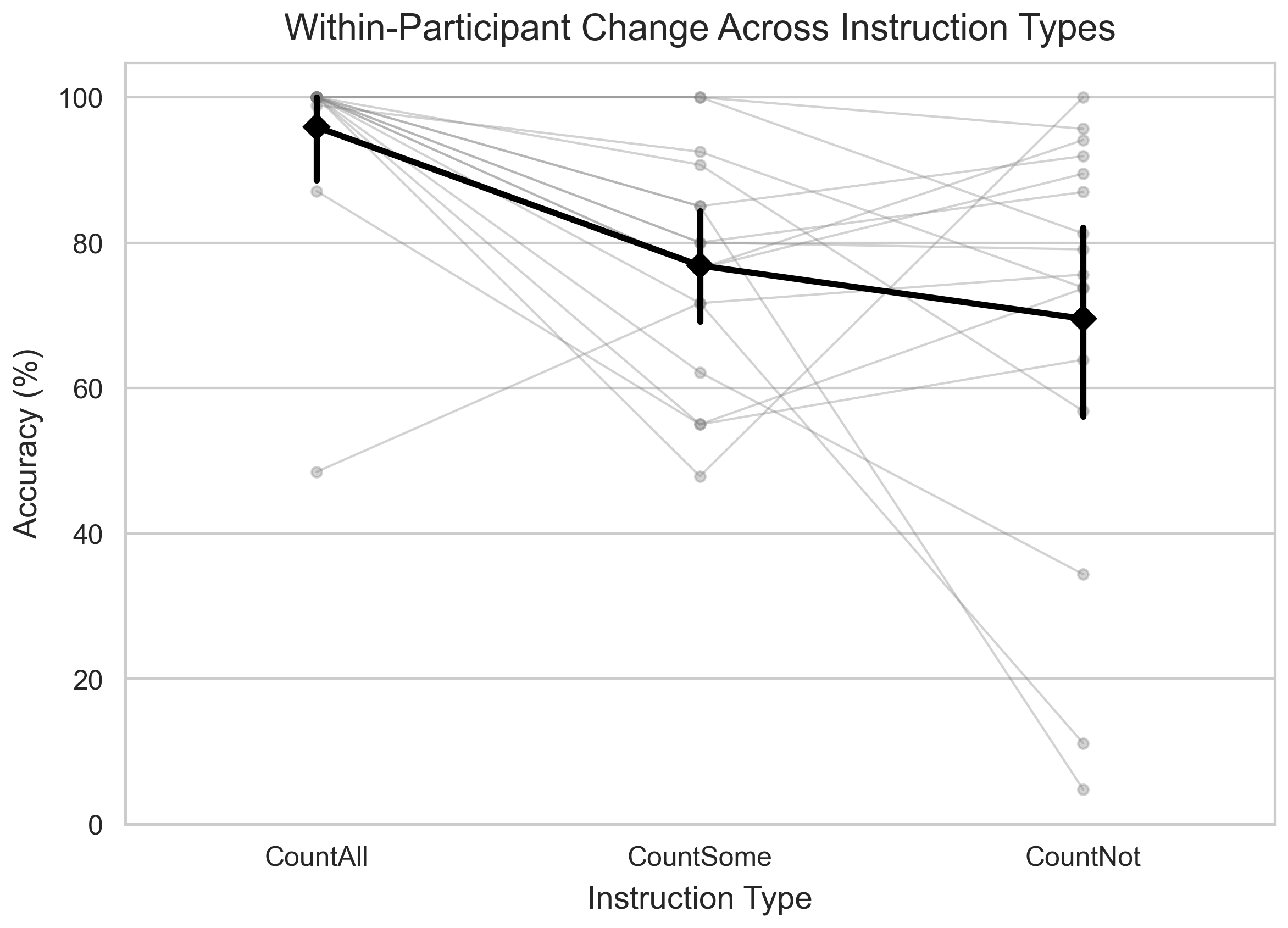}
        \caption{Change across Instruction Types. A steep decline in accuracy is evident, with some participants showing catastrophic drops.}
        \label{fig:spaghetti_instruction_acc_p2}
    \end{subfigure}
    \hfill % Adds horizontal space between the subfigures
    \begin{subfigure}[b]{0.49\textwidth}
        \centering
        \includegraphics[width=\textwidth]{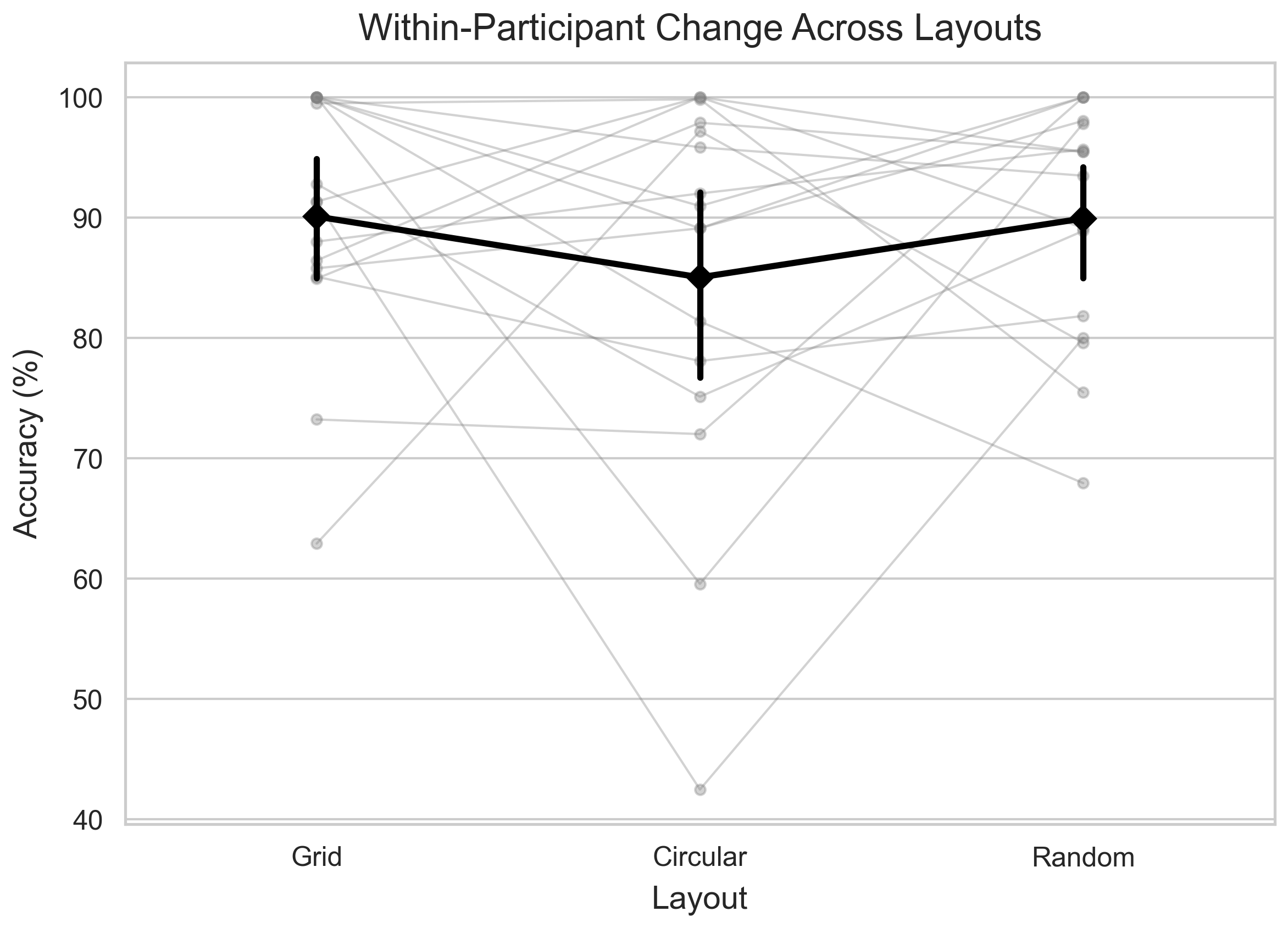}
        \caption{Change across Layouts. Individual performance is highly chaotic with no clear mean trend.}
        \label{fig:spaghetti_layout_acc_p2}
    \end{subfigure}
    \caption{Within-participant spaghetti plots for Counting Accuracy in Phase 2, illustrating the powerful and consistent effect of instruction type (a) versus the inconsistent and negligible effect of spatial layout (b).}
    \label{fig:phase2_spaghetti_plots_acc}
\end{figure}

The within-participant plots in Figure~\ref{fig:phase2_spaghetti_plots_acc} starkly illustrate the dominant influence of the instruction type. The plot for instruction (Figure~\ref{fig:spaghetti_instruction_acc_p2}) shows a steep and consistent decline in accuracy for a majority of participants, with several individuals exhibiting a catastrophic performance drop in the demanding `CountNot` condition. In sharp contrast, the plot for layout (Figure~\ref{fig:spaghetti_layout_acc_p2}) reveals highly chaotic and inconsistent individual trajectories, with a mean trend that shows no systematic effect, confirming that layout had a negligible impact on accuracy when dealing with complex objects.

% Note: Ensure you have \usepackage{graphicx} and \usepackage{subcaption} in your document preamble.

\begin{figure}[ht!]
    \centering
    \begin{subfigure}[b]{0.49\textwidth}
        \centering
        \includegraphics[width=\textwidth]{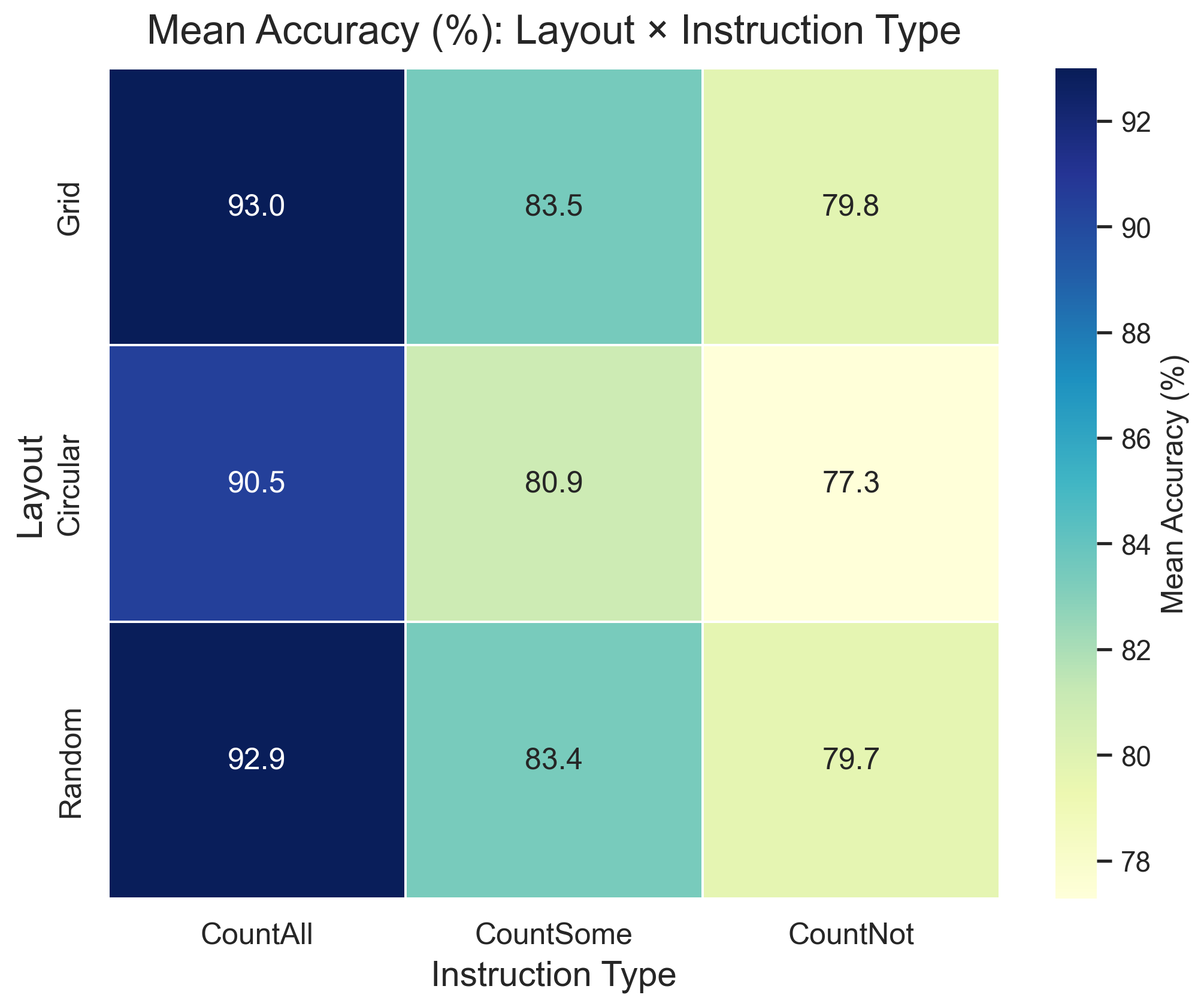}
        \caption{Heatmap of mean accuracy (\%) for Phase 2. The dominant visual pattern is the vertical gradient, showing a sharp drop in accuracy from 'CountAll' to 'CountNot', with minimal differences horizontally across layouts.}
        \label{fig:interaction_acc_heatmap_p2}
    \end{subfigure}
    \hfill % Adds horizontal space between the subfigures
    \begin{subfigure}[b]{0.49\textwidth}
        \centering
        \includegraphics[width=\textwidth]{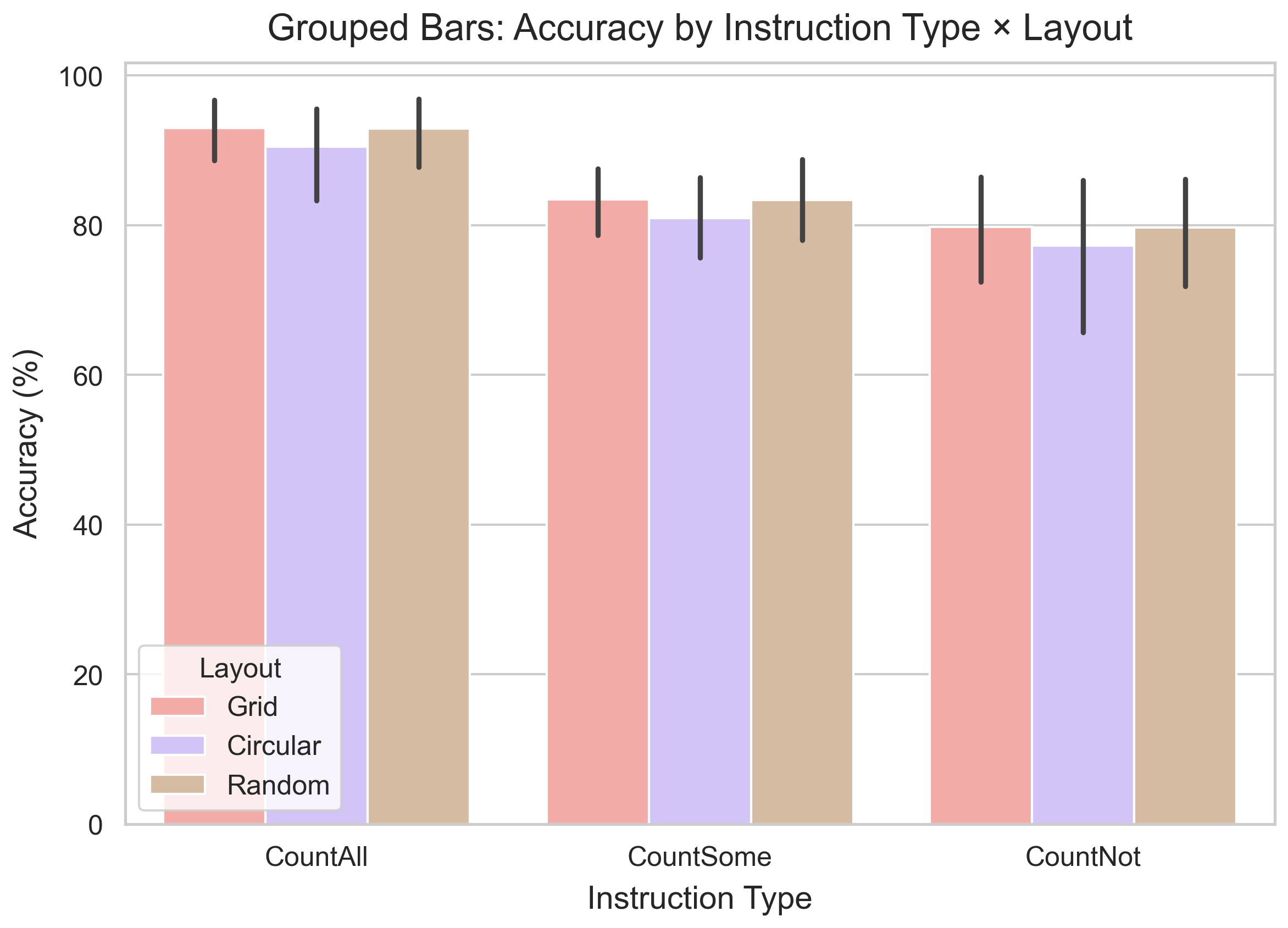}
        \caption{Grouped bar chart showing mean accuracy for each condition in Phase 2. The chart visually confirms that the effect of instruction type on accuracy is far more pronounced than the effect of layout.}
        \label{fig:interaction_acc_bars_p2}
    \end{subfigure}
    \caption{Interaction effects on Counting Accuracy in Phase 2, visualized as (a) a heatmap of mean values and (b) a grouped bar chart with error bars.}
    \label{fig:phase2_accuracy_interactions}
\end{figure}

The interaction heatmap in Figure~\ref{fig:interaction_acc_heatmap_p2} quantifies these effects, confirming that the drop in accuracy was almost entirely driven by the instruction type, regardless of layout. For instance, mean accuracy in the 'CountAll' instruction was high and nearly identical for the 'Grid' (93.0\%) and 'Random' (92.9\%) layouts. Similarly, in the most difficult 'CountNot' condition, accuracy was consistently low across all layouts ('Grid'=79.8\%, 'Random'=79.7\%). The grouped bar chart in Figure~\ref{fig:interaction_acc_bars_p2} further reinforces this point, visually showing that the bars representing different layouts within each instruction group are of very similar height, with their error bars overlapping significantly. The most prominent visual feature of the chart is the overall step-down in accuracy from the 'CountAll' group to the 'CountNot' group.

% Note: Ensure you have \usepackage{graphicx} and \usepackage{subcaption} in your document preamble.

% \begin{figure}[ht!]
%     \centering
%     \includegraphics[width=0.7\textwidth]{Figures/Phase2/Accuracy/13_heatmap_means_layout_by_instruction_acc_p2.png}
%     \caption{Heatmap of mean accuracy (\%) for Phase 2. The dominant visual pattern is the vertical gradient, showing a sharp drop in accuracy from 'CountAll' to 'CountNot', with minimal differences horizontally across layouts.}
%     \label{fig:interaction_acc_heatmap_p2}
% \end{figure}

% \begin{figure}[ht!]
%     \centering
%     \includegraphics[width=0.8\textwidth]{Figures/Phase2/Accuracy/14_grouped_bars_interaction_acc_p2.png}
%     \caption{Grouped bar chart showing mean accuracy for each condition in Phase 2. The chart visually confirms that the effect of instruction type on accuracy is far more pronounced than the effect of layout.}
%     \label{fig:interaction_acc_bars_p2}
% \end{figure}

\paragraph{Discussion}
The accuracy results from Phase 2 provide overwhelming support for \textbf{H1(b)}, while demonstrating that the trend predicted in \textbf{H2(b)} does not hold under conditions of high stimulus complexity. The severe decline in accuracy during selective counting tasks reveals the substantial cognitive cost of semantic filtering. The task was no longer a simple perceptual exercise as in Phase 1; it became a demanding cognitive challenge. Each item required a slower, more effortful semantic judgment ("Is this a vehicle?"), which significantly increased the opportunity for error. 

The catastrophic failures observed in the 'CountNot' condition suggest that the cognitive load for some participants surpassed a critical threshold. The dual task of holding a complex semantic category (e.g., "vehicles") in working memory while simultaneously identifying and counting objects from all other diverse categories (animals, furniture, tools, etc.) proved to be an exceptionally difficult, error-prone, and for some, an intractable problem.

Crucially, the effect of the spatial layout on accuracy, which was a subtle trend in Phase 1, was rendered almost entirely insignificant in Phase 2. This suggests a shift in the primary cognitive bottleneck. In Phase 1, errors were more likely to be mechanical (e.g., losing one's place), and a structured layout could help mitigate this. In Phase 2, however, the errors were predominantly cognitive (e.g., miscategorising an object). The main challenge was no longer navigating the space but processing each item. As such, the organisational benefits of the 'Grid' layout became marginal. When the difficulty of identifying \textit{what} an object is becomes the dominant challenge, the structure of \textit{where} it is located becomes a secondary concern. This finding highlights that in complex, real-world enumeration tasks, the cognitive demands imposed by the task's goal can completely overshadow the influence of the environment's physical structure.

\subsubsection{Statistical Significance}

\paragraph{Methodology}
To rigorously assess the effects of our experimental manipulations in Phase 2, we conducted a statistical analysis on the data from the 16 participants. Consistent with Phase 1, we employed non-parametric tests to accommodate the data's distribution. The Friedman test was used as the primary method for our within-subjects comparisons, with the Kruskal-Wallis test serving as a confirmatory measure. Significant main effects were further investigated using post-hoc Wilcoxon signed-rank tests for pairwise comparisons, with p-values adjusted via the Holm-Bonferroni method to control for multiple comparisons.

\paragraph{Effect of Task Intent (Instruction Type)}
The statistical analysis for Phase 2 revealed that task intent had a powerful and highly significant effect on both speed and accuracy, confirming it as the dominant factor influencing performance.
\begin{itemize}
    \item For \textbf{Task Completion Time}, both the Friedman test ($\chi^2(2)=18.375, p < 0.001$) and the Kruskal-Wallis test ($H=20.629, p < 0.001$) indicated a highly significant effect of the instruction type. Post-hoc Wilcoxon tests confirmed that the 'CountAll' condition was significantly faster than both the 'CountSome' condition ($p_{holm} $<$ 0.001$) and the 'CountNot' condition ($p_{holm} = 0.001$). Interestingly, the difference in completion time between the two selective tasks, 'CountSome' and 'CountNot', was not statistically significant ($p_{holm} = 0.980$).
    \item For \textbf{Accuracy}, the results were similarly robust. The instruction type had a highly significant effect, as shown by both the Friedman test ($\chi^2(2)=18.000, p < 0.001$) and the Kruskal-Wallis test ($H=19.239, p < 0.001$). The post-hoc Wilcoxon tests revealed that participants were significantly more accurate in the 'CountAll' condition compared to both 'CountSome' ($p_{holm} = 0.009$) and 'CountNot' ($p_{holm} = 0.002$). As with completion time, the difference in accuracy between 'CountSome' and 'CountNot' was not statistically significant ($p_{holm} = 0.528$).
\end{itemize}

\paragraph{Effect of Spatial Layout}
Consistent with the findings from Phase 1, the spatial layout of the real-world images did not have a statistically significant impact on participant performance.
\begin{itemize}
    \item For \textbf{Task Completion Time}, neither the Friedman test ($\chi^2(2)=1.125, p=0.570$) nor the Kruskal-Wallis test ($H=0.460, p=0.795$) found a significant difference among the three layouts.
    \item Similarly for \textbf{Accuracy}, the Friedman test ($\chi^2(2)=0.667, p=0.717$) and the Kruskal-Wallis test ($H=0.843, p=0.656$) were also not significant. The post-hoc tests confirmed that no pairwise comparisons between the layouts were statistically significant for either performance metric.
\end{itemize}

\paragraph{Discussion}
The statistical results from Phase 2 provide powerful and conclusive support for \textbf{Hypothesis H1}. The highly significant effect of the instruction type on both speed and accuracy underscores the immense cognitive cost associated with semantic filtering of complex, real-world objects. The post-hoc analysis reveals a crucial insight: while both selective counting tasks ('CountSome' and 'CountNot') were significantly more difficult than the baseline 'CountAll' task, they were not statistically different from each other. This suggests that the primary cognitive hurdle was the act of selective filtering itself. Once this demanding process of semantic identification was required, the added logical step of exclusion ('CountNot') over inclusion ('CountSome') did not produce a further statistically reliable drop in performance.

Furthermore, the data provide a clear refutation of \textbf{Hypothesis H2} in the context of complex stimuli. The consistent lack of a significant layout effect on either time or accuracy reinforces the conclusion drawn from the descriptive data: when the cognitive bottleneck is the semantic processing of each individual item, the influence of the environment's spatial structure becomes negligible. The cognitive load required to identify \textit{what} each object is overshadows any potential benefits or hindrances of \textit{where} it is located. This finding strongly suggests that for complex, real-world visual enumeration, top-down task demands are the paramount driver of performance, largely overriding the influence of bottom-up environmental features.

\subsubsection{Gaze Distribution}

\begin{figure}[ht!]
    \centering
    \includegraphics[width=\textwidth]{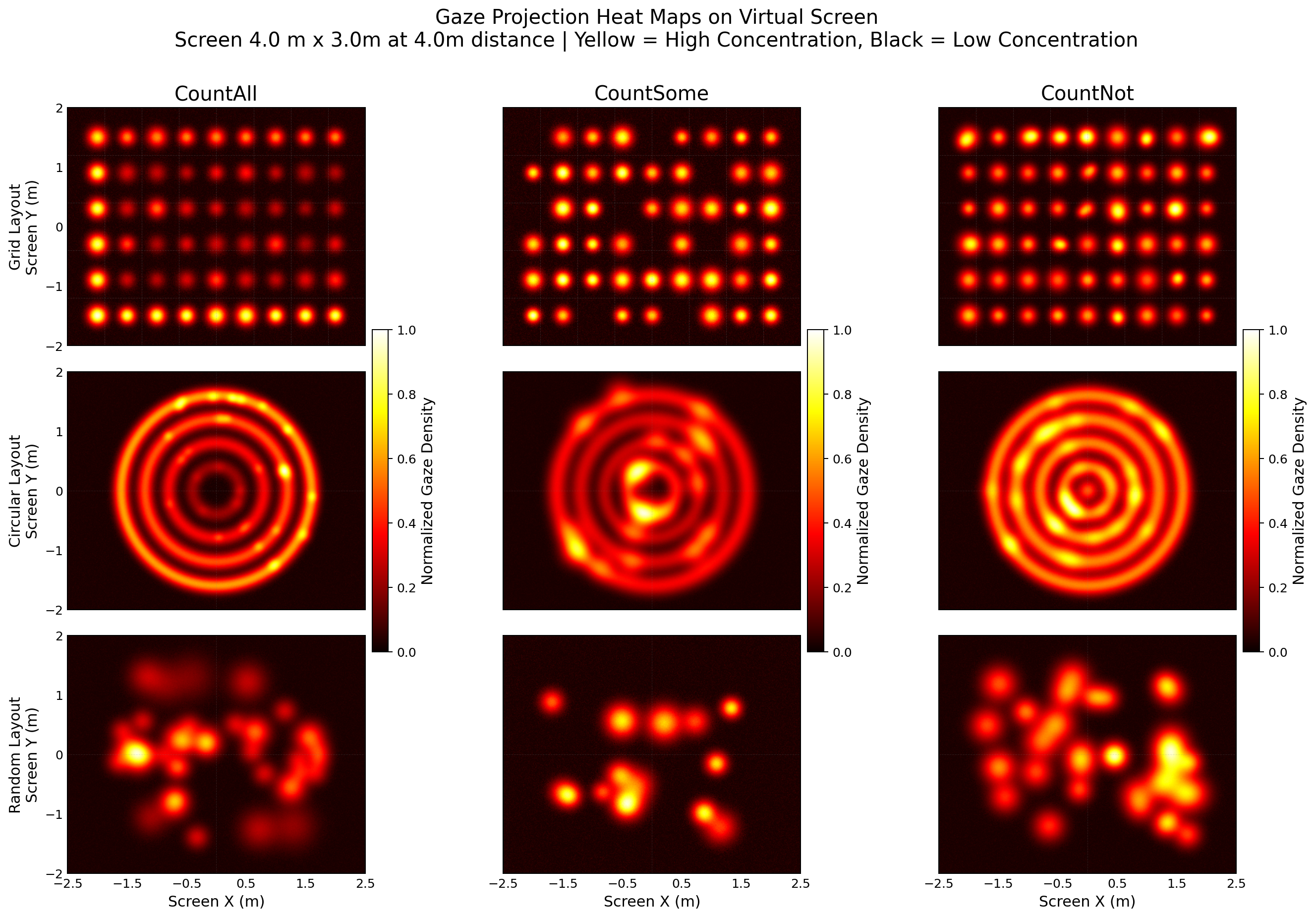}
    \caption{Aggregated gaze density heatmaps for Phase 2. Despite substantially longer task times and lower accuracy, the overall spatial strategies remained consistent with Phase 1: 'CountSome' elicited a sparse, selective gaze pattern, while 'CountAll' and 'CountNot' prompted an exhaustive search.}
    \label{fig:gaze_heatmaps_p2}
\end{figure}

\paragraph{Observations}
The analysis of aggregated gaze data for Phase 2 revealed that the fundamental visual search strategies employed by participants were remarkably consistent with those observed in Phase 1, despite the significant differences in stimulus complexity and performance. The gaze density heatmaps, presented in Figure~\ref{fig:gaze_heatmaps_p2}, show that the high-level patterns of attentional allocation were preserved.

As in the first phase, the task intent was the primary driver of search selectivity. Across all three layouts, the 'CountSome' instruction consistently produced sparse heatmaps, with high-intensity gaze concentrated on a subset of target item locations while distractors were largely ignored. In contrast, both the 'CountAll' and 'CountNot' instructions resulted in exhaustive gaze coverage, with nearly every item location showing a bright hotspot. This indicates that participants visually scanned almost the entire scene for these two conditions. The underlying spatial layout—'Grid', 'Circular', or 'Random'—continued to provide the structural scaffold for these search patterns, shaping the overall form of the gaze distribution.

\paragraph{Discussion}
The gaze distribution results from Phase 2 provide further compelling support for \textbf{Hypothesis H2}, reinforcing the findings from the initial phase. The clear dichotomy between the selective gaze patterns of 'CountSome' and the exhaustive patterns of 'CountAll' and 'CountNot' demonstrates that top-down task goals consistently and powerfully control the spatial allocation of attention.

The most crucial insight, however, comes from comparing these heatmaps to those from Phase 1. The striking similarity in the high-level visual strategies is a critical finding. It suggests that the dramatic increase in task completion time and the sharp decline in accuracy observed in Phase 2 were not caused by a breakdown in participants' ability to execute a search plan. They did not get lost in the layouts, nor did their fundamental scanning strategy change. Instead, the performance cost must be attributed to the cognitive operations occurring \textit{at each point of fixation}.

This analysis allows us to decouple the spatial component of the search (the "where") from the cognitive processing component (the "what"). The heatmaps show us \textit{where} participants allocated their cognitive resources, and this spatial allocation strategy remained robust. The performance data (time and accuracy) tells us \textit{how much effort} was required at each of these locations. The additional time and errors in Phase 2 stemmed from the prolonged and more difficult process of semantic recognition and categorisation that occurred at each bright spot on the heatmap. This powerfully reinforces our earlier conclusion: the primary cognitive bottleneck in the enumeration of complex, real-world scenes is not the visual search for items, but the semantic processing required to understand them.

\subsubsection{Cognitive Load}

\paragraph{Observations}
The shift from abstract shapes to real-world objects in Phase 2 resulted in a substantial and consistent increase in the self-reported cognitive load across all conditions. Participants' subjective ratings of mental demand, captured on a 1-5 scale, clearly indicated that they found the task of enumerating complex images to be significantly more challenging than counting simple shapes.

\begin{figure}[ht!]
    \centering
    \begin{subfigure}[b]{0.49\textwidth}
        \centering
        \includegraphics[width=\textwidth]{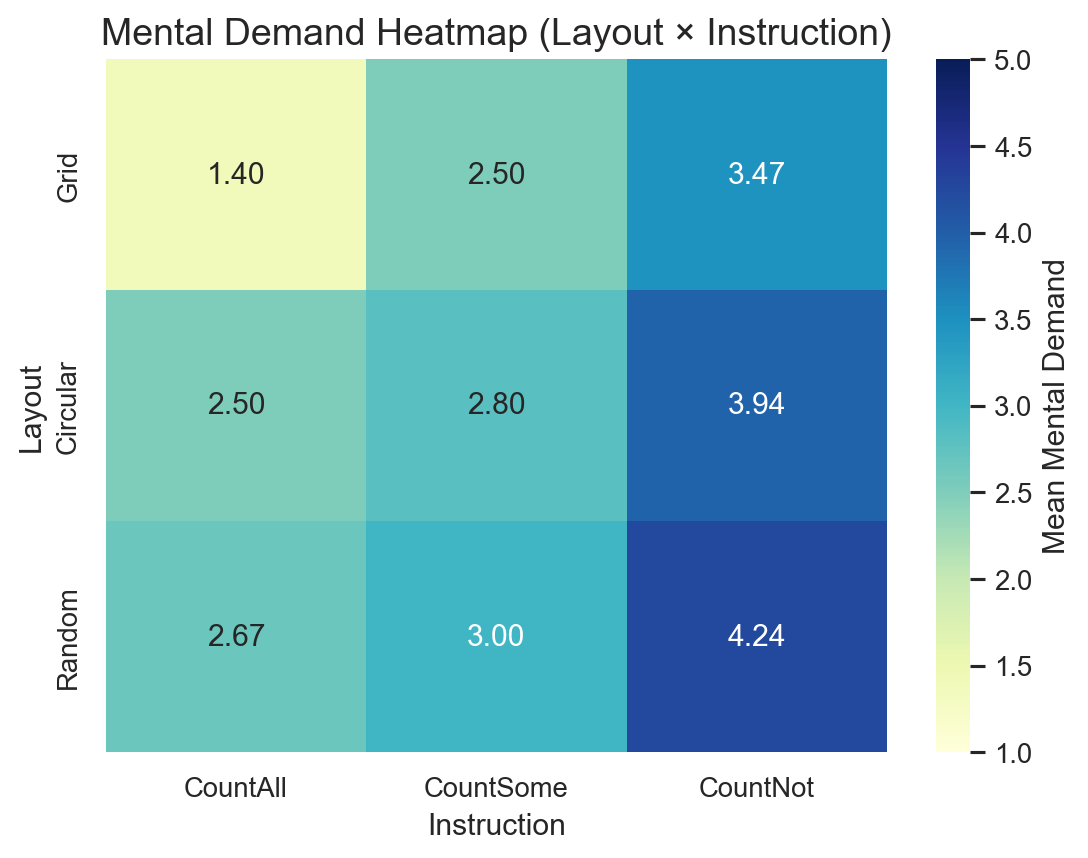}
        \caption{Heatmap of mean self-reported mental demand for Phase 2. The ratings are substantially higher than in Phase 1, with the 'Random-CountNot' condition approaching the scale's maximum value.}
        \label{fig:mental_demand_heatmap_p2}
    \end{subfigure}
    \hfill % Adds horizontal space between the subfigures
    \begin{subfigure}[b]{0.49\textwidth}
        \centering
        \includegraphics[width=\textwidth]{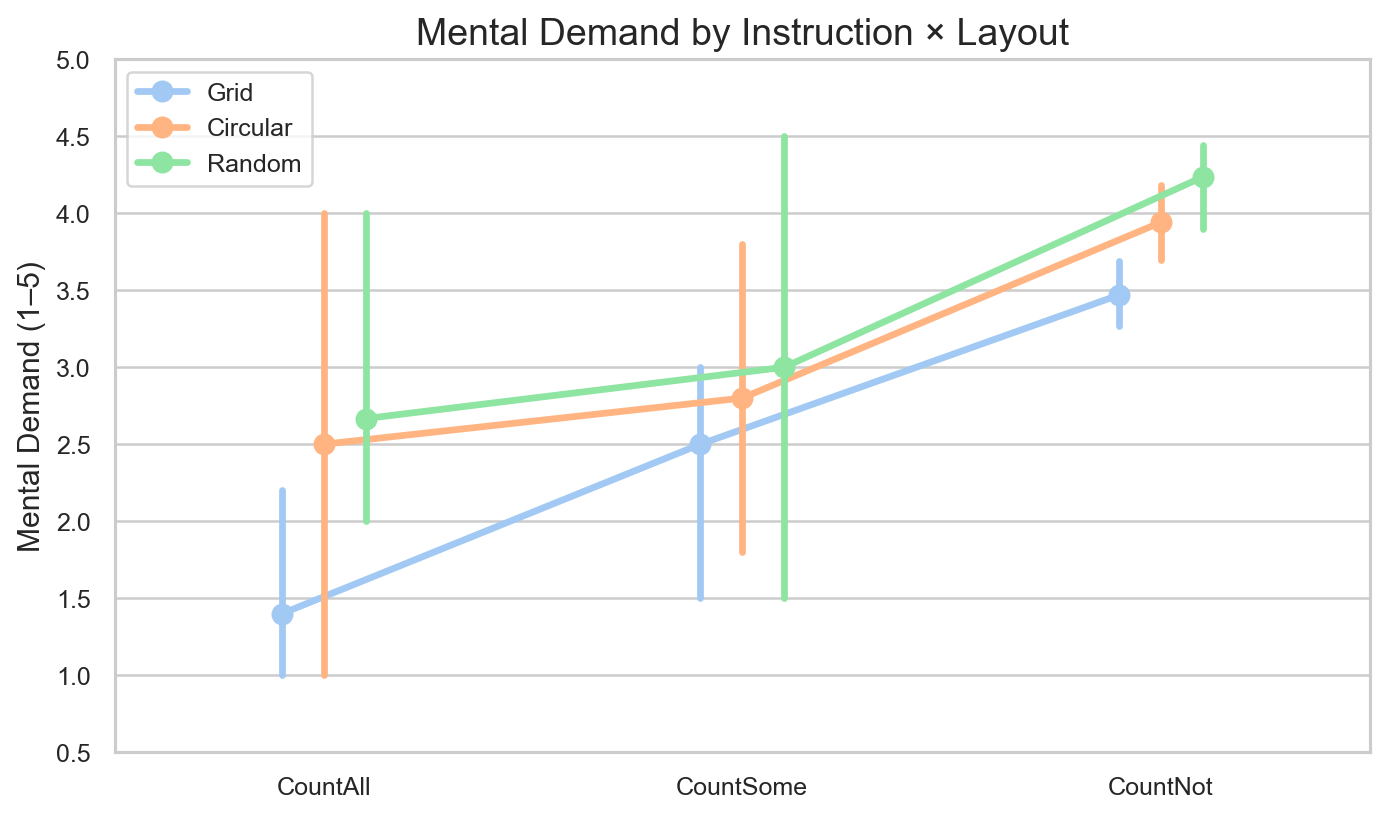}
        \caption{Interaction line plot for mental demand in Phase 2. The steep upward slope of all lines demonstrates the powerful effect of instruction type on cognitive load when dealing with complex, real-world images.}
        \label{fig:mental_demand_lines_p2}
    \end{subfigure}
    \caption{Visualizations of self-reported cognitive load in Phase 2, showing the interaction between instruction type and spatial layout as (a) a heatmap and (b) a line plot.}
    \label{fig:phase2_cognitiveload_visuals}
\end{figure}

The primary driver of this increased cognitive load was the task intent. The interaction plot in Figure~\ref{fig:mental_demand_lines_p2} shows a steep and consistent rise in perceived mental demand from the 'CountAll' instruction, to 'CountSome', and finally to 'CountNot'. This trend was more pronounced than in Phase 1, with the 'Random-CountNot' condition receiving an average rating of 4.24, approaching the maximum of the scale. The heatmap in Figure~\ref{fig:mental_demand_heatmap_p2} quantifies this, showing that every condition involving selective counting was rated as more demanding than any of the 'CountAll' conditions.

The effect of spatial layout on cognitive load was also more distinct in this phase. The 'Grid' layout was consistently rated as the least mentally demanding, while the 'Random' layout was rated as the most demanding across all three instruction types. The line plot (Figure~\ref{fig:mental_demand_lines_p2}) clearly shows the line for 'Grid' positioned below the other two, while the line for 'Random' is positioned at the top, indicating a clear hierarchy of perceived difficulty based on the environment's structure. The error bars are again notably large, particularly for the more difficult conditions, signifying considerable individual differences in how taxing the tasks were perceived to be.

% Note: Ensure you have \usepackage{graphicx} and \usepackage{subcaption} in your document preamble.

% \begin{figure}[ht!]
%     \centering
%     \includegraphics[width=0.8\textwidth]{Figures/Phase2/CognitiveLoad/02_heatmap_mental_p2.png}
%     \caption{Heatmap of mean self-reported mental demand for Phase 2. The ratings are substantially higher than in Phase 1, with the 'Random-CountNot' condition approaching the scale's maximum value.}
%     \label{fig:mental_demand_heatmap_p2}
% \end{figure}

% \begin{figure}[ht!]
%     \centering
%     \includegraphics[width=\textwidth]{Figures/Phase2/CognitiveLoad/01_interaction_lines_mental_p2.png}
%     \caption{Interaction line plot for mental demand in Phase 2. The steep upward slope of all lines demonstrates the powerful effect of instruction type on cognitive load when dealing with complex, real-world images.}
%     \label{fig:mental_demand_lines_p2}
% \end{figure}

\paragraph{Discussion}
The cognitive load ratings from Phase 2 provide powerful validation for our hypotheses and offer a clear window into the subjective experience of the participants. The results provide the strongest support yet for \textbf{H1(c)} and align perfectly with the trends predicted in \textbf{H2}. The substantially higher ratings across the board confirm that the cognitive work required to enumerate real-world images is far greater than for simple shapes.

This increase in perceived mental effort is a direct consequence of the added layer of semantic processing. Participants' ratings accurately reflected the increased difficulty shown in the objective performance data (longer times and lower accuracy). The mental demand of the 'CountNot' task, for example, was exceptionally high because it combined the difficult cognitive act of inhibiting a complex semantic category (e.g., "vehicles") with the already demanding task of searching and counting in a cluttered scene. This aligns perfectly with the objective data, where this condition produced the longest completion times and the most errors.

Furthermore, the amplification of the layout's effect on perceived load is also telling. When the per-item processing is simple (as in Phase 1), the extra effort of navigating a 'Random' layout is noticeable but modest. However, when the per-item processing is itself highly demanding (as in Phase 2), the additional burden of a difficult search becomes much more salient. The total cognitive load—the sum of the effort for searching and the effort for identifying—becomes punishingly high in the 'Random-CountNot' condition. In conclusion, the subjective load data from Phase 2 confirms that participants were keenly aware of the increased task difficulty. Their ratings provide convergent evidence that the primary challenge in real-world enumeration is not just the act of counting itself, but the cognitively expensive process of semantic identification and filtering that must precede it.

\subsubsection{Recall Performance}

\paragraph{Observations}
The most striking finding in Phase 2 was a dramatic overall decline in participants' ability to recall items from memory compared to Phase 1. The introduction of complex, real-world objects appeared to severely tax the cognitive resources available for memory encoding.

\begin{figure}[ht!]
    \centering
    \begin{subfigure}[b]{0.49\textwidth}
        \centering
        \includegraphics[width=\textwidth]{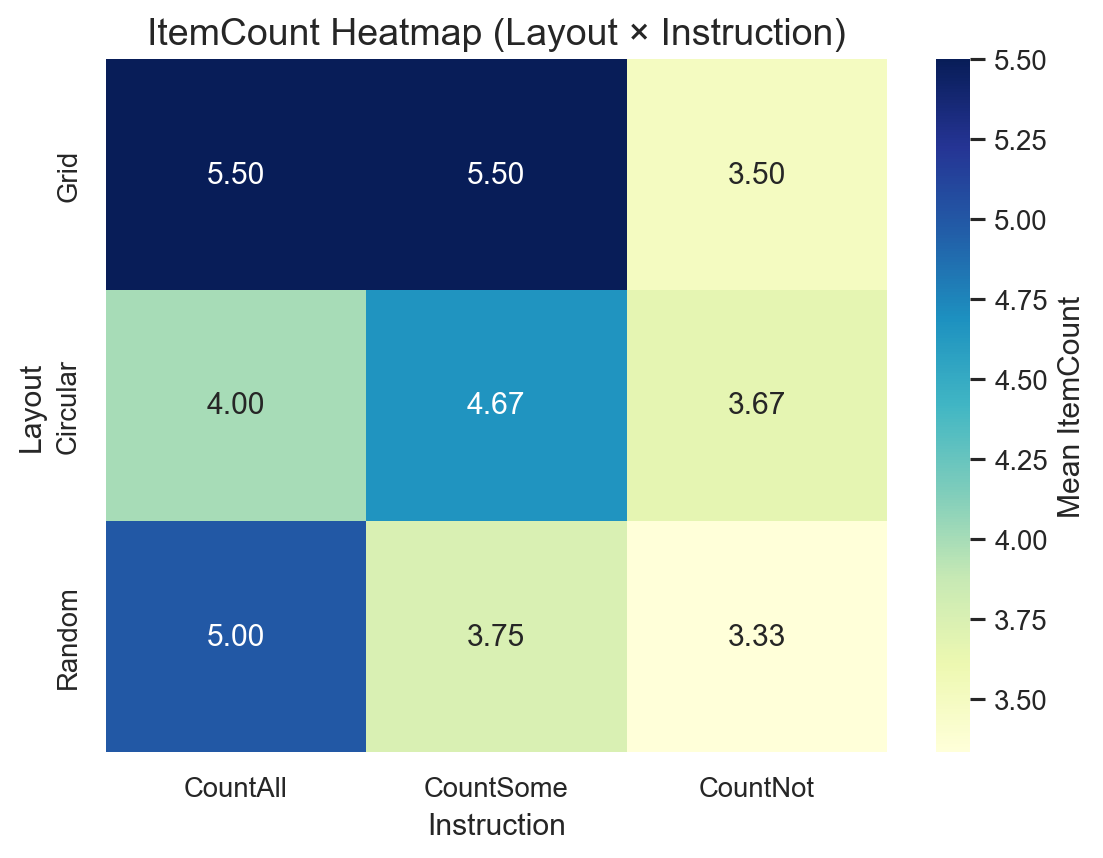}
        \caption{Heatmap of mean item recall for Phase 2. The overall number of items remembered is significantly lower than in Phase 1, with the best condition averaging only 5.50 recalled items.}
        \label{fig:recall_heatmap_p2}
    \end{subfigure}
    \hfill % Adds horizontal space between the subfigures
    \begin{subfigure}[b]{0.49\textwidth}
        \centering
        \includegraphics[width=\textwidth]{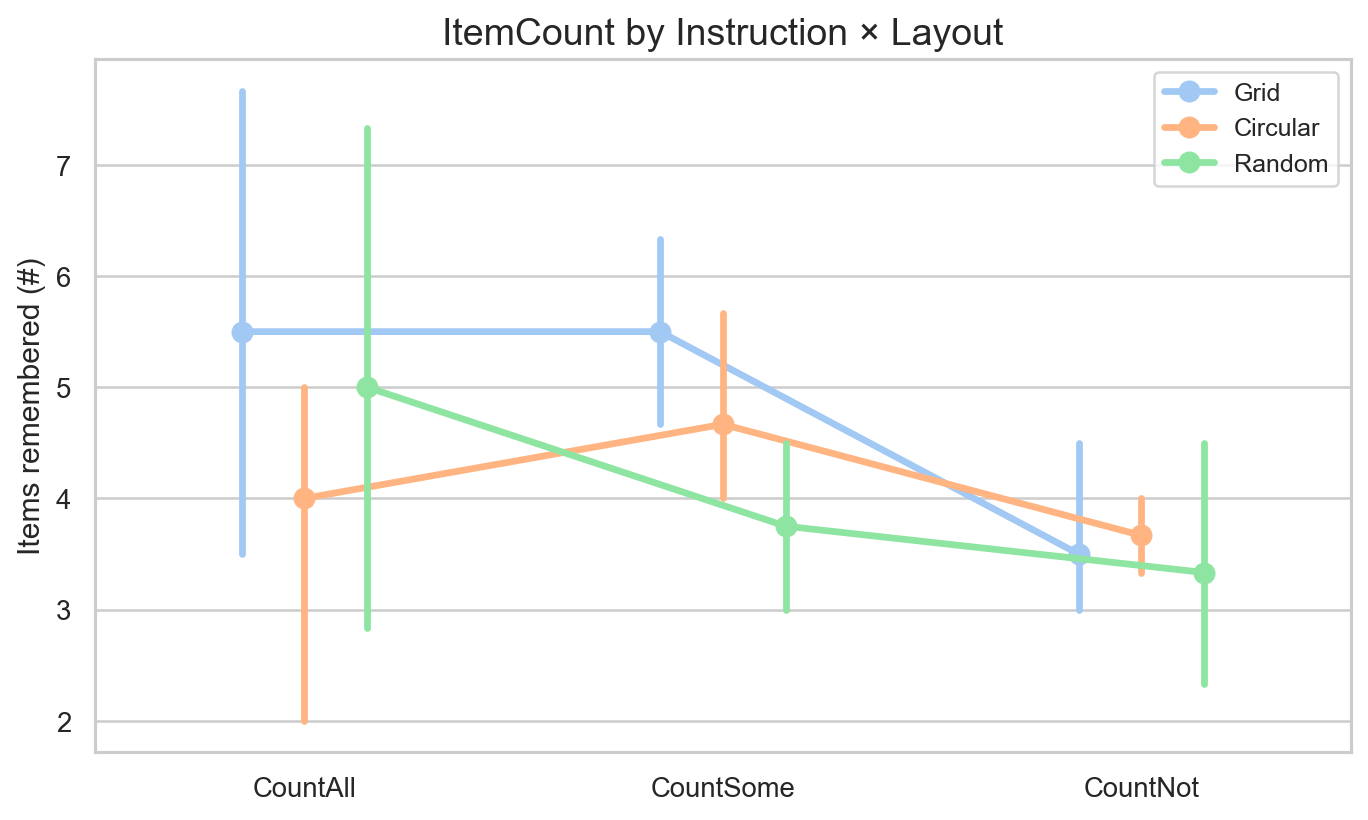}
        \caption{Interaction line plot for recall performance in Phase 2. The plot shows a general decline in recall with more complex instructions, but the effect is weaker and less consistent than in Phase 1, with a notable flattening of the trend for the 'Circular' layout.}
        \label{fig:recall_lines_p2}
    \end{subfigure}
    \caption{Visualizations of memory recall performance in Phase 2, showing the interaction between instruction type and spatial layout as (a) a heatmap and (b) a line plot.}
    \label{fig:phase2_recall_visuals}
\end{figure}

As shown in the heatmap in Figure~\ref{fig:recall_heatmap_p2}, the mean number of items recalled was substantially lower across all conditions. The best-performing condition, 'Grid-CountAll', yielded an average recall of only 5.50 items, a sharp drop from the 13.60 items remembered in the same condition in Phase 1. The interaction plot in Figure~\ref{fig:recall_lines_p2} illustrates that while a general downward trend in recall from 'CountAll' to 'CountNot' was present for the 'Grid' and 'Random' layouts, the effect was much weaker than in Phase 1, and the 'Circular' layout showed an almost flat trend. The 'Grid' layout generally led to the highest recall, while the 'Random' layout was often associated with the poorest performance, but the patterns were less consistent than in the first phase. The large error bars also indicate continued high variability in memory performance among participants.

% Note: Ensure you have \usepackage{graphicx} and \usepackage{subcaption} in your document preamble.

% \begin{figure}[ht!]
%     \centering
%     \includegraphics[width=0.8\textwidth]{Figures/Phase2/MemoryRecall/01_heatmap_itemcount_p2.png}
%     \caption{Heatmap of mean item recall for Phase 2. The overall number of items remembered is significantly lower than in Phase 1, with the best condition averaging only 5.50 recalled items.}
%     \label{fig:recall_heatmap_p2}
% \end{figure}

% \begin{figure}[ht!]
%     \centering
%     \includegraphics[width=\textwidth]{Figures/Phase2/MemoryRecall/02_interaction_itemcount_p2.png}
%     \caption{Interaction line plot for recall performance in Phase 2. The plot shows a general decline in recall with more complex instructions, but the effect is weaker and less consistent than in Phase 1, with a notable flattening of the trend for the 'Circular' layout.}
%     \label{fig:recall_lines_p2}
% \end{figure}

\paragraph{Discussion}
The memory recall results from Phase 2 provide a powerful and nuanced validation of \textbf{Hypothesis H3}. While the specific performance trends predicted in \textbf{H3(a)} were weaker and less consistent, the data strongly supports the underlying principle of a link between attention and memory by demonstrating the profound impact of cognitive overload.

The massive overall drop in recall performance suggests that the high cognitive load of the primary task—identifying, categorising, and filtering semantically rich images—consumed the vast majority of participants' available cognitive resources. This left very few resources available for the secondary, and perhaps optional, task of encoding the items into memory for later recall. This phenomenon can be understood as a form of \textbf{cognitive overload leading to memory suppression}. The mental effort required to simply execute the counting task correctly was so substantial that it actively interfered with the consolidation of memories, even for items that were directly attended to.

This overload effect explains why the clear and distinct trends observed in Phase 1 became muddled in Phase 2. The system was operating closer to a floor effect; when overall recall is poor, the differentiating effects of layout or instruction type naturally become smaller and less reliable. The cognitive system was so taxed by the primary task that the subtler benefits of a 'Grid' layout or the added burden of a 'CountNot' instruction had a diminished impact on the already-impaired memory encoding process. In conclusion, the recall data from Phase 2 offers a stark demonstration of the limits of our cognitive capacity. It powerfully supports the core idea of \textbf{H3} by showing that attention and memory are inextricably linked through a shared pool of limited cognitive resources. When a task completely saturates these resources, our ability to remember what we have seen is severely compromised.

\subsection{Summary of Findings}
The comprehensive analysis of both experimental phases reveals a cohesive narrative about the cognitive architecture of visual enumeration in complex, large-field environments. By systematically comparing the enumeration of simple abstract shapes with that of complex real-world objects, we have been able to deconstruct the influence of task intent, environmental structure, and stimulus complexity on performance, attention, and memory.

The most unequivocal finding across both phases was the powerful and dominant role of \textbf{task intent}. The instructions given to participants proved to be the single most significant factor determining the difficulty and outcome of the enumeration task. The baseline 'CountAll' instruction consistently served as the least demanding condition, yielding the fastest performance, highest accuracy, and lowest subjective cognitive load. The introduction of any form of selective filtering—be it inclusion ('CountSome') or exclusion ('CountNot')—imposed a substantial and statistically significant cognitive cost. The 'CountNot' task, in particular, consistently emerged as the most challenging, demonstrating that the cognitive act of inhibiting a category while counting all others is a profoundly difficult and error-prone process.

In contrast, the influence of \textbf{spatial layout} was found to be secondary and modulatory. While there was a consistent trend indicating that structured layouts ('Grid', 'Circular') facilitated more efficient performance than the unstructured 'Random' layout, this effect was not statistically significant in either phase. This suggests that while environmental structure can provide a helpful scaffold for visual search, its benefits are easily overshadowed by the top-down cognitive demands of the task itself. The visual search strategies, as revealed by the gaze heatmaps, were indeed shaped by the layout, but the ultimate performance was dictated more by the complexity of the decisions required at each fixation point.

The central story of this research lies in the \textbf{amplification effect of stimulus complexity}. The transition from Phase 1 (abstract shapes) to Phase 2 (real-world objects) acted as a magnifying glass, amplifying the cognitive costs observed in the baseline condition. The introduction of a mandatory semantic processing step—the act of recognizing and categorising each object—dramatically increased the time and effort required for the selective tasks. The penalty for filtering items became much more severe, and the drop in accuracy became catastrophic for some participants, because the per-item decision was no longer a simple feature check but a complex semantic judgment. Critically, the high-level gaze strategies remained remarkably consistent across both phases. This allows us to infer that the performance degradation in Phase 2 was not due to a failure in visual search, but was almost entirely attributable to the increased cognitive load of processing the \textit{meaning} of each item.

Finally, the results paint a clear picture of the relationship between \textbf{cognitive load and memory}. Memory recall was found to be inversely proportional to the difficulty of the primary counting task. In Phase 1, where the cognitive load was manageable, attention patterns directly predicted memory: exhaustive attention led to better recall. In Phase 2, however, the immense cognitive load required for the enumeration task appeared to cause a general suppression of memory encoding. Overall recall performance collapsed, suggesting that the cognitive resources were so completely consumed by the primary task of semantic filtering and counting that very few were left available for the secondary process of forming lasting memories. This provides a stark demonstration that memory is not a passive recording of what we see, but an active and resource-intensive process that falters when our cognitive system is overloaded.

\subsection{Limitations and Future Work}
While the present study provides a detailed and multi-faceted investigation into visual enumeration in immersive environments, it is important to acknowledge its limitations, which in turn illuminate promising directions for future research.

First, the characterisation of stimuli in our two experimental phases, while systematic, was not exhaustive. In Phase 1, we deliberately simplified the abstract stimuli to vary only in \textbf{colour} to establish a clean cognitive baseline. This did not, however, account for other fundamental visual features such as \textbf{shape}, \textbf{size}, or \textbf{orientation}. Real-world abstract enumeration often involves filtering by these characteristics. Future work could systematically deconstruct these features. For instance, a study could directly compare the cognitive load and search strategies for colour-based filtering versus shape-based filtering to understand if different feature dimensions impose unique cognitive costs.

Second, the entire experiment was conducted within a \textbf{static environment}. The objects, while presented in a 3D space, were stationary. This contrasts with many real-world scenarios where enumeration must be performed on dynamic scenes containing moving objects, such as counting vehicles in traffic or animals in a field. The introduction of motion would represent a significant step towards greater ecological validity. Future research could investigate how object motion—including speed, trajectory, and temporary occlusion—affects attentional allocation, counting accuracy, and the ability to maintain a running tally in working memory.

Third, although Phase 2 utilised real-world object images, they were presented as discrete, decontextualized "posters" floating in a neutral virtual space. This is a step beyond a 2D screen but still falls short of a fully \textbf{integrated and naturalistic scene}, where objects are embedded within a coherent environment (e.g., furniture within a room, tools on a workbench). Future studies could leverage the full potential of VR by creating complete 3D environments. Tasking participants with enumerating objects in a richly rendered virtual kitchen or a cluttered garage would introduce more realistic challenges, such as partial occlusion, complex lighting, and background distractions, providing deeper insights into real-world enumeration.

Fourth, our participant sample was drawn from a university population, which typically represents a narrow demographic in terms of age and technological familiarity. The cognitive strategies and performance levels observed in this study may not be generalisable to other populations, such as \textbf{children or older adults}, whose attentional capacities and working memory functions may differ. Future work should aim to recruit a more diverse sample to investigate how enumeration strategies and the susceptibility to cognitive overload evolve across the human lifespan.

Finally, the counting task itself, while immersive, remained an explicit and somewhat abstract instruction-based activity. In the real world, enumeration is often embedded within a broader, more \textbf{meaningful goal-oriented context}. Future research could enhance participant engagement and motivation by embedding the enumeration task within a gamified or purpose-driven scenario. For example, framing the task as an inventory audit for a virtual store, a biodiversity survey in a simulated ecosystem, or a puzzle-solving challenge could elicit different levels of engagement and potentially alter the cognitive strategies participants employ.

%============================================================
\section{Conclusion}
This research was motivated by a fundamental question: how do humans perform the seemingly simple act of visual counting in complex, realistic environments that extend beyond the confines of a traditional computer screen? To answer this, we developed an immersive virtual reality system, RAVEN-VR, and conducted a comprehensive two-phase experiment. We systematically manipulated the core factors that define real-world enumeration complexity: the intrinsic \textbf{characteristics} of the items (from simple abstract shapes to complex real-world objects), the cognitive \textbf{intent} of the task (ranging from exhaustive to highly selective counting), and the spatial \textbf{distribution} of the items (from structured to unstructured layouts). By measuring performance, eye movements, and memory, we sought to build a more holistic and ecologically valid model of the human counting process.

Our findings reveal a clear hierarchy of cognitive challenges. The most powerful determinant of performance was not the structure of the environment, but the cognitive demands of the task's intent. The requirement to filter items based on their semantic meaning imposed a severe and immediate cost, dramatically increasing task time while reducing accuracy and elevating mental load. We discovered that this cost is massively amplified by stimulus complexity; the cognitive effort required to identify and categorise real-world objects was the primary bottleneck, capable of overwhelming the cognitive system. This intense focus on "what" to count overshadowed the influence of "where" the items were located, rendering the benefits of a structured layout secondary.

Ultimately, this research demonstrates that real-world enumeration is a delicate and resource-limited interplay between external visual search and internal cognitive processing. The most crucial takeaway is that the high cognitive load demanded by a primary task, such as identifying and filtering complex objects, comes at a direct and significant cost to our secondary cognitive functions, most notably our ability to form lasting memories of what we have seen. This work shows that counting is not merely a matter of finding objects in space, but a cognitively demanding task of assigning meaning to them—an effort where the struggle to understand can eclipse the ability to remember.

%==============================================================
\backmatter
\bmhead{Acknowledgements}

The authors would like to extend their sincere gratitude to all the individuals who willingly participated in this study. Their time, effort, and engagement were invaluable to the completion of this research.

\section*{Declarations}

\begin{itemize}
\item \textbf{Funding} \\
Not applicable. The authors did not receive any specific funding for this work.

\item \textbf{Conflict of interest/Competing interests} \\
The authors declare that they have no conflict of interest or competing interests.

\item \textbf{Ethics approval and consent to participate} \\
The experimental protocol was reviewed and approved by the Institutional Human Ethics Committee (IHEC). All procedures performed in studies involving human participants were in accordance with the ethical standards of the institutional and/or national research committee and with the 1964 Helsinki declaration and its later amendments or comparable ethical standards. Written informed consent was obtained from all individual participants included in the study.

\item \textbf{Consent for publication} \\
Participants were informed that the data collected would be anonymized and used for research purposes, including publication in academic journals and presentations at scientific conferences. Consent for the publication of anonymized data was included as part of the informed consent process.

\item \textbf{Data availability} \\
The datasets and code collected and/or generated during the current study are available and will be provided by the corresponding author upon reasonable request.

\item \textbf{Materials availability} \\
The RAVEN-VR application developed for this study is licensed for research and educational purposes.

% \item \textbf{Author contribution} \\
% Sankar Balasubramanian: Conceptualization, Methodology, Software, Investigation, Data Curation, Formal Analysis, Writing – Original Draft. Devottama Sen: Conceptualization, Data Analysis, Writing – Review \& Editing. All authors read and approved the final manuscript.
\end{itemize}

\begin{appendices}

\section{Post-Trial Questionnaire Design}\label{secA1}

A detailed questionnaire was administered via Google Forms after each experimental trial to capture a comprehensive range of subjective data regarding participant experience, strategy, and memory. The forms utilized conditional branching, presenting participants with a unique sequence of questions based on the specific \textbf{Layout Type} and \textbf{Count Instruction Type} they had just experienced. This appendix outlines the core components and logic of the questionnaires used for Phase 1 (Abstract Shapes) and Phase 2 (Real-World Objects).

\subsection{Core Components of the Questionnaire}
Both questionnaires were structured around several key modules designed to probe different aspects of the cognitive experience.

\subsubsection{Participant and Trial Demographics}
Each session began with collecting basic participant information (Name, Age, Gender). The first question of each post-trial block required the selection of the Layout Type, which then directed the participant to a layout-specific block of questions.

\subsubsection{Layout-Specific Probes}
This module assessed the participants' strategy and perceived difficulty associated with the specific spatial arrangement they encountered. Key questions included:
\begin{itemize}
    \item \textbf{Strategy Assessment:} Participants were asked about their counting strategy via multiple-choice questions tailored to each layout. For the \textbf{Grid}, options included "I counted row by row" and "I counted column by column". For the \textbf{Circular} layout, options included "I counted ring by ring (Clockwise)" and "I started counting from the centre outward". For the \textbf{Random} layout, options included "I tried grouping images visually" and "I counted from left to right or vice versa".
    \item \textbf{Perceived Difficulty:} A five-point Likert scale question, "How easy was it to count the items in the [...] Layout?", was used to gauge subjective difficulty for each condition.
    \item \textbf{Layout-Specific Challenges:} Questions targeted potential issues unique to each layout, such as whether the scattered placement in the \textbf{Random} layout made them "more likely to miss or double-count" items.
\end{itemize}

\subsubsection{Instruction-Specific Memory Probes}
After the layout section, participants selected the Count Instruction Type, which branched to a block of questions designed to measure memory for both task-relevant (target) and task-irrelevant (distractor) items.
\begin{itemize}
    \item \textbf{Phase 1 (Abstract Shapes):} The questions probed memory for specific colours. For example, after a `CountAll` trial, participants were asked, "...how many RED dots do you remember seeing?". After a `Count If Not Contains` trial, they were asked how many of the ignored coloured dots they "ended up noticing anyway".
    \item \textbf{Phase 2 (Real-World Objects):} The memory probes were more semantic. After a `Count If It Contains [Animals]` trial, participants were asked to recall the number of specific types of animals they saw (e.g., ZEBRA, DONKEY, DEER). Similarly, after a `Count If Not Contains [Vehicles]` trial, they were probed on their memory for specific ignored vehicles (e.g., TRAIN, SHIP, CYCLE).
\end{itemize}

\subsubsection{General Cognitive and Self-Report Assessment}
The final module of the questionnaire contained a series of general questions that were consistent across all conditions. These aimed to capture a holistic view of the participant's cognitive state.
\begin{itemize}
    \item \textbf{Memory Recall:} An open-ended question asked participants to "Name at least 3 items you distinctly remember seeing" or "Specify at least 3 dots (by colour or position) which you recall".
    \item \textbf{Cognitive Load:} This was measured using multiple questions, including a direct five-point rating of "How mentally demanding did you find this layout?" and questions about mental effort, such as "How would you describe your mental effort during the task?" with options like "Manageable - I had to stay focused" and "Overwhelming - I lost track often".
    \item \textbf{Confidence:} Participants rated their confidence in their counting accuracy on a five-point scale.
    \item \textbf{Influencing Factors:} A checklist question asked participants to identify factors that "significantly influenced your ability to count accurately," with options including "Layout Complexity," "Item Similarity," and "Mental Fatigue".
    \item \textbf{Open-Ended Feedback:} The questionnaire concluded with fields for participants to describe any specific difficulties or provide general comments.
\end{itemize}

The complete questionnaires used in this study can be accessed via the following links:
\begin{itemize}
    \item \textbf{Phase 1 (Dots):} \url{https://forms.gle/niTs8fiFYWw8Pgb6A}
    \item \textbf{Phase 2 (Objects):} \url{https://forms.gle/fcCpnZ6qXwrYoJ9x8}
\end{itemize}

\end{appendices}

\bibliography{2_bibliography}% common bib file

\begin{thebibliography}{}
\renewcommand{\doi}[1]{\url{https://doi.org/#1}}
\bibcommenthead

\bibitem [\protect \citeauthoryear {%
Ansari%
, Lyons%
, van Eimeren%
\BCBL {}\ \BBA {} Xu%
}{%
Ansari%
\ \protect \BOthers {.}}{%
{\protect \APACyear {2007}}%
}]{%
Ansari2007-mm}
\APACinsertmetastar {%
Ansari2007-mm}%
\begin{APACrefauthors}%
Ansari, D.%
, Lyons, I.M.%
, van Eimeren, L.%
\BCBL {} Xu, F.%
\end{APACrefauthors}%
\unskip\
\newblock
\APACrefYearMonthDay{2007}{{\APACmonth{11}}}{}.
\newblock
{\BBOQ}\APACrefatitle {Linking visual attention and number processing in the brain: the role of the temporo-parietal junction in small and large symbolic and nonsymbolic number comparison} {Linking visual attention and number processing in the brain: the role of the temporo-parietal junction in small and large symbolic and nonsymbolic number comparison}.{\BBCQ}
\newblock
\APACjournalVolNumPages{J. Cogn. Neurosci.}{19}{11}{1845--1853,}
\newblock

\newblock

\PrintBackRefs{\CurrentBib}

\bibitem [\protect \citeauthoryear {%
Chen%
, Paul%
\BCBL {}\ \BBA {} Reeve%
}{%
Chen%
\ \protect \BOthers {.}}{%
{\protect \APACyear {2022}}%
}]{%
Chen2022}
\APACinsertmetastar {%
Chen2022}%
\begin{APACrefauthors}%
Chen, J.%
, Paul, J.M.%
\BCBL {} Reeve, R.%
\end{APACrefauthors}%
\unskip\
\newblock
\APACrefYearMonthDay{2022}{}{}.
\newblock
{\BBOQ}\APACrefatitle {Manipulation of attention affects subitizing performance: A systematic review and meta-analysis} {Manipulation of attention affects subitizing performance: A systematic review and meta-analysis}.{\BBCQ}
\newblock
\APACjournalVolNumPages{Neuroscience and Biobehavioral Reviews}{139}{}{104753,}
\newblock

\newblock

\PrintBackRefs{\CurrentBib}

\bibitem [\protect \citeauthoryear {%
Chiossi%
\ \protect \BOthers {.}}{%
Chiossi%
\ \protect \BOthers {.}}{%
{\protect \APACyear {2024}}%
}]{%
Chiossi2024}
\APACinsertmetastar {%
Chiossi2024}%
\begin{APACrefauthors}%
Chiossi, F.%
, Gruenefeld, U.%
, Hou, B.J.%
, Newn, J.%
, Ou, C.%
, Liao, R.%
\BDBL {}Mayer, S.%
\end{APACrefauthors}%
\unskip\
\newblock
\APACrefYearMonthDay{2024}{}{}.
\newblock
{\BBOQ}\APACrefatitle {Understanding the Impact of the Reality-Virtuality Continuum on Visual Search Using Fixation-Related Potentials and Eye Tracking Features} {Understanding the impact of the reality-virtuality continuum on visual search using fixation-related potentials and eye tracking features}.{\BBCQ}
\newblock
\APACjournalVolNumPages{Proc. ACM Hum.-Comput. Interact.}{8}{}{281:1--281:33,}
\newblock

\newblock

\PrintBackRefs{\CurrentBib}

\bibitem [\protect \citeauthoryear {%
Dehaene%
}{%
Dehaene%
}{%
{\protect \APACyear {1992}}%
}]{%
Dehaene1992-xz}
\APACinsertmetastar {%
Dehaene1992-xz}%
\begin{APACrefauthors}%
Dehaene, S.%
\end{APACrefauthors}%
\unskip\
\newblock
\APACrefYearMonthDay{1992}{{\APACmonth{08}}}{}.
\newblock
{\BBOQ}\APACrefatitle {Varieties of numerical abilities} {Varieties of numerical abilities}.{\BBCQ}
\newblock
\APACjournalVolNumPages{Cognition}{44}{1-2}{1--42,}
\newblock

\newblock

\PrintBackRefs{\CurrentBib}

\bibitem [\protect \citeauthoryear {%
Feigenson%
, Dehaene%
\BCBL {}\ \BBA {} Spelke%
}{%
Feigenson%
\ \protect \BOthers {.}}{%
{\protect \APACyear {2004}}%
}]{%
Feigenson2004-uh}
\APACinsertmetastar {%
Feigenson2004-uh}%
\begin{APACrefauthors}%
Feigenson, L.%
, Dehaene, S.%
\BCBL {} Spelke, E.%
\end{APACrefauthors}%
\unskip\
\newblock
\APACrefYearMonthDay{2004}{{\APACmonth{07}}}{}.
\newblock
{\BBOQ}\APACrefatitle {Core systems of number} {Core systems of number}.{\BBCQ}
\newblock
\APACjournalVolNumPages{Trends Cogn. Sci.}{8}{7}{307--314,}
\newblock

\newblock

\PrintBackRefs{\CurrentBib}

\bibitem [\protect \citeauthoryear {%
Herten%
, Otto%
\BCBL {}\ \BBA {} Wolf%
}{%
Herten%
\ \protect \BOthers {.}}{%
{\protect \APACyear {2017}}%
}]{%
Herten2017}
\APACinsertmetastar {%
Herten2017}%
\begin{APACrefauthors}%
Herten, N.%
, Otto, T.%
\BCBL {} Wolf, O.T.%
\end{APACrefauthors}%
\unskip\
\newblock
\APACrefYearMonthDay{2017}{}{}.
\newblock
{\BBOQ}\APACrefatitle {The role of eye fixation in memory enhancement under stress--An eye tracking study} {The role of eye fixation in memory enhancement under stress--an eye tracking study}.{\BBCQ}
\newblock
\APACjournalVolNumPages{Neurobiology of Learning and Memory}{140}{}{134--144,}
\newblock

\newblock

\PrintBackRefs{\CurrentBib}

\bibitem [\protect \citeauthoryear {%
Hesse%
}{%
Hesse%
}{%
{\protect \APACyear {2016}}%
}]{%
Hesse2016}
\APACinsertmetastar {%
Hesse2016}%
\begin{APACrefauthors}%
Hesse, P.N.%
\end{APACrefauthors}%
\unskip\
\newblock
\APACrefYear{2016}.
\unskip\
\newblock
\APACrefbtitle {An Eye on Numbers: The Processing of Numerical Information in the Context of Visual Perception} {An eye on numbers: The processing of numerical information in the context of visual perception}\ \APACtypeAddressSchool {\BUPhD}{}{}.
\unskip\
\newblock
\APACaddressSchool {}{Philipps-Universit{\"a}t Marburg}.
\PrintBackRefs{\CurrentBib}

\bibitem [\protect \citeauthoryear {%
Jevons%
}{%
Jevons%
}{%
{\protect \APACyear {1871}}%
}]{%
Jevons1871}
\APACinsertmetastar {%
Jevons1871}%
\begin{APACrefauthors}%
Jevons, W.S.%
\end{APACrefauthors}%
\unskip\
\newblock
\APACrefYearMonthDay{1871}{}{}.
\newblock
{\BBOQ}\APACrefatitle {The power of numerical discrimination} {The power of numerical discrimination}.{\BBCQ}
\newblock
\APACjournalVolNumPages{Nature}{3}{}{281--282,}
\newblock

\newblock

\PrintBackRefs{\CurrentBib}

\bibitem [\protect \citeauthoryear {%
Kaufman%
, Lord%
, Reese%
\BCBL {}\ \BBA {} Volkmann%
}{%
Kaufman%
\ \protect \BOthers {.}}{%
{\protect \APACyear {1949}}%
}]{%
Kaufman1949}
\APACinsertmetastar {%
Kaufman1949}%
\begin{APACrefauthors}%
Kaufman, E.L.%
, Lord, M.W.%
, Reese, T.W.%
\BCBL {} Volkmann, J.%
\end{APACrefauthors}%
\unskip\
\newblock
\APACrefYearMonthDay{1949}{}{}.
\newblock
{\BBOQ}\APACrefatitle {The discrimination of visual number} {The discrimination of visual number}.{\BBCQ}
\newblock
\APACjournalVolNumPages{American Journal of Psychology}{62}{}{498--525,}
\newblock

\newblock

\PrintBackRefs{\CurrentBib}

\bibitem [\protect \citeauthoryear {%
Logan%
\ \BBA {} Zbrodoff%
}{%
Logan%
\ \BBA {} Zbrodoff%
}{%
{\protect \APACyear {2010}}%
}]{%
Logan2010}
\APACinsertmetastar {%
Logan2010}%
\begin{APACrefauthors}%
Logan, G.D.%
\BCBT {}\ \BBA {} Zbrodoff, N.J.%
\end{APACrefauthors}%
\unskip\
\newblock
\APACrefYearMonthDay{2010}{}{}.
\newblock
{\BBOQ}\APACrefatitle {Eye movements in enumeration} {Eye movements in enumeration}.{\BBCQ}
\newblock
\APACjournalVolNumPages{Psychonomic Bulletin \& Review}{17}{}{671--678,}
\newblock

\newblock

\PrintBackRefs{\CurrentBib}

\bibitem [\protect \citeauthoryear {%
Mazza%
\ \BBA {} Caramazza%
}{%
Mazza%
\ \BBA {} Caramazza%
}{%
{\protect \APACyear {2012}}%
}]{%
Mazza2012}
\APACinsertmetastar {%
Mazza2012}%
\begin{APACrefauthors}%
Mazza, V.%
\BCBT {}\ \BBA {} Caramazza, A.%
\end{APACrefauthors}%
\unskip\
\newblock
\APACrefYearMonthDay{2012}{}{}.
\newblock
{\BBOQ}\APACrefatitle {Perceptual Grouping and Visual Enumeration} {Perceptual grouping and visual enumeration}.{\BBCQ}
\newblock
\APACjournalVolNumPages{PLoS ONE}{7}{}{e50862,}
\newblock

\newblock

\PrintBackRefs{\CurrentBib}

\bibitem [\protect \citeauthoryear {%
Mock%
, Nuerk%
\BCBL {}\ \BBA {} Moeller%
}{%
Mock%
\ \protect \BOthers {.}}{%
{\protect \APACyear {2016}}%
}]{%
Mock2016}
\APACinsertmetastar {%
Mock2016}%
\begin{APACrefauthors}%
Mock, A.%
, Nuerk, H\BHBI C.%
\BCBL {} Moeller, K.%
\end{APACrefauthors}%
\unskip\
\newblock
\APACrefYearMonthDay{2016}{}{}.
\newblock
{\BBOQ}\APACrefatitle {The processing of numerical information in the context of visual perception} {The processing of numerical information in the context of visual perception}.{\BBCQ}
\newblock
\APACjournalVolNumPages{Psychological Research}{80}{}{334--359,}
\newblock

\newblock

\PrintBackRefs{\CurrentBib}

\bibitem [\protect \citeauthoryear {%
Paul%
, Reeve%
\BCBL {}\ \BBA {} Forte%
}{%
Paul%
\ \protect \BOthers {.}}{%
{\protect \APACyear {2020}}%
}]{%
Paul2020}
\APACinsertmetastar {%
Paul2020}%
\begin{APACrefauthors}%
Paul, J.M.%
, Reeve, R.%
\BCBL {} Forte, J.D.%
\end{APACrefauthors}%
\unskip\
\newblock
\APACrefYearMonthDay{2020}{}{}.
\newblock
{\BBOQ}\APACrefatitle {Enumeration strategy differences revealed by eye tracking} {Enumeration strategy differences revealed by eye tracking}.{\BBCQ}
\newblock
\APACjournalVolNumPages{Cognition}{198}{}{104204,}
\newblock

\newblock

\PrintBackRefs{\CurrentBib}

\bibitem [\protect \citeauthoryear {%
Piazza%
, Mechelli%
, Butterworth%
\BCBL {}\ \BBA {} Price%
}{%
Piazza%
\ \protect \BOthers {.}}{%
{\protect \APACyear {2002}}%
}]{%
Piazza2002-lu}
\APACinsertmetastar {%
Piazza2002-lu}%
\begin{APACrefauthors}%
Piazza, M.%
, Mechelli, A.%
, Butterworth, B.%
\BCBL {} Price, C.J.%
\end{APACrefauthors}%
\unskip\
\newblock
\APACrefYearMonthDay{2002}{{\APACmonth{02}}}{}.
\newblock
{\BBOQ}\APACrefatitle {Are subitizing and counting implemented as separate or functionally overlapping processes?} {Are subitizing and counting implemented as separate or functionally overlapping processes?}{\BBCQ}
\newblock
\APACjournalVolNumPages{Neuroimage}{15}{2}{435--446,}
\newblock

\newblock

\PrintBackRefs{\CurrentBib}

\bibitem [\protect \citeauthoryear {%
Pylyshyn%
}{%
Pylyshyn%
}{%
{\protect \APACyear {1989}}%
}]{%
Pylyshyn1989}
\APACinsertmetastar {%
Pylyshyn1989}%
\begin{APACrefauthors}%
Pylyshyn, Z.W.%
\end{APACrefauthors}%
\unskip\
\newblock
\APACrefYearMonthDay{1989}{}{}.
\newblock
{\BBOQ}\APACrefatitle {FINST spatial-index model} {Finst spatial-index model}.{\BBCQ}
\newblock
\APACjournalVolNumPages{Cognition}{}{}{,}
\newblock

\newblock

\PrintBackRefs{\CurrentBib}

\bibitem [\protect \citeauthoryear {%
Pylyshyn%
}{%
Pylyshyn%
}{%
{\protect \APACyear {2001}}%
}]{%
Pylyshyn2001}
\APACinsertmetastar {%
Pylyshyn2001}%
\begin{APACrefauthors}%
Pylyshyn, Z.W.%
\end{APACrefauthors}%
\unskip\
\newblock
\APACrefYearMonthDay{2001}{}{}.
\newblock
{\BBOQ}\APACrefatitle {Visual indexes, preconceptual objects, and situated vision} {Visual indexes, preconceptual objects, and situated vision}.{\BBCQ}
\newblock
\APACjournalVolNumPages{Cognition}{80}{}{127--158,}
\newblock

\newblock

\PrintBackRefs{\CurrentBib}

\bibitem [\protect \citeauthoryear {%
Railo%
, Koivisto%
, Revonsuo%
\BCBL {}\ \BBA {} Hannula%
}{%
Railo%
\ \protect \BOthers {.}}{%
{\protect \APACyear {2008}}%
}]{%
Railo2008}
\APACinsertmetastar {%
Railo2008}%
\begin{APACrefauthors}%
Railo, H.%
, Koivisto, M.%
, Revonsuo, A.%
\BCBL {} Hannula, M.M.%
\end{APACrefauthors}%
\unskip\
\newblock
\APACrefYearMonthDay{2008}{}{}.
\newblock
{\BBOQ}\APACrefatitle {The role of attention in subitizing} {The role of attention in subitizing}.{\BBCQ}
\newblock
\APACjournalVolNumPages{Cognition}{107}{}{82--104,}
\newblock

\newblock

\PrintBackRefs{\CurrentBib}

\bibitem [\protect \citeauthoryear {%
Schleifer%
\ \BBA {} Landerl%
}{%
Schleifer%
\ \BBA {} Landerl%
}{%
{\protect \APACyear {2011}}%
}]{%
Schleifer2011}
\APACinsertmetastar {%
Schleifer2011}%
\begin{APACrefauthors}%
Schleifer, P.%
\BCBT {}\ \BBA {} Landerl, K.%
\end{APACrefauthors}%
\unskip\
\newblock
\APACrefYearMonthDay{2011}{}{}.
\newblock
{\BBOQ}\APACrefatitle {Subitizing and counting in typical and atypical development} {Subitizing and counting in typical and atypical development}.{\BBCQ}
\newblock
\APACjournalVolNumPages{Developmental Science}{14}{}{280--292,}
\newblock

\newblock

\PrintBackRefs{\CurrentBib}

\bibitem [\protect \citeauthoryear {%
Simon%
\ \BBA {} Vaishnavi%
}{%
Simon%
\ \BBA {} Vaishnavi%
}{%
{\protect \APACyear {1996}}%
}]{%
Simon1996}
\APACinsertmetastar {%
Simon1996}%
\begin{APACrefauthors}%
Simon, T.J.%
\BCBT {}\ \BBA {} Vaishnavi, S.%
\end{APACrefauthors}%
\unskip\
\newblock
\APACrefYearMonthDay{1996}{}{}.
\newblock
{\BBOQ}\APACrefatitle {Subitizing and counting depend on different attentional mechanisms: Evidence from visual enumeration in afterimages} {Subitizing and counting depend on different attentional mechanisms: Evidence from visual enumeration in afterimages}.{\BBCQ}
\newblock
\APACjournalVolNumPages{Perception \& Psychophysics}{58}{}{915--926,}
\newblock

\newblock

\PrintBackRefs{\CurrentBib}

\bibitem [\protect \citeauthoryear {%
Sophian%
\ \BBA {} Crosby%
}{%
Sophian%
\ \BBA {} Crosby%
}{%
{\protect \APACyear {2007}}%
}]{%
Sophian}
\APACinsertmetastar {%
Sophian}%
\begin{APACrefauthors}%
Sophian, C.%
\BCBT {}\ \BBA {} Crosby, M.E.%
\end{APACrefauthors}%
\unskip\
\newblock
\APACrefYearMonthDay{2007}{}{}.
\newblock
{\BBOQ}\APACrefatitle {What Eye Fixation Patterns Tell Us About Subitizing} {What eye fixation patterns tell us about subitizing}.{\BBCQ}
\newblock
\APACjournalVolNumPages{Developmental Review}{27}{4}{455--473,}
\newblock

\newblock

\PrintBackRefs{\CurrentBib}

\bibitem [\protect \citeauthoryear {%
Tai%
}{%
Tai%
}{%
{\protect \APACyear {2004}}%
}]{%
Tai2004}
\APACinsertmetastar {%
Tai2004}%
\begin{APACrefauthors}%
Tai, Y\BHBI C.%
\end{APACrefauthors}%
\unskip\
\newblock
\APACrefYear{2004}.
\unskip\
\newblock
\APACrefbtitle {Visual enumeration: Subitizing and its nature} {Visual enumeration: Subitizing and its nature}\ \APACtypeAddressSchool {\BUPhD}{}{}.
\unskip\
\newblock
\APACaddressSchool {}{University of Illinois at Urbana-Champaign}.
\PrintBackRefs{\CurrentBib}

\bibitem [\protect \citeauthoryear {%
Trick%
\ \BBA {} Pylyshyn%
}{%
Trick%
\ \BBA {} Pylyshyn%
}{%
{\protect \APACyear {1993}}%
}]{%
Trick1993}
\APACinsertmetastar {%
Trick1993}%
\begin{APACrefauthors}%
Trick, L.M.%
\BCBT {}\ \BBA {} Pylyshyn, Z.W.%
\end{APACrefauthors}%
\unskip\
\newblock
\APACrefYearMonthDay{1993}{}{}.
\newblock
{\BBOQ}\APACrefatitle {What enumeration studies can show us about spatial attention} {What enumeration studies can show us about spatial attention}.{\BBCQ}
\newblock
\APACjournalVolNumPages{Journal of Experimental Psychology: Human Perception and Performance}{19}{}{331--351,}
\newblock

\newblock

\PrintBackRefs{\CurrentBib}

\bibitem [\protect \citeauthoryear {%
Trick%
\ \BBA {} Pylyshyn%
}{%
Trick%
\ \BBA {} Pylyshyn%
}{%
{\protect \APACyear {1994}}%
}]{%
Trick1994}
\APACinsertmetastar {%
Trick1994}%
\begin{APACrefauthors}%
Trick, L.M.%
\BCBT {}\ \BBA {} Pylyshyn, Z.W.%
\end{APACrefauthors}%
\unskip\
\newblock
\APACrefYearMonthDay{1994}{{\APACmonth{01}}}{}.
\newblock
{\BBOQ}\APACrefatitle {Why are small and large numbers enumerated differently? A limited-capacity preattentive stage in vision} {Why are small and large numbers enumerated differently? a limited-capacity preattentive stage in vision}.{\BBCQ}
\newblock
\APACjournalVolNumPages{Psychol. Rev.}{101}{1}{80--102,}
\newblock

\newblock

\PrintBackRefs{\CurrentBib}

\bibitem [\protect \citeauthoryear {%
Watson%
, Maylor%
\BCBL {}\ \BBA {} Bruce%
}{%
Watson%
\ \protect \BOthers {.}}{%
{\protect \APACyear {2007}}%
}]{%
Watson2007}
\APACinsertmetastar {%
Watson2007}%
\begin{APACrefauthors}%
Watson, D.G.%
, Maylor, E.A.%
\BCBL {} Bruce, L.A.M.%
\end{APACrefauthors}%
\unskip\
\newblock
\APACrefYearMonthDay{2007}{}{}.
\newblock
{\BBOQ}\APACrefatitle {Saccadic eye movements are not necessary for efficient subitizing} {Saccadic eye movements are not necessary for efficient subitizing}.{\BBCQ}
\newblock
\APACjournalVolNumPages{Perception \& Psychophysics}{69}{}{1118--1129,}
\newblock

\newblock

\PrintBackRefs{\CurrentBib}

\bibitem [\protect \citeauthoryear {%
Wilder%
, Kowler%
, Schnitzer%
, Gersch%
\BCBL {}\ \BBA {} Dosher%
}{%
Wilder%
\ \protect \BOthers {.}}{%
{\protect \APACyear {2009}}%
}]{%
Wilder2009}
\APACinsertmetastar {%
Wilder2009}%
\begin{APACrefauthors}%
Wilder, J.D.%
, Kowler, E.%
, Schnitzer, B.S.%
, Gersch, T.M.%
\BCBL {} Dosher, B.A.%
\end{APACrefauthors}%
\unskip\
\newblock
\APACrefYearMonthDay{2009}{}{}.
\newblock
{\BBOQ}\APACrefatitle {Attention during active visual tasks: Counting, pointing, or simply looking} {Attention during active visual tasks: Counting, pointing, or simply looking}.{\BBCQ}
\newblock
\APACjournalVolNumPages{Vision Research}{49}{}{1017--1031,}
\newblock

\newblock

\PrintBackRefs{\CurrentBib}

\end{thebibliography}
%% if required, the content of the .bbl file can be included here once the bbl is generated

\end{document}